\documentclass[journal,doublecolumn]{IEEEtran}

\usepackage[mathscr]{eucal}
\usepackage[cmex10]{amsmath}
\usepackage{epsfig,epsf,psfrag}
\usepackage{amssymb,amsmath,amsthm,amsfonts,latexsym}
\usepackage{amsmath,graphicx,bm,xcolor,url,overpic}
\usepackage{subfig} 
\usepackage{fixltx2e}
\usepackage{array}
\usepackage{verbatim}
\usepackage{bm}
\usepackage{algorithmic}
\usepackage{algorithm}
\usepackage{verbatim}
\usepackage{textcomp}
\usepackage{mathrsfs}
\usepackage{epstopdf}

\usepackage{mathtools, nccmath, relsize}

\newcommand{\openone}{\leavevmode\hbox{\small1\normalsize\kern-.33em1}}

\catcode`~=11 \def\UrlSpecials{\do\~{\kern -.15em\lower .7ex\hbox{~}\kern .04em}} \catcode`~=13 

\allowdisplaybreaks[1]

\newcommand{\nn}{\nonumber}

\newcommand{\calA}{\mathcal{A}}
\newcommand{\calB}{\mathcal{B}}

\newcommand{\calD}{\mathcal{D}}
\newcommand{\calE}{\mathcal{E}}
\newcommand{\calF}{\mathcal{F}}

\newcommand{\calP}{\mathcal{P}}

\newcommand{\calT}{\mathcal{T}}

\newcommand{\calX}{\mathcal{X}}

\newcommand{\calZ}{\mathcal{Z}}

\newcommand{\bP}{\mathbf{P}}

\newcommand{\bQ}{\mathbf{Q}}

\newcommand{\bT}{\mathbf{T}}

\newcommand{\bx}{\mathbf{x}}

\newcommand{\by}{\mathbf{y}}

\newcommand{\rmc}{\mathrm{c}}

\newcommand{\rme}{\mathrm{e}}

\newcommand{\rmG}{\mathrm{G}}

\newcommand{\rmL}{\mathrm{L}}

\newcommand{\rmR}{\mathrm{R}}

\newcommand{\bbP}{\mathbb{P}}

\newcommand{\bbR}{\mathbb{R}}

\newcommand{\bbZ}{\mathbb{Z}}

\DeclareMathAlphabet{\mathbsf}{OT1}{cmss}{bx}{n}
\DeclareMathAlphabet{\mathssf}{OT1}{cmss}{m}{sl}

\DeclareSymbolFont{bsfletters}{OT1}{cmss}{bx}{n}  
\DeclareSymbolFont{ssfletters}{OT1}{cmss}{m}{n}
\DeclareMathSymbol{\bsfGamma}{0}{bsfletters}{'000}
\DeclareMathSymbol{\ssfGamma}{0}{ssfletters}{'000}
\DeclareMathSymbol{\bsfDelta}{0}{bsfletters}{'001}
\DeclareMathSymbol{\ssfDelta}{0}{ssfletters}{'001}
\DeclareMathSymbol{\bsfTheta}{0}{bsfletters}{'002}
\DeclareMathSymbol{\ssfTheta}{0}{ssfletters}{'002}
\DeclareMathSymbol{\bsfLambda}{0}{bsfletters}{'003}
\DeclareMathSymbol{\ssfLambda}{0}{ssfletters}{'003}
\DeclareMathSymbol{\bsfXi}{0}{bsfletters}{'004}
\DeclareMathSymbol{\ssfXi}{0}{ssfletters}{'004}
\DeclareMathSymbol{\bsfPi}{0}{bsfletters}{'005}
\DeclareMathSymbol{\ssfPi}{0}{ssfletters}{'005}
\DeclareMathSymbol{\bsfSigma}{0}{bsfletters}{'006}
\DeclareMathSymbol{\ssfSigma}{0}{ssfletters}{'006}
\DeclareMathSymbol{\bsfUpsilon}{0}{bsfletters}{'007}
\DeclareMathSymbol{\ssfUpsilon}{0}{ssfletters}{'007}
\DeclareMathSymbol{\bsfPhi}{0}{bsfletters}{'010}
\DeclareMathSymbol{\ssfPhi}{0}{ssfletters}{'010}
\DeclareMathSymbol{\bsfPsi}{0}{bsfletters}{'011}
\DeclareMathSymbol{\ssfPsi}{0}{ssfletters}{'011}
\DeclareMathSymbol{\bsfOmega}{0}{bsfletters}{'012}
\DeclareMathSymbol{\ssfOmega}{0}{ssfletters}{'012}

\newcommand{\hatC}{\hat{C}}

\newcommand{\tilP}{\tilde{P}}

\newcommand{\tilQ}{\tilde{Q}}

\newcommand{\tilT}{\tilde{T}}

\newcommand{\tilx}{\tilde{x}}
\newcommand{\tilX}{\tilde{X}}

\def\fndot{\, \cdot \,}

\DeclareMathOperator*{\argmax}{arg\,max}
\DeclareMathOperator*{\argmin}{arg\,min}

\DeclareMathOperator{\supp}{supp}

\usepackage{thmtools, thm-restate}
\newtheorem{theorem}{Theorem}
\newtheorem{lemma}{Lemma}

\newtheorem{corollary}[lemma]{Corollary}
\newtheorem{definition}{Definition}

\newtheorem{remark}{Remark}

\usepackage{epstopdf,bbm} 
 
\usepackage{amsmath}
\usepackage{amssymb,amsfonts}
\usepackage{graphicx}
\usepackage{amsthm}
\usepackage{bbm}
\usepackage{textcomp}
\usepackage{booktabs}
\usepackage{float}
\usepackage{caption}
\usepackage{extarrows}

\newcommand{\tilPjCleft}{\tilP_j^-}
\newcommand{\tilPjCright}{\tilP_j^+}
\newcommand{\LD}{\mathrm{L}}
\newcommand{\upchi}{\raisebox{2pt}{$\chi$}}

\usepackage{algorithm}           
\usepackage{algorithmic} 

\usepackage{enumerate}
\def\BibTeX{{\rm B\kern-.05em{\sc i\kern-.025em b}\kern-.08em
		T\kern-.1667em\lower.7ex\hbox{E}\kern-.125emX}}
	
\usepackage{cite}

\usepackage[level]{datetime}

\usepackage{ulem}
\usepackage{xifthen}
\usepackage{pdfpages}
\usepackage{transparent}

\DeclareSubrefFormat{parens}{#1(#2)}
\usepackage{color}

\allowdisplaybreaks

\makeatletter 
\newcommand\figcaption{\def\@captype{figure}\caption} 
\newcommand\tabcaption{\def\@captype{table}\caption} 
\makeatother

\usepackage[font=footnotesize,labelfont=bf]{caption}

\title{Optimal Change-Point Detection with Training Sequences in the Large and Moderate Deviations Regimes}

\author{Haiyun He, {\em Student Member, IEEE}, Qiaosheng Zhang, and Vincent~Y.~F.~Tan, {\em Senior Member, IEEE}  \thanks{This work is partially funded by a Singapore National Research Foundation (NRF) Fellowship (R-263-000-D02-281).} \thanks{This paper was  presented in part at the International Symposium on Information Theory and Its Applications (ISITA)  2020.} \thanks{H.~He, Q.~Zhang and V.~Y.~F.~Tan are with the Department of Electrical and Computer Engineering, National University of Singapore (Emails:  haiyun.he@u.nus.edu, elezqiao@nus.edu.sg, and  vtan@nus.edu.sg). } }

\begin{document}
	
\maketitle

\begin{abstract}
%
This paper investigates a novel offline change-point detection problem from an information-theoretic perspective. In contrast to most related works, we assume that the knowledge of the underlying pre- and post-change distributions are not known and can only be learned from the training sequences which are available. We further require the probability of the \emph{estimation error} to decay either exponentially or sub-exponentially fast (corresponding respectively to the large and moderate deviations regimes in information theory parlance). Based on the training sequences as well as the test sequence consisting of a single change-point, we design a change-point estimator and further show that this estimator is optimal by establishing matching (strong) converses. This leads to a full characterization of the optimal confidence width (i.e., half the width of the confidence interval within which the true change-point is located at with high probability) as a function of the undetected error, under both the large and moderate deviations regimes.
\end{abstract}

\begin{IEEEkeywords}
Change-point detection, Training sequences, Error exponent, Moderate deviations regime, Optimal confidence width
\end{IEEEkeywords}

\section{Introduction and Motivation}
The change-point detection (CPD) problem consists in finding changes in the underlying statistical model of data sequences that are modelled as time series. This problem has a plethora of applications in industrial systems \cite{marti2015anomaly}, medical diagnoses~\cite{zhang2010detecting}, environmental monitoring \cite{reeves2007review}, speech processing~\cite{harchaoui2009regularized}, finance,  economics, and so on~\cite{tartakovsky2014sequential}. The CPD problems can be divided into two main types: \emph{offline} CPD and \emph{online} CPD \cite{truong2019selective}; the latter is also known as \emph{sequential} CPD. This depends on whether the data sequence is fixed or obtained in a real-time setting. Offline CPD is a problem that is studied in, for example,  anomaly detection problems such as  detecting climate change based on existing and known statistics. Online CPD is studied in, for example,  signal segmentation problems such as  extracting information from streaming audio signals. 
In classical CPD problems, researchers either assume that the underlying distributions are known  \cite{page1954continuous,page1955test,moustakides1986optimal,siegmund2013sequential,basseville1993detection} or they assume that there is only access to a sequence of test data samples without knowledge of any underlying distributions \cite{mcgilchrist1975note,pettitt1979non,brodsky2013nonparametric}. However, in recent times, there has been a rising trend to adopt a modern statistical approach that makes use of \emph{training sequences} to augment classical problems, including learning of the change-point(s) in the test sequence. To the best of the authors' knowledge,  existing works have not considered this problem setup. 

Different from the online setting where   authors of~\cite{veeravalli2014quickest,unnikrishnan2011minimax,brodsky2013nonparametric}   are primarily concerned with the tradeoff between average detection delay and the probability of false alarm, in the offline setting, which is what this paper focuses on,   authors are instead concerned with the tradeoff between the  {\em confidence width} (i.e., half the width of the confidence interval within which the true change-point is located at with high probability) and the \emph{estimation (equivalently, undetected) error probability} (i.e., the probability that the distance between the estimated and true change-points is larger than the confidence width). They mainly derived the {\em order of growth} of the confidence width between the estimated and true change-points. For example, the results in \cite{baranowski2016narrowest,garreau2018consistent,wang2020univariate,padilla2019optimal,padilla2019change} indicated that if we require the error probability in detecting the (single) true change-point within an $n$-length data sequence to decay polynomially fast, there exists an algorithm or strategy such that the confidence width is  $O(\log n)$ as $n\to\infty$. The authors \cite{yao1989least,boysen2009consistencies,zou2014nonparametric,matteson2014nonparametric} as well as \cite[Chapter~1]{csorgo1997limit} showed that if we only require that the error probability vanishes (at an arbitrarily slow speed)  then there exists algorithms for which the confidence width is of order $O(1)$. 
	These works either do not provide closed-form expressions for the implied constants in these asymptotic results (in terms of the underlying distributions) or do not prove the converse of the derived convergence rate. This is, in part, what this paper sets out to do, albeit for a different regime of undetected error probability. In the spirit of classical {\em information-theoretic} problems \cite{Csi97}, we demand that the undetected error probability decays exponentially (or subexponentially) fast with a certain error exponent $\lambda>0$. Consequently, we show that the optimal confidence width scales as $\Theta(n)$ in the exponentially decaying case, and $\Theta(n^{1-t})$ (for $t \in (0,1/2)$) in the subexponentially decaying case. We also characterize the exact pre-constant in the $\Theta(\cdot)$ notation. En route to this endeavor, we show that the  pre-constant factor is optimal by deriving a strong converse in the sense of Wolfowitz~\cite{WolfowitzBook}. Our problem setting, at least from a theoretical perspective, should be of sufficient interest to the information theory community who seek to understand the fundamental performance (e.g. the error exponents) of practical problems.


In this paper, we study the offline CPD problem with a single change-point when the underlying distributions are unknown. We  are instead equipped with test and training sequences of the pre- and post-change distributions. This setting is applicable to many real-life situations. A simple example is detecting when a light switches between on or off in a room with a sensor. The training sequences can be obtained by taking a collection of samples when the light is on, and taking another collection of samples when the light is off. Based on the training sequences, we can then design an estimator in the sensor to detect when the status of the light changes.  
 In this paper, we assume that the test sequence is such that  its first and second parts are independently and identically distributed (i.i.d.) according to two unknown pre- and post-change distributions. Two labelled training sequences that are sampled i.i.d.\ from the pre- and post-change distributions    are  also provided to the learner. Our problem setup is simple but can give fundamental and insightful results. Our objective is to deduce  the fundamental performance limits of the CPD problem, i.e., the asymptotically optimal confidence width between the estimated and true change-points. 

\subsection{Main Contributions}
We formulate an offline single-CPD problem of finding the optimal confidence width between the estimated and true change-points without the knowledge of the underlying distributions but given training sequences and a test sequence with a single change-point; the extension  to multiple change-points is feasible.  In this setup, we assume that the length  of the test sequence is proportional to the length of training sequences. Our main contributions are as follows.

Firstly, inspired by  Gutman~\cite{gutman1989asymptotically}, we derive a type-based estimator (i.e., an estimator based on empirical distributions) under two different asymptotic regimes, namely the regimes in which the worst-case undetected error probability decays exponentially fast and sub-exponentially fast, which is more stringent than the requirements in existing works \cite{baranowski2016narrowest,garreau2018consistent,wang2020univariate,padilla2019optimal,padilla2019change}. These two regimes are respectively known as the {\em large} and {\em moderate deviations}~\cite{AW14} regimes. Moreover, we allow for the rejection option (also known as the erasure option) \cite{Forney68}, used when one is not sufficiently confident to decide which point is the true change-point. Since erasures are much less costly compared to undetected errors, we assume that the asymptotic worst-case erasure probability is upper bounded by a constant~$\epsilon\in [0,1)$.

Secondly, we derive the asymptotically optimal confidence width under both the large and moderate deviations regimes by showing that the above-mentioned type-based estimator is asymptotically optimal. By an appropriate use of the Berry-Esseen theorem, we also prove the strong converse, i.e., the optimal confidence width in both asymptotic regimes do not depend on the upper bound on the worst-case erasure probability $\epsilon$.

Finally, we study the dependence of the optimal confidence width on various parameters, e.g., the ratio between the lengths of training and test sequences, the distance between the two underlying pre- and post-change distributions and the exponent of the undetected error probability. In the moderate deviations regime, the optimal confidence width is a function of a \emph{symmetrized} version of the chi-square distance, which implies that this new divergence-like quantity finds an operational meaning based on the CPD problem. We note that  the usual asymmetric version of the chi-square distance has found operational interpretations in information-theoretic contexts such as covert communications~\cite{WWZ16, Bloch16, TL19}.

\subsection{Related Works}\label{Sec:related works}
The CPD problem has been studied by numerous authors for many years and it would be futile to survey all existing works.  Here, we attempt to highlight the most significant ones, partitioned according to whether the underlying distributions are known or not. We also survey some works on information-theoretic limits of  statistical classification. 

\paragraph{Known distributions}
In the early years in which the CPD problem was studied, authors commonly assumed that the distributions are known. In the offline scenario, Smith \cite{smith1975bayesian} adopted a Bayesian approach to infer the change-point in a sequence with known or partially-known underlying parameters. Some other probabilistic Bayesian approaches for offline CPD studied in \cite{chib1998estimation,barry1993bayesian} were based on retrospective segmentation strategies. Sen and Srivastava \cite{sen1975tests} adopted a non-Bayesian approach to detect a change in mean of normal distributions. Lavielle \cite{lavielle1999detection} studied the problem of detecting multiple changes by using maximum likelihood estimation.  In the online setting, the first works date back to the 1950s, done by Page \cite{page1954continuous,page1955test}. The author considered detecting a change of known  parameters and proposed the cumulative sums (CUSUM) algorithm that can be applied to both offline and online CPD problems. Later, Page's optimal stopping time result was further generalized in various directions by Moustakides~\cite{moustakides1986optimal}. 
Papers \cite{pollak1985optimal,lrden1971procedures} provided optimal results for online CPD problems in non-Bayesian settings.

\paragraph{Unknown distributions}  In the 1970s, there was an increasing number of works that began to study CPD problem when the distributions were unknown.  In the offline setting, 
 Pettitt \cite{pettitt1979non} introduced non-parametric techniques for the CPD problem and analyzed them based on different types of observations. Matteson and James~\cite{matteson2014nonparametric} derived a non-parametric method for detecting multiple change-points. Empirical likelihood tests were also proposed to detect single and multiple change-points in \cite{einmahl2003empirical,zou2014nonparametric}. Harchaoui and Capp{\'e} \cite{harchaoui2007retrospective} proposed a kernel-based retrospective algorithm to detect multiple change-points in a sequence. For the online case,   McGilchrist and Woodyer \cite{mcgilchrist1975note} proposed a distribution-free CUSUM algorithm.    Shiryaev \cite{shiryaev1963optimum} provided an optimal method in quickest CPD where the unknown parameter varies over time according to a Markov process. Lai \cite{lai1995sequential} proposed a nearly optimal window-limited generalized likelihood ratio test that can be implemented on parallel processors for distribution-free sequential CPD. 
 Xie, Wang, and Thompson \cite{xie2015sketching} studied detecting change-points in high-dimensional signal vectors using statistics based on the generalized likelihood ratio. 
 However, they are not aware of using training sequences to augment the performance of the CPD problem. Perhaps the work that is most related to the current one is that by Gruner and Johnson~\cite{gruner96}  in which the authors considered two sets of strings and attempted to detect the time when their distributions started to change from being the same to being different (also see point~3 in Sec.~\ref{sec:concl}). A type-based generalized likelihood ratio test similar to that of Gutman~\cite{gutman1989asymptotically} was used. However, while interesting numerical experiments on distributed detection problems are performed in \cite{gruner96}, no theoretical guarantees are provided.



\paragraph{Information-theoretic limits for classification} 
This paper is mainly inspired by existing works on classification with test and training sequences.
Gutman \cite{gutman1989asymptotically} was the first to propose asymptotically optimal type-based tests for the binary and multiple hypothesis testing problems. Merhav and Ziv \cite{merhav1991bayesian} derived a Bayesian approach for classification of Markov sources with unknown parameters.  Unnikrishnan \cite{unnikrishnan2014asymptotically} extended Gutman's results to matching multiple sequences to source sequences and proposed a symmetric type-based test compared to Gutman's test. Zhou, Tan, and Motani~\cite{zhou2019second} proved that Gutman's results are second-order asymptotically optimal. Recently, He, Zhou, and Tan \cite{he2020distributed} proposed an asymptotically optimal type-based test for the distributed detection problem with test and training sequences. Some of the proof techniques used in this paper leverage the techniques introduced in these papers.


\section{Problem Formulation}
In this section, we start by  introducing the offline single-CPD problem with test and training sequences.  Let $\calX$ be a finite set. We assume that there is a sequence of  observations $X^n=(X_1,\ldots,X_n)\in\calX^n$, in which there is a single change-point $C\in[1:n]$ (or $C\in[n]$ for short), where $[a:b]:=\{a,a+1,\ldots,b\}$ for any pair of integers $a\leq b$ and $[a]:=[1:a]$  \cite{el2011network}. Let $\alpha:=C/n$ denote the relative location of the change-point in the sequence.  Before the change occurs at time $C$,  the samples $X^C=(X_1,\ldots,X_C)$ are i.i.d.\ according to distribution $P_1\in\calP(\calX)$. After the change occurs, the samples $X_{C+1}^n=(X_{C+1},\ldots,X_n)$ are i.i.d.\ according to distribution $P_2\in\calP(\calX)$. We assume that $P_1\ne P_2$ and $\supp(P_1)=\supp(P_2)=|\calX|$. Unlike the traditional CPD problem \cite{page1954continuous, page1955test}, here we assume that $P_1, P_2$ are unknown. Instead, we have access to two sets of training sequences. The first (resp.\ second) training sequence $Y_1^N=(Y_{1,1},\ldots,Y_{1,N})\in\calX^N$ (resp.\ $Y_2^N=(Y_{2,1},\ldots,Y_{2,N})\in\calX^N$   is i.i.d.\ according to $P_1$  (resp.\ $P_2$). We assume $N=\lceil rn \rceil$ for some constant $r\in\bbR_+$ (that does not depend on $n$ or $N$); that is, $N$ is linear in $n$ and grows as $n$ grows. In the following, for the sake of brevity, we ignore  integer constraints on $n$ and $N$ and simply write $N=rn$. The parameter $r$ is the ratio between the lengths of the training sequences and test sequence. We expect that the performance  of CPD improves as $r$ grows as we have more information about the underlying distributions as $r\to\infty$. In addition, here we restrict ourselves to discrete finite alphabet $\calX$ since our proof techniques are based on the method of types~\cite{csiszar1998method}. In principle, it is also possible to generalize the method of types to continuous alphabets via discrete approximation (please see~\cite{csiszar1998method} for a detailed discussion); however, such generalization is non-trivial, and it might be of interests for future works.

Using $(Y_1^N,Y_2^N)$ and $X^n$,  an \emph{estimator} (known as a test in \cite{pettitt1979non}) $\gamma: \calX^{n+2N}\mapsto [n] \cup \{\rme\}$ is used either to declare that one of $n$ points of  the test sequence is the change-point or to declare that an ``erasure'' has occurred when we are not sufficiently  confident in declaring which of the points corresponds to the change-point.  
Given any true change-point $C\in[n]$, let us define the set of all test and training sequences $(x^n, y_1^N, y_2^N)$ that results in an \emph{undetected error} as 
\begin{align}
	\calE_C:=\big\{&(x^n,y_1^N,y_2^N)\in\calX^{n+2N}: \nn\\
	&\gamma(x^n,y_1^N,y_2^N)\notin [C\pm\Delta]\cup \{\rme\}\big\},
\end{align}
where $\Delta$ represents the \emph{confidence width} (i.e., half the width of the confidence interval within which the true change-point is with high probability)  and  $[a\pm b]:=[a-b:a+b]$. The set of test and training sequences that leads to an \emph{erasure event} is defined as
\begin{align}
	\calE_\rme:=\{(x^n,y_1^N,y_2^N)\in\calX^{n+2N}:\gamma(x^n,y_1^N,y_2^N)=\rme\}.
\end{align}

For any true change-point $C$ and any tuple of distributions $(P_1,P_2,\tilP_1,\tilP_2)\in\calP(\calX)^4$, we define $\bbP_C$ and $\tilde{\bbP}_C$ to be probability measures satisfying $\bbP_C\circ(X^C, X_{C+1}^n ,Y_1^N,Y_2^N)^{-1}=P_1^C\times P_2^{n-C}\times P_1^N\times P_2^N$ and $\tilde{\bbP}_C\circ(X^C, X_{C+1}^n ,Y_1^N,Y_2^N)^{-1}=\tilP_1^C\times \tilP_2^{n-C}\times \tilP_1^N\times \tilP_2^N$ respectively. That is, $P_1^C\times P_2^{n-C}\times P_1^N\times P_2^N$ is the pushforward measure of $\bbP_C$ corresponding to the random vector (a measurable function) $(X^C, X_{C+1}^n, Y_1^N, Y_2^N)$. Then the {\em undetected error probability} is defined as
\begin{align}
\bbP_{C}\{\calE_C\} :=\Pr\big\{ \gamma(X^n,Y_1^N,Y_2^N)\notin [C\pm\Delta]\cup \{\rme\} \big\},\label{eqn:undet}
\end{align} 
and the {\em erasure probability} is defined as 
\begin{align}
\bbP_C\{\calE_\rme\}:=\Pr\big\{ \gamma(X^n,Y_1^N,Y_2^N)=\rme \big\},\label{eqn:era}
\end{align}
 where in \eqref{eqn:undet} and \eqref{eqn:era}, $ (X^n,Y_1^N,Y_2^N) $ is distributed as 
$ X^C\sim P_1^C,X_{C+1}^n\sim P_2^{n-C}, Y_1^N\sim P_1^N$, and  $Y_2^N\sim P_2^N$.

We observe that the error and erasure probabilities depend on $n$ (the length of the test sequence), $N=rn$ (the length of the training sequences), $\Delta$ (the confidence width) as well as the generating distributions $P_1$ and $P_2$. We can further define the performance of any  estimator $\gamma$ as follows.
\begin{definition}[Good Estimator]\label{Def:estimator}
	For any $\Delta\in [0,n/2 )$, $r\in\bbR_+$, $\lambda\in\bbR_+$, $\epsilon\in[0,1)$, and $t\in[0,1/2)$, given any pair  of distributions $(P_1,P_2)\in\calP(\calX)^2$, an estimator $\gamma: \calX^{n+2N}\mapsto [n]\cup\{\rme\}$ is said to be {\em  $(n,\Delta,r,\lambda,\epsilon,t)$-good} if 
	\begin{align}
		\max_{C\in[n]}\bbP_C\{\calE_\rme\}&\leq \epsilon,   \label{Eq: estimator metric_erasure}
	\end{align}
	and for all $(\tilP_1, \tilP_2)\in\calP(\calX^2)$, 
	\begin{align}
		\max_{C\in[n]}\tilde{\bbP}_C\{\calE_C\} &\leq \exp(-n^{1-t}\lambda). \label{Eq: estimator metric}
	\end{align}
\end{definition}

When we design an estimator for the CPD problem, since we do not know the underlying distributions $(P_1,P_2)$ of the test and training sequences, we cannot design one with respect to a particular pair of distributions. Thus, we are interested in designing a \emph{universal} estimator such that the worst-case undetected error probability has good performance for all pairs of possible distributions $(\tilP_1,\tilP_2)$ and at the same time, constrain the erasure probability with respect to a particular pair of distributions $(P_1,P_2)$. That is, we can design an estimator based on $\lambda$ regardless of the underlying distributions. This is similar in spirit to Gutman's work~\cite{gutman1989asymptotically} as well as several other follow-up works~\cite{zhou2019second, he2020distributed}. 

Note that the   definition  above corresponds to different asymptotic regimes for the undetected error probability as we vary $t$. If $t=0$, the undetected error probability in \eqref{Eq: estimator metric}  is required to decay {\em exponentially fast}; this corresponds to the large deviations regime. If instead $t\in (0,1/2)$, the undetected error probability in \eqref{Eq: estimator metric} is required to decay {\em subexponentially fast} with rate $\exp(- \Theta(n^{1-t} ))$; this corresponds to the moderate deviations regime, which has been popularized in the information theory literature by Altu\u{g} and Wagner~\cite{AW14} among others. Although in standard moderate deviations regime, $t$ is allowed to vary between $(0,1)$, here we restrict $t\in (0,1/2)$ due to technical limitations in the proofs. Compared to existing works \cite{baranowski2016narrowest,garreau2018consistent,wang2020univariate,padilla2019optimal,padilla2019change,yao1989least,csorgo1997limit,boysen2009consistencies,zou2014nonparametric,matteson2014nonparametric} in which the upper bound in \eqref{Eq: estimator metric} is replaced by a polynomially decaying sequence or that the  undetected error probability is only required to vanish (at an arbitrarily slow speed), our formulation is one in which the constraint on the undetected error probability is more explicit and indeed far more stringent. However, since there exists a natural tradeoff between the optimal confidence width and the undetected error probability, consequently, it is also natural to expect that the optimal confidence width in our setting is larger. In fact, it is of order $\Theta(n)$ when $t=0$ (or $\Theta(n^{1-t/2})$ when $t\in(0,1/2)$) in \eqref{Eq: estimator metric}; see Definition~\ref{def:opt_res}. Our main contribution in this paper is to characterize the optimal confidence width {\em exactly  up to the  pre-constant term} in $\Theta(\cdot)$. 


In this CPD problem, our goal is to design an estimator that is $(n,\Delta,r,\lambda,\epsilon,t)$-good. Intuitively, increasing the confidence width $\Delta$ allows us to design estimators with smaller error probabilities, and vice versa. Thus,  keeping all other parameters $(r, \lambda, \epsilon,t)$ fixed, we are primarily interested in the smallest $\Delta$ such that \eqref{Eq: estimator metric_erasure} and \eqref{Eq: estimator metric} hold.
\begin{definition}[Optimal Normalized Confidence Width (NCW)]  \label{def:opt_res}
	Fix parameters $r\in\bbR_+, \lambda\in \bbR_+,\epsilon \in [0,1)$ and $t \in [0,1/2)$. We say that $\bar{\Delta}$ is a  {\em $(r,\lambda,\epsilon,t)$-achievable normalized confidence width (NCW)} if there exists a sequence of $(n,\Delta_n, r,\lambda, \epsilon_n,t)$-good estimators such that 
	\begin{align}\label{Eq:normalize Delta}
		\limsup_{n\to\infty}\epsilon_n \le\epsilon,\quad\mbox{and}\quad
		\limsup_{n\to\infty}\frac{\Delta_n}{n^{1- t/2} }&\le\bar{\Delta}.
	\end{align} 
	The {\em $t$-optimal NCW}
	\begin{align}
		\bar{\Delta}_t^* (r,\lambda,\epsilon) := \inf\left\{ \bar{\Delta} :  \bar{\Delta} \mbox{ is } (r,\lambda,\epsilon,t)\text{{-achievable}}  \right\}. 
	\end{align}
	If $t=0$ (corresponding to the large deviations regime), we simply write $\bar{\Delta}_t^*(\fndot)$ as $\bar{\Delta}^*(\fndot)$  (instead of $\bar{\Delta}_0^*(\fndot)$)   and call the corresponding quantity the {\em optimal NCW}.

\end{definition}

We now comment on why we restrict our attention to $\Delta < n/2$  in Definition \ref{Def:estimator}. Note that if $\Delta\geq n/2$, there exists a $C\in[n]$ (e.g. $C={n}/{2}$) such that $[C\pm\Delta ]^{\rmc}\cap [n]=\emptyset$. Then we can easily design a trivial estimator (e.g. $\gamma(X^n,Y_1^N,Y_2^N)={n}/{2}$) such that $\max_{C\in[n]}\bbP_C\{\calE_C\}=0$ and $\max_{C\in[n]}\bbP_C\{\calE_\rme\}=0$. Hence, if  $\Delta\geq  n/2$, the problem is vacuous and thus we only consider the case  in which $\Delta<  n/2$.


For any pair of distributions $(Q,\tilQ)\in\calP(\calX)^2$, let us define the chi-square distance between $Q$ and $\tilQ$ as
\begin{align}
	\upchi_2(Q \|\tilQ ) :=\sum_{x\in\calX}\frac{( Q(x)-\tilQ (x))^2}{\tilQ (x)}.
\end{align}
Note that $\upchi_2(Q\|\tilQ)<\infty$ for all pairs of $(Q, \tilQ)\in\calP(\calX)^2$ such that $\supp(Q)=\supp(\tilQ)=|\calX|$.

 For any $a\in\bbR_+$, let us further define the {\em generalized Jensen-Shannon divergence}
\begin{align}
\! \! 	\mathrm{GJS}(Q,\tilQ,a ) \!:=\! a D\bigg(Q\bigg\|\frac{a Q\! +\!  \tilQ}{a+1}\bigg)\! +\! D\bigg(\tilQ \bigg\|\frac{ a Q \! +\!  \tilQ }{a+1}\bigg). \label{Eq:def GJS}
\end{align}
The $\mathrm{GJS}(\fndot)$ quantifies, in an $a$-weighted manner, the distance between $Q$ and $\tilQ$. This quantity has featured prominently in information-theoretic decision problems in which there are training and test sequences \cite{gutman1989asymptotically, zhou2019second, he2020distributed, mahdi19}.
\begin{lemma}\label{Lem:GJS increase in a}
	For any pair of $(Q,\tilQ)\in\calP(\calX)^2$ such that $Q\ne \tilQ$,  $\mathrm{GJS}(Q,\tilQ,a )$ is increasing in $a$.
\end{lemma}
Lemma \ref{Lem:GJS increase in a} can be obtained by calculating the first and second order derivatives of \eqref{Eq:def GJS} and the proof is omitted for brevity.

For any set of distributions $(Q_1,Q_2,\tilQ_1,\tilQ_2)\in\calP(\calX)^4$, any  $\beta\in(0,1)$ and $ r\in\bbR_+$, we define the following linear combination of generalized Jensen-Shannon divergences as follows
\begin{align}
&\label{eqn:defLD}\LD(Q_1,Q_2,\tilQ_1,\tilQ_2,\beta,r) \nn\\  &:=r\, \mathrm{GJS}\left(Q_1,\tilQ_1,\frac{\beta}{r}\right)+r\, \mathrm{GJS}\left(Q_2,\tilQ_2,\frac{1-\beta}{r}\right).
\end{align}
 The function $\LD(\fndot)$, on the other hand, additionally quantifies, in a $(\beta,r)$-weighted manner, the sum of the distances between $Q_1$ and $\tilQ_1$ as well as $Q_2$ and $\tilQ_2$. 


\begin{lemma}\label{Lem:Gmin increase}
	For any $r\in\bbR_+$ and any pair of $(P_1,P_2)\in\calP(\calX)^2$, $\mathrm{GJS}(tP_1+(1-t)P_2,P_2,\frac{1}{r})$ and $\mathrm{GJS}((1-t )P_1+tP_2,P_1,\frac{1}{r})$ are strictly increasing and convex functions of $t\in[0,1]$.
\end{lemma}
 The proof of Lemma \ref{Lem:Gmin increase} is provided in Appendix \ref{proof of monotonicity}.

\begin{figure*}
 		\vspace{-2cm}
 		\centering
 		\subfloat[Illustration of $\Delta_\rmL(\bar{\Delta})$.]{
 		\begin{minipage}[c][1\width]{
 				0.34\textwidth}
 			\centering
 			\includegraphics[width=1\textwidth]{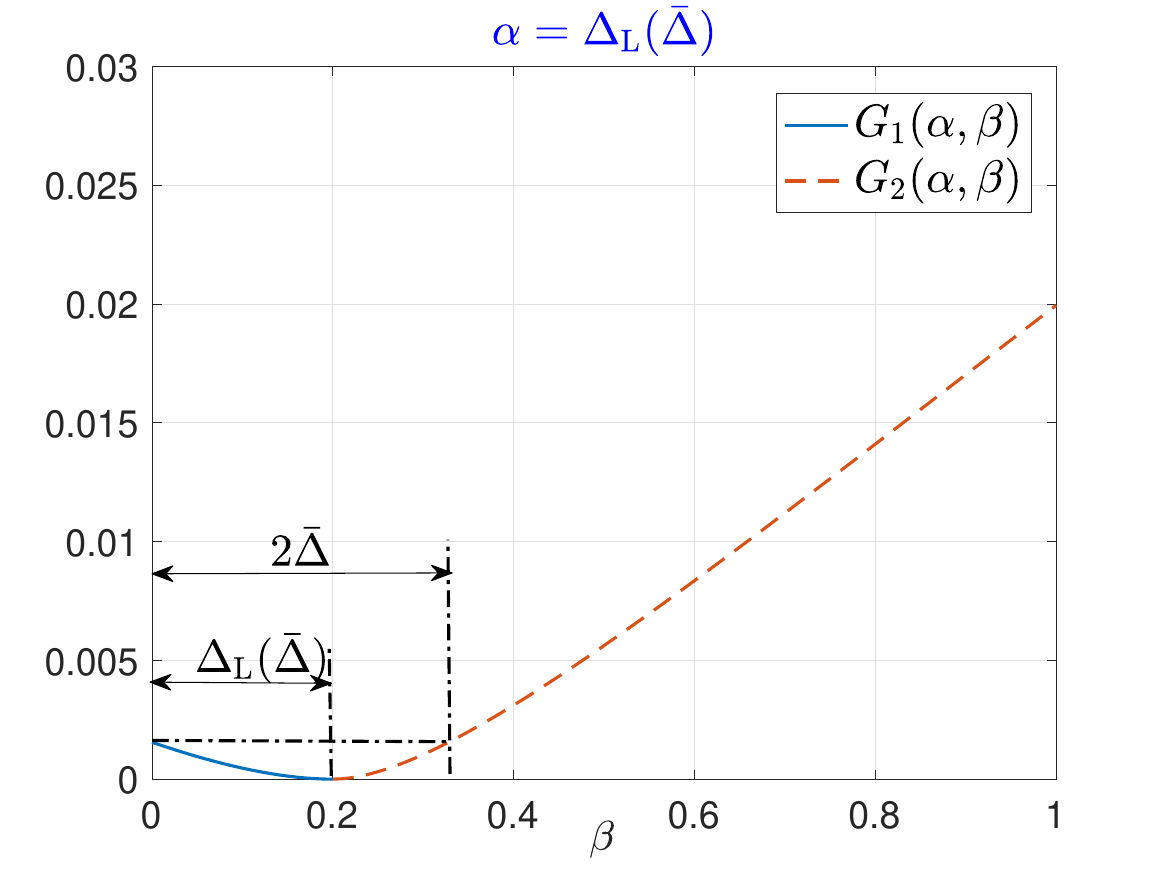}
 			\vspace{-2cm}
 			\label{fig:Delta_L}
 		\end{minipage}}
 	\hspace{-1.5em}
 	\subfloat[Illustration of $\iota(\alpha)$. When $\alpha$ increases, both the curves will translate to the right and vice versa.]{
 		\begin{minipage}[c][1\width]{
 				0.34\textwidth}
 			\centering
 			\includegraphics[width=1\textwidth]{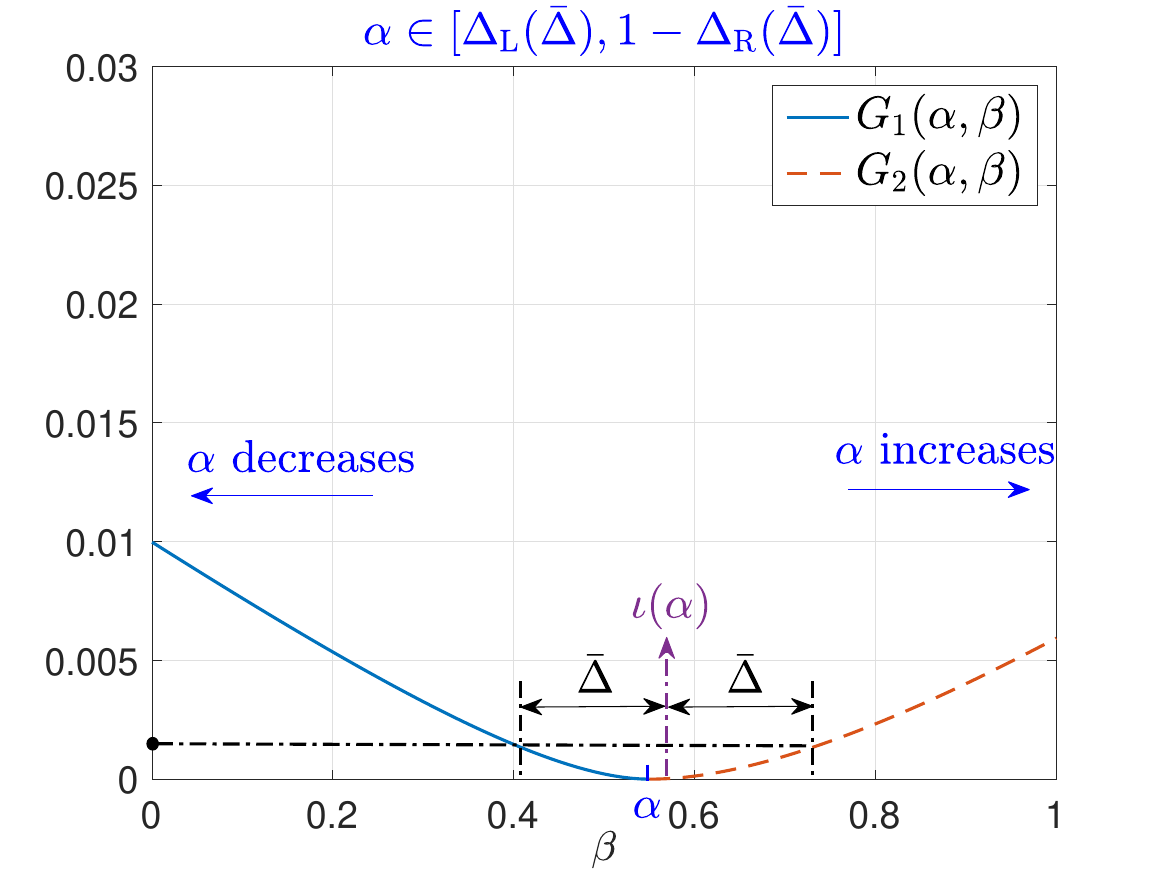}
 			\vspace{-2cm}
 			\label{fig:alpha middle}
 		\end{minipage}}
 		\hspace{-1.5em}
 		\subfloat[Illustration of $\Delta_\rmR(\bar{\Delta})$.]{
 		\begin{minipage}[c][1\width]{
 				0.34\textwidth}
 			\centering
 			\includegraphics[width=1\textwidth]{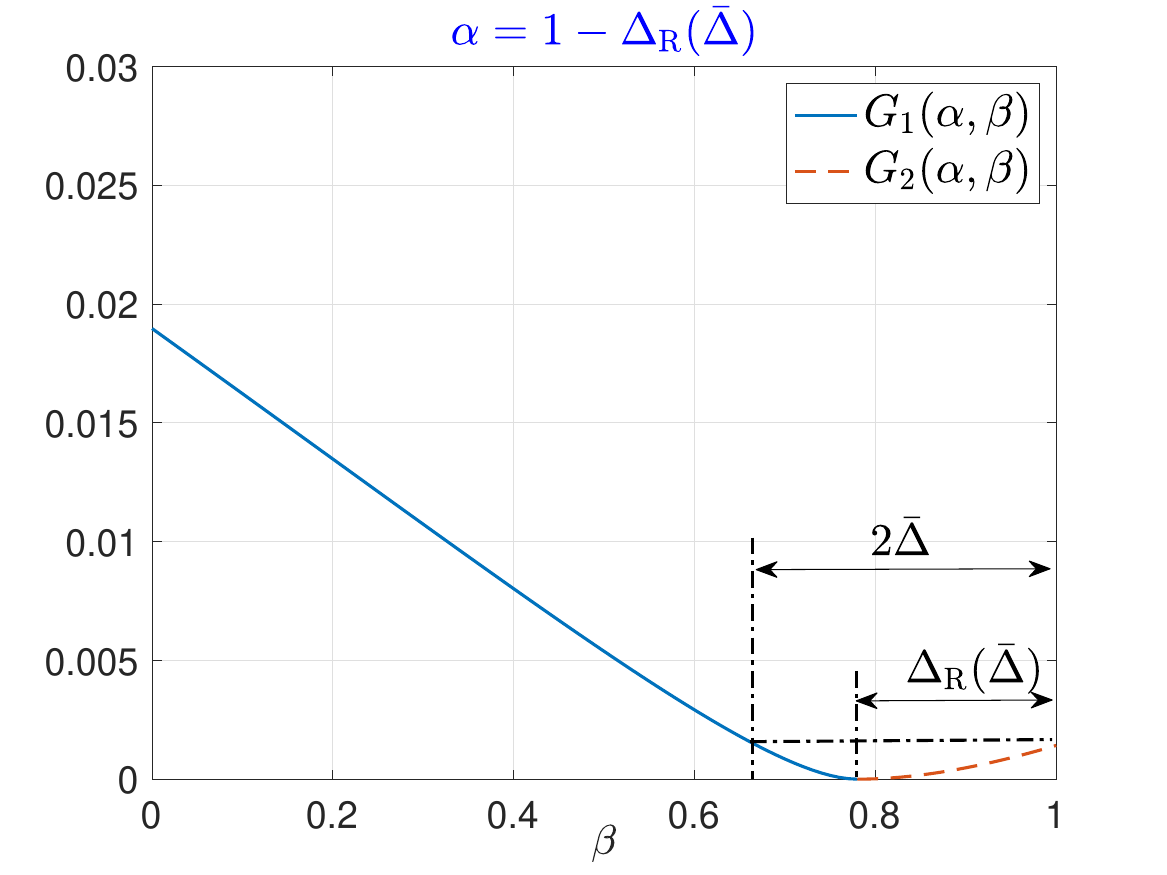}
 			\vspace{-2cm}
 			\label{fig:Delta_R}
 		\end{minipage}}
 		\caption{$G_1(\alpha,\beta)$ and $G_2(\alpha,\beta)$ versus $\beta$ under different values of $\alpha$ when $P_1=\mathrm{Bern}(0.6)$, $P_2=\mathrm{Bern}(0.2)$, $r=10$ and $\bar{\Delta}=0.175$. 
 			(a)~$\Delta_\rmL(\bar{\Delta})$ is the value that satisfies $G_1(\Delta_\rmL(\bar{\Delta}),0)=G_2(\Delta_\rmL(\bar{\Delta}),2\bar{\Delta})$ (i.e., \eqref{Eq:def Delta_L}). ~(b)~$\iota(\alpha)$ is the value that satisfies $G_1(\alpha,\iota(\alpha)-\bar{\Delta})=G_2(\alpha,\iota(\alpha)+\bar{\Delta})$ (i.e. \eqref{Eq:def iota}). ~(c)~$\Delta_\rmR(\bar{\Delta})$ is the value that satisfies $G_1(1-\Delta_\rmR(\bar{\Delta}),1-2\bar{\Delta})=G_2(1-\Delta_\rmR(\bar{\Delta}),1)$ (i.e., \eqref{Eq:def Delta_R}).}
 		\label{Fig:CPD_illu_Delta_L_R}
 	\end{figure*}

	For any $\alpha\in[0,1]$ and any $\beta \in[\alpha,1]$, define
	\begin{align}
		G_1(\alpha,\beta)\!:=\!\mathrm{GJS}\bigg(\frac{(\alpha\!-\!\beta)P_1+(1\!-\!\alpha)P_2}{1-\beta},P_2,\frac{1\!-\!\beta}{r} \bigg),
	\end{align}
and for any $\beta \in [0,\alpha]$, define
	\begin{align}
		G_2(\alpha,\beta)&:=\mathrm{GJS}\bigg(\frac{\alpha P_1+(\alpha-\beta)P_2}{\beta},P_1,\frac{\beta}{r} \bigg).
	\end{align}
 	Given any $\bar{\Delta}\in[0,1/2)$, let $\Delta_\rmL(\bar{\Delta})\in[0,2\bar{\Delta}]$ be the unique number such that
 	\begin{align}
 		G_1(\Delta_\rmL(\bar{\Delta}),0)=G_2(\Delta_\rmL(\bar{\Delta}),2\bar{\Delta}) \label{Eq:def Delta_L}
 	\end{align}
 	and $\Delta_\rmR(\bar{\Delta})\in[0,2\bar{\Delta}]$ be the unique number such that
 	\begin{align}
 		G_1(1-\Delta_\rmR(\bar{\Delta}),1-2\bar{\Delta})=G_2(1-\Delta_\rmR(\bar{\Delta}),1). \label{Eq:def Delta_R}
 	\end{align}
 The illustrations of $\Delta_\rmL(\bar{\Delta})$ and $\Delta_\rmR(\bar{\Delta})$ are given in Figure \ref{Fig:CPD_illu_Delta_L_R}.
 	\begin{corollary}\label{coro:Delta L R increase}
 		$\Delta_\rmL(\bar{\Delta})$ and $\Delta_\rmR(\bar{\Delta})$ are both monotonically increasing in $\bar{\Delta}\in [0,1/2)$.
 	\end{corollary}
 	Corollary \ref{coro:Delta L R increase} can be deduced from Lemma \ref{Lem:GJS increase in a} and Lemma \ref{Lem:Gmin increase}.

Furthermore, for any $\alpha\in[\Delta_\rmL(\bar{\Delta}),1-\Delta_\rmR(\bar{\Delta})]$, let $\iota(\alpha)$ be such that
\begin{align}
	G_1(\alpha,\iota(\alpha)-\bar{\Delta})=G_2(\alpha,\iota(\alpha)+\bar{\Delta}); \label{Eq:def iota}
\end{align}
for any $\alpha\in[0,\Delta_{\rmL}(\bar{\Delta}))$, let $\iota(\alpha)=\bar{\Delta}$; for any $\alpha\in(1-\Delta_{\rmR}(\bar{\Delta}),1]$, let $\iota(\alpha)=1-\bar{\Delta}$. The illustration of $\iota(\alpha)$ is given in Figure \subref*{fig:alpha middle}.

Given any $r\in\bbR_+$ and any pair of $(P_1,P_2)\in\calP(\calX)^2$, let us define the function $\rmG_{\min}:[0,1/2)\to\bbR_+$ as follows
\begin{align}
	\rmG_{\min}(\bar{\Delta}):=r \min_{\alpha\in[\Delta_\rmL(\bar{\Delta}) ,1-\Delta_\rmR(\bar{\Delta})]}  G_1(\alpha,\iota(\alpha)-\bar{\Delta}) 
	\label{Eq: def Gmin}.
\end{align}
\begin{corollary}\label{coro:Gmin increase}
	For any $r\in\bbR_+$, $\rmG_{\min}(\bar{\Delta})$ is a strictly increasing function of $\bar{\Delta}\in[0,1/2)$.
\end{corollary}
The proof of Corollary \ref{coro:Gmin increase} is provided in Appendix \ref{proof of coro:Gmin increase}. Note that $\rmG_{\min}(0)=0$. From these facts, we deduce that the inverse  $\rmG_{\min}^{-1}(y)$ for $y\in \big[0,\rmG_{\min}( 1/2 )\big)$ exists and is also  strictly increasing in $y$.

\section{Main Results}

The following theorem constitutes our main result and presents a single-letter expression for the optimal NCW in both asymptotic regimes.  

\begin{theorem}\label{Thm:optimal resolution}
	For any $r\in\bbR_+$,   $\epsilon\in [0,1)$, any pair of distributions $(P_1,P_2)\in\calP(\calX)^2$, 
	the optimal NCW is  
	\begin{align}\label{Eq:LDR Delta^*}
		\bar{\Delta}^*(r,\lambda,\epsilon)=\begin{cases}
		\rmG_{\min}^{-1}(\lambda), & \displaystyle\lambda\in\bigg(0,\rmG_{\min}\Big(\frac{1}{2}\Big)\bigg),\\
		\displaystyle\frac{1}{2}, & \text{otherwise};
		\end{cases}
	\end{align}
	In the moderate deviations regime, for any $P_1\ne P_2$, the  $t$-optimal NCW for any $t\in(0,1/2)$ and $\lambda>0$ is given in \eqref{Eq:MDR Delta^*} at the top of next page.
	\begin{figure*}[t]
	\begin{align}\label{Eq:MDR Delta^*}
		\bar{\Delta}_t^*(r,\lambda,\epsilon)=\max_{\alpha\in[0,1]}\frac{
			\sqrt{\lambda}\big(\sqrt{\alpha(\alpha+r)\upchi_2(P_1\|P_2)}+\sqrt{(1-\alpha)(1-\alpha+r)\upchi_2(P_2\|P_1)}~ \big)}{\sqrt{2r\upchi_2(P_1\|P_2)\upchi_2(P_2\|P_1)}}.
	\end{align}\hrulefill
	\end{figure*}
\end{theorem}

The proof of Theorem \ref{Thm:optimal resolution} is given in Section \ref{proof of Thm 1}. Several remarks are in order.

\begin{figure*}[t]
	\centering
	\includegraphics[width=0.9\textwidth]{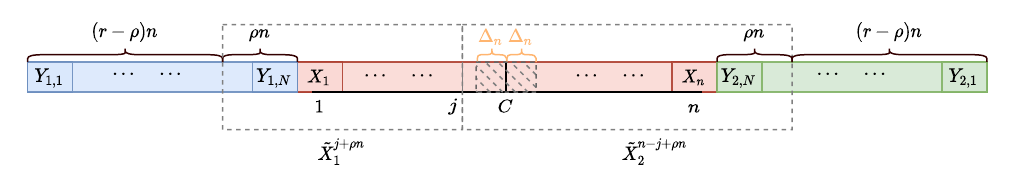}
	\caption{Illustration of concatenating test and parts of training sequences.}
	\label{fig:rho diagram}
\end{figure*}

First, we define the estimation scheme. Let us define $\rho\in(0,r)$ such that $\rho n\in \bbZ_+$. In the following, for the sake of brevity, we omit the integer constraint on $\rho n$ and simply write $\rho\in(0,r)$. For any tuple of sequences $(X^n,Y_1^N,Y_2^N)\in\calX^{n+2N}$ and any $j\in[n]$, construct two new test vectors 
\begin{align}
	\tilX_1^{j+\rho n}&=(Y_{1,N-\rho n+1}^N,X^j), \label{Eq:tilX_1}\\
	\tilX_2^{n-j+\rho n}&=(X_{j+1}^{n},Y_{2,N-\rho n+1}^N) \label{Eq:tilX_2}
\end{align}
by concatenating different parts of $X^n,Y_1^N,Y_2^N$ as illustrated in Figure \ref{fig:rho diagram}, and let 
\begin{equation}
\bT_{j}^{\rho}:=(T_{\tilX_1^{j+\rho n}},T_{\tilX_2^{n-j+\rho n}},T_{Y_1^{N-\rho n}},T_{Y_2^{N-\rho n}} )
\end{equation}
 denote the tuple of their sub-types. We recall that the {\em type}~\cite{Csi97} or {\em empirical distribution} of $Z^n = (Z_1,\ldots, Z_n)\in\calZ^n$ is  $T_{Z^n}(z) : =\frac{1}{n}\sum_{i=1}^n\mathbbm{1}\{ Z_i = z\}$ for all $z\in\calZ$. 
In the achievability proof (estimation scheme) of Theorem \ref{Thm:optimal resolution}, given any confidence width $\Delta_n\in[0:n/2-1]$,  we make use of the estimator $\gamma(X^n,Y_1^N,Y_2^N)$ that first computes 
\begin{align}
\! 	S_i\! :=\! \min_{j\in[i\pm\Delta_n]\cap[n]}\!  (1\!+\! 2\rho)\LD\Big(\bT_j^\rho,\frac{j+\rho n}{n+2\rho n} ,\frac{r-\rho}{1\! +\! 2\rho} \Big), ~\forall i\in[n]. \label{eqn:Si}
\end{align}
The estimator is defined as
\begin{align}\label{Eq:estimator}
	\gamma(X^n,Y_1^N,Y_2^N)=\left\{
	\begin{array}{rl}
		\rme, & \text{if } S_i \leq \lambda, \forall i\in[n],\\
		I, & \text{if } \exists i\in[n] \text{ s.t. } S_i>\lambda,	
	\end{array}\right. 
\end{align}
where $I$ is any $i\in[n]$ such that $S_i>\lambda$. 
We show that this estimator is asymptotically optimal by taking $\rho\approx n^{-1/4}$ (we specify the exact value of $\rho$ in Section V) in large deviations regime and also in moderate deviations regime. In the latter regime, we replace the threshold $\lambda$  in~\eqref{Eq:estimator} by $\lambda n^{-t}$. This estimator is based on the partial types and types of the test and training sequences; this is what we call a {\em type-based estimator}. Intuitively, the estimator declares a point $k$ to be the change-point when $T_{\tilX_1^{k+\rho n}}$ is deemed to be sufficiently close enough to $T_{Y_1^{N-\rho n}}$ and  $T_{\tilX_2^{n-k+\rho n}}$ is similarly  to be  sufficiently close enough to $T_{Y_2^{N-\rho n}}$, where the proximity is measured in terms of the function $\LD$, defined in \eqref{eqn:defLD}. If no such point exists, then the estimator declares an ``erasure''.

Second, Theorem \ref{Thm:optimal resolution} implies that when the undetected error probability decays exponentially fast with exponent $\lambda$ (i.e., $t=0$ in \eqref{Eq: estimator metric}), the asymptotically optimal confidence width is of order $\Theta(n)$ and the pre-constant term is specified by \eqref{Eq:LDR Delta^*}; when the undetected error decays subexponentially as $\exp(-n^{1-t}\lambda)$ (for $t\in (0,1/2)$), the asymptotically optimal confidence width is of order $\Theta(n^{1-t/2})$ and the pre-constant term is specified by~\eqref{Eq:MDR Delta^*}. The former result  is similar in spirit to those by Garreau and Arlot~\cite{garreau2018consistent} and Carlstein~\cite[Theorem~2]{carlstein1988nonparametric} who showed for a general class nonparametric distributions that, in the absence of training sequences, a confidence width of order $\Theta(n)$ is achievable when the error probability decays exponentially fast; however, exact pre-constant terms were not derived in \cite{carlstein1988nonparametric} and lower bounds for the NCW (the converse parts) were also not derived in both works. Although in \cite{baranowski2016narrowest,garreau2018consistent,wang2020univariate,padilla2019optimal,padilla2019change,yao1989least,csorgo1997limit,boysen2009consistencies,zou2014nonparametric,matteson2014nonparametric}, confidence widths of order $O(\log n)$ or even $O(1)$ were shown to be achievable, these results were derived under the less stringent constraint that the error probability decays at most polynomially fast or simply vanishes arbitrarily slow.
Thus, it is reasonable that our derived optimal confidence width is larger than those in \cite{baranowski2016narrowest,garreau2018consistent,wang2020univariate,padilla2019optimal,padilla2019change,yao1989least,csorgo1997limit,boysen2009consistencies,zou2014nonparametric,matteson2014nonparametric}. Moreover, the statements in Theorem~\ref{Thm:optimal resolution} consist of accompanying {\em strong converses} (in the spirit of Wolfowitz~\cite{WolfowitzBook}) while the results in \cite{baranowski2016narrowest,garreau2018consistent,wang2020univariate,padilla2019optimal,padilla2019change,yao1989least,csorgo1997limit,boysen2009consistencies,zou2014nonparametric,matteson2014nonparametric} do not include their impossibility counterparts.

\begin{figure}[t]
\hspace{.1in}
\begin{tabular}{c}
	\hspace{-.25in} \includegraphics[width=.5\textwidth]{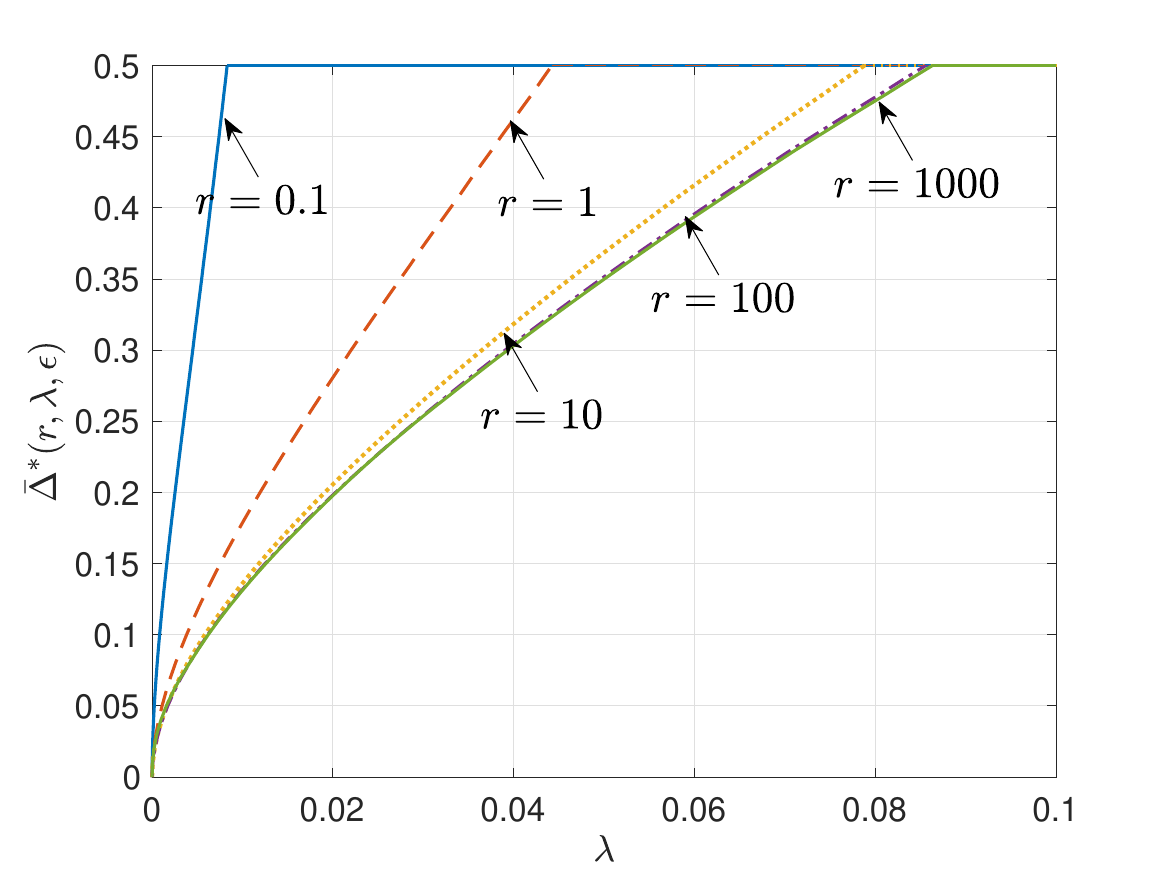} \\
	\hspace{-.25in} {\footnotesize \textbf{(a)} Large deviations regime}\\ \hspace{-.23in}\includegraphics[width=.5\textwidth]{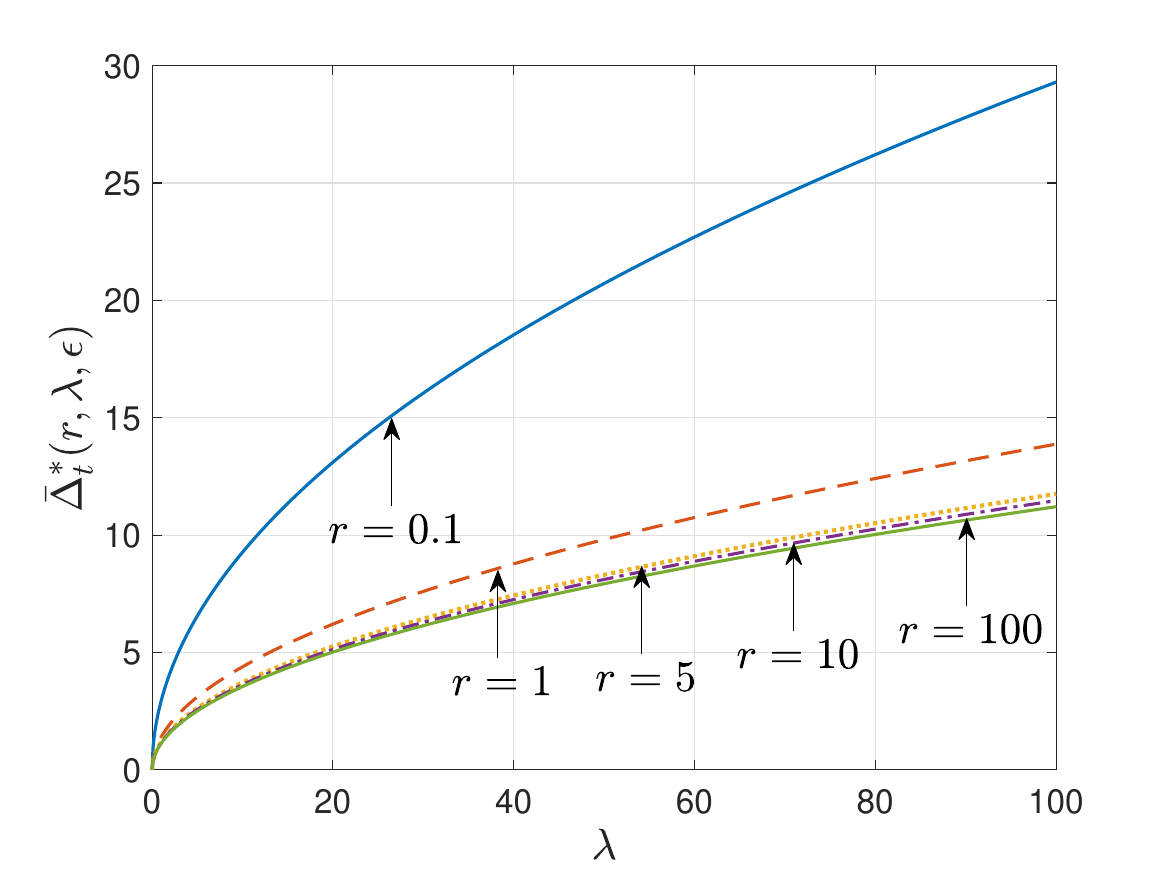}\\
	\hspace{-.25in}  {\footnotesize \textbf{(b)} Moderate deviations regime}
\end{tabular}	
	\caption{(a) Large deviations regime: $\bar{\Delta}^*(r,\lambda,\epsilon)$ versus $\lambda$ under different values of $r$ when $P_1=\mathrm{Bern}(0.6),~ P_2=\mathrm{Bern}(0.2)$. (b) Moderate deviations regime: $\bar{\Delta}_t^*(r,\lambda,\epsilon)$ versus $\lambda$ under different values of $r$ when $P_1=\mathrm{Bern}(0.6),~ P_2=\mathrm{Bern}(0.2)$.}
	\label{fig:CPD_r}
\end{figure}

\begin{figure}[t]
	\hspace{.1in}
	\begin{tabular}{c}
		\hspace{-.25in} \includegraphics[width=.5\textwidth]{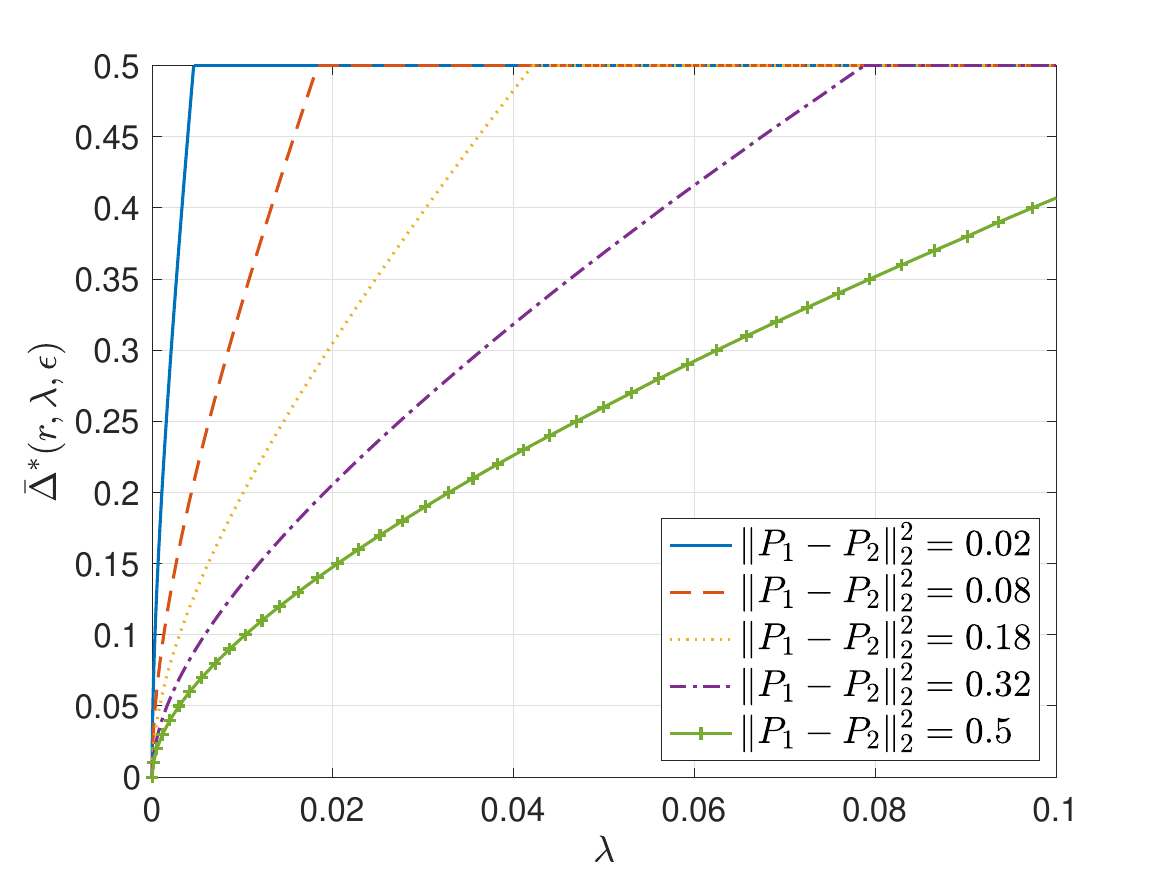}\\
		\hspace{-.25in} {\footnotesize \textbf{(a)} Large deviations regime} \\
		\hspace{-.23in} \includegraphics[width=.5\textwidth]{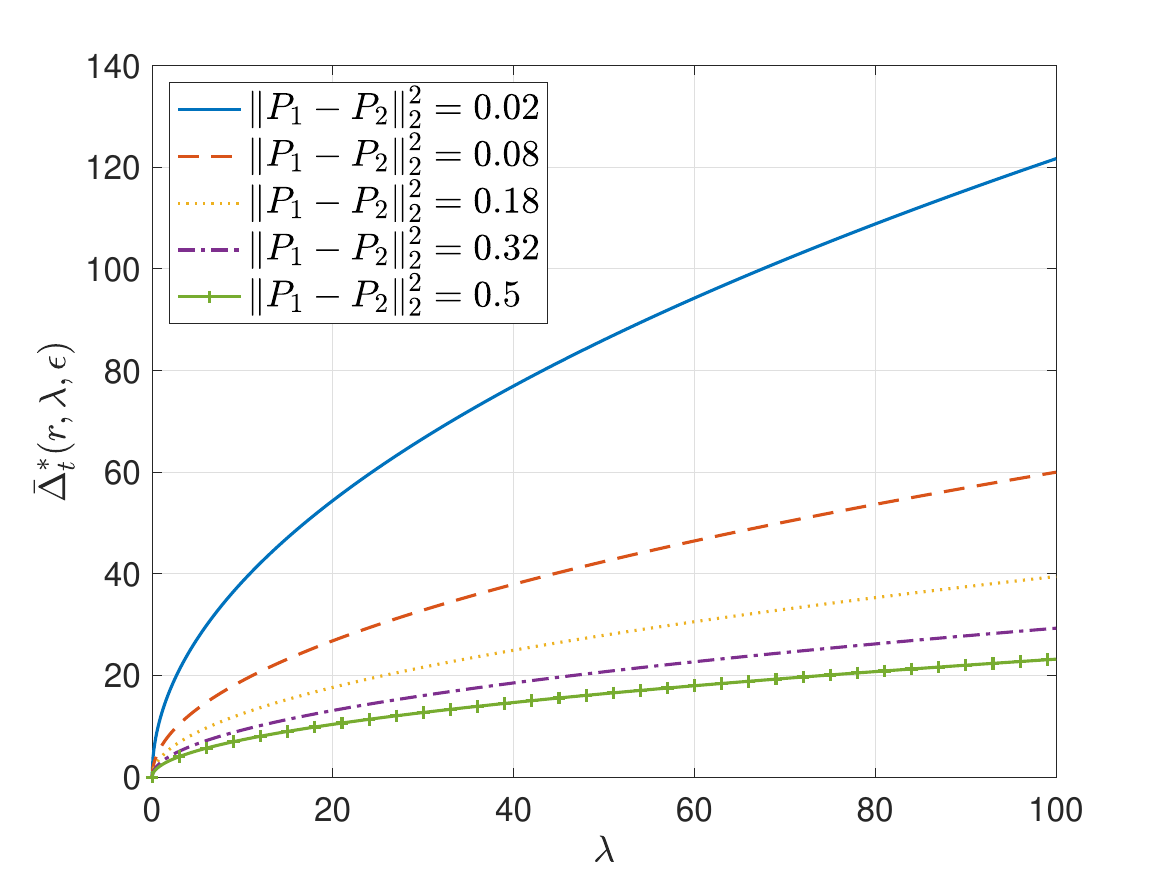}\\
		\hspace{-.25in}  {\footnotesize \textbf{(b)} Moderate deviations regime}
	\end{tabular}
	\caption{(a) Large deviations regime: $\bar{\Delta}^*(r,\lambda,\epsilon)$ versus $\lambda$ under different pairs of $(P_1,P_2)$ when $r=10$. (b) Moderate deviations regime: $\bar{\Delta}_t^*(r,\lambda,\epsilon)$ versus $\lambda$ under different pairs of $(P_1,P_2)$ when $r=10$. Both (a) and (b) are under different pairs of $(P_1,P_2)$ where $P_1=\mathrm{Bern}(0.6)$ and $P_2=\mathrm{Bern}(0.6-e)$ for $e\in\{0.1,0.2,0.3,0.4,0.5\}$.}
	\label{fig:CPD_P1P2}
\end{figure}

Third, it can be seen from Figures \ref{fig:CPD_r}--\ref{fig:CPD_P1P2} that for any $t\in[0,1/2)$,  $\bar{\Delta}_t^*(r,\lambda,\epsilon)$ increases as $\lambda$ increases, $r$ decreases, and the distance between $P_1$ and $P_2$ decreases. The observations can be intuitively explained as follows. 
\begin{itemize}
	\item As $\lambda$ increases, the requirement in~\eqref{Eq: estimator metric} concerning the worst-case undetected error probability $\max_{C\in [n]}\mathbb{P}\{ \calE_C \}$ becomes more stringent and thus the optimal NCW between the true and estimated change-points increases. This is illustrated in Figures \ref{fig:CPD_r}--\ref{fig:CPD_P1P2}.
	
	
	\item As $r$ decreases, the length of the training sequences relative to the test sequence decreases, and thus less knowledge about distributions $P_1$ and $P_2$ can be learned from the training sequences. Therefore,  to maintain the same undetected error exponent $\lambda$ and erasure probability $\epsilon$, the confidence width $\Delta$ should be enlarged correspondingly. This is illustrated in Figure~\ref{fig:CPD_r}.
	
	\item As the distance between $P_1$ and $P_2$ decreases, it is harder to distinguish between them and thus the accuracy of detection decreases, leading to a larger confidence width. This is illustrated in Figure~\ref{fig:CPD_P1P2}.
\end{itemize}
We can also see that
 $\bar{\Delta}_t^*(r,\lambda,\epsilon)=0$ when $\lambda=0$, which means that if the confidence width is of order $o(n^{1-t/2})$, the undetected error probabilities cannot decay faster than $\exp(-n^{1-t}\lambda)$. Note that $\bar{\Delta}_t^*(r,\lambda,\epsilon)$  for any $t\in [0,1/2)$ is independent of $\epsilon$, which implies that strong converses hold.

Fourth, we analyze the impact of the availability of training data on the optimal NCW by considering the scenarios in which $r$ assumes its extremal values. We note that as $r\to\infty$, in  the large deviations regime, for any $\lambda\in\big(0,\rmG_{\min} ( 1/{2} )\big)$, let 
\begin{align}
\bar{\Delta}^*(\lambda,\epsilon) &=\lim_{r\to\infty}\bar{\Delta}^*(r,\lambda,\epsilon),\\
\tilde{\Delta}_{\rmL}(\bar{\Delta}^*(\lambda,\epsilon))&=\lim_{r\to\infty}\Delta_{\rmL}(\bar{\Delta}^*(r,\lambda,\epsilon)),\quad\mbox{and}\\ 
\tilde{\Delta}_{\rmR}(\bar{\Delta}^*(\lambda,\epsilon))&=\lim_{r\to\infty}\Delta_{\rmR}(\bar{\Delta}^*(r,\lambda,\epsilon)).
\end{align}
Then $\tilde{\Delta}_{\rmL}(\bar{\Delta}^*(\lambda,\epsilon))$ and  $\tilde{\Delta}_{\rmR}(\bar{\Delta}^*(\lambda,\epsilon))$ satisfy
\begin{align}
	&D\big(\tilde{\Delta}_{\rmL}(\bar{\Delta}^*(\lambda,\epsilon))P_1+(1-\tilde{\Delta}_{\rmL}(\bar{\Delta}^*(\lambda,\epsilon)))P_2 \| P_2 \big) \nn\\
	&\!\!\!\!=D\bigg(\frac{\tilde{\Delta}_{\rmL}(\bar{\Delta}^*(\lambda,\epsilon))P_1\!+\!(2\bar{\Delta}^*(\lambda,\epsilon)\!-\!\tilde{\Delta}_{\rmL}(\bar{\Delta}^*(\lambda,\epsilon)))P_2}{2\bar{\Delta}^*(\lambda,\epsilon)} \bigg\| P_1 \bigg)
\end{align}
and
\begin{align}
	&D((1-\tilde{\Delta}_{\rmR}(\bar{\Delta}^*(\lambda,\epsilon))P_1)+\tilde{\Delta}_{\rmR}(\bar{\Delta}^*(\lambda,\epsilon)P_2 \| P_1)\nn\\
	&\!\!\!\!=D\bigg(\frac{(2\bar{\Delta}^*(\lambda,\epsilon)\!-\!\tilde{\Delta}_\rmR(\bar{\Delta}^*(\lambda,\epsilon))) P_1\!+\!\Delta_\rmR(\bar{\Delta}^*(\lambda,\epsilon)) P_2}{2\bar{\Delta}^*(\lambda,\epsilon)} \bigg\| P_2\bigg).
\end{align}
Let $\tilde{\iota}(\alpha)$ be the unique solution to
\begin{align}
	&D\bigg(\frac{(\alpha-\tilde{\iota}(\alpha)+\bar{\Delta}^*(\lambda,\epsilon))P_1+(1-\alpha)P_2}{1-\tilde{\iota}(\alpha)+\bar{\Delta}^*(\lambda,\epsilon)} \bigg\| P_2\bigg)\nn\\
	&=D\bigg(\frac{\alpha P_1+(\tilde{\iota}(\alpha)+\bar{\Delta}^*(\lambda,\epsilon)-\alpha)P_2}{\tilde{\iota}(\alpha)+\bar{\Delta}^*(\lambda,\epsilon)} \bigg\| P_1\bigg)
\end{align}
and let
\begin{align}
	 &\tilde{\rmG}_{\min}(\bar{\Delta}^*(\lambda,\epsilon)):=\lim\limits_{r\to\infty}\rmG_{\min}(\bar{\Delta}^* (r,\lambda,\epsilon))\\*
	=&\min_{\alpha\in[\tilde{\Delta}_\rmL(\bar{\Delta}^*(\lambda,\epsilon)) ,1-\tilde{\Delta}_\rmR(\bar{\Delta}^*(\lambda,\epsilon))]} \nn\\ &D\bigg(\frac{(\alpha-\tilde{\iota}(\alpha)+\bar{\Delta}^*(\lambda,\epsilon)) P_1+(1-\alpha)P_2}{1-\tilde{\iota}(\alpha)+\bar{\Delta}^*(\lambda,\epsilon)} \bigg\| P_2\bigg).
\end{align}
Then we have $\bar{\Delta}^*(\lambda,\epsilon)=\tilde{\rmG}_{\min}^{-1}(\lambda)$. In the moderate deviations regime, 
\begin{align}
	&\lim\limits_{r\to\infty}\bar{\Delta}_t^*(r,\lambda,\epsilon) \nn\\*
	&= \max_{\alpha\in[0,1]}\frac{
		\sqrt{\lambda}\big(\sqrt{\alpha\upchi_2(P_1\|P_2)}+\sqrt{(1-\alpha)\upchi_2(P_2\|P_1)}~ \big)}{\sqrt{2\upchi_2(P_1\|P_2)\upchi_2(P_2\|P_1)}}.
\end{align}
As $r\to\infty$, the length of the training sequences far exceeds that of the test sequence, implying that we can estimate the underlying distributions $P_1$ and $P_2$ arbitrarily accurately (e.g., using their types). Thus, as $r\to\infty$, the optimal NCW converges to a fixed value given by the test in which the distributions $P_1$ and $P_2$ are known; this is also illustrated in Figure \ref{fig:CPD_r}. On the other hand, as $r\to 0$, in large deviations regime, if $\lambda\to 0$, $\bar{\Delta}^*(r,\lambda,\epsilon)\to 0$; otherwise, $\bar{\Delta}^*(r,\lambda,\epsilon)= 1/2$ for all $\lambda\in(0,\infty)$; in moderate deviations regime, we have $(\alpha+r)/r \to\infty$ and $(1-\alpha+r)/r \to\infty$, and if $\lambda\to 0$, $\lim_{r\to0}\bar{\Delta}_t^*(r,\lambda,\epsilon)<\infty$; if $\lambda$ assumes a positive value, $\lim_{r\to0}\bar{\Delta}_t^*(r,\lambda,\epsilon)=\infty$. These limiting scenarios can be intuitively explained as follows. When the lengths of the training sequences are significantly shorter than that of the test sequence, the error probabilities cannot vanish exponentially fast if the  confidence width is smaller than the maximal value of ${n}/{2}$ or sub-exponentially fast if the  confidence width is of order $o(n^{1-t/2})$.

Finally, in the moderate deviations regime, we see that  the optimal NCW $\bar{\Delta}^*_t(r,\lambda,\epsilon)$ is a function of the symmetrized chi-square distance $\sqrt{\upchi_2(P_1\|P_2)\upchi_2(P_2\|P_1)}$. Thus, by studying the CPD problem according to our unique setup, we assign an operational interpretation of  $\sqrt{\upchi_2(P_1\|P_2)\upchi_2(P_2\|P_1)}$. Since the chi-squared distance $\upchi_2(P_1\|P_2)$ is, in general, asymmetric, our moderate deviations result shows that the symmetrized version is   operationally meaningful in problems such as CPD with training samples. 


\section{Conclusion and Future Works} \label{sec:concl}
In this paper, we derived the optimal NCW for the CPD problem as a function of the ratio of the lengths of the training to test sequences, the exponent of the undetected error probability, and the distance between the distributions among other parameters. We proposed an asymptotically optimal type-based estimator in~\eqref{Eq:estimator}. We also proved strong converse statements, namely, that the optimal NCWs in both the large and moderate deviation regimes are independent of the bound on the erasure probability. Our results provide new insights on the fundamental limits of the CPD problem when side information in the form of training data is available. Our moderate deviations result  demonstrates the operational significance of the symmetrized chi-square distance $\sqrt{\upchi_2(P_1\|P_2)\upchi_2(P_2\|P_1)}$.

This work opens up a multitude of research directions, some of which are listed as follows.
\begin{enumerate}
\item While $\bar{\Delta}_t^*(r,\lambda,\epsilon)$ for any $t\in [0,1/2)$ is independent of $\epsilon\in [0,1)$---a strong converse statement---a natural question beckons. What are the {\em second-order terms}~\cite{TanBook,zhou2019second} of the {\em non-normalized} optimal confidence widths in both regimes? We believe a more intricate and careful analysis that is largely based on the use of various strengthenings of the central limit theorem may provide satisfactory answers. These second-order terms would, in general, depend on $\epsilon$ and shed light on the finite length performance of optimal tests.
\item Other   problem settings can also be explored. For example, the techniques herein do not directly extend over to the more practical setting of {\em online CPD problem} with training sequences. For this setting, we may need to leverage ideas from  sequential hypothesis testing or sequential classification; see~\cite{mahdi19}.  Another setting that is worth investigating is that of detecting multiple change-points. For this proposed extension, we expect   the majority of techniques here to carry through.
\item Finally, the techniques contained herein may be utilized to provide theoretical guarantees for the setting considered in Gruner and Johnson~\cite{gruner96}. In that problem, the authors, motivated by problems in distributed detection, consider two sets of sequences $\{\bx_1,\bx_2,\ldots, \bx_{C-1},\bx_C,\ldots, \bx_N\}\subset \calX^{nN}$ and $\{\by_1,\by_2,\ldots, \by_{C-1},\by_C,\ldots, \by_N\}\subset \calX^{nN}$. At times $i\le C-1$, the vectors $\bx_i$ and $\by_i$ have the same distribution. At times $i\geq C$, the vectors $\bx_i$ and $\by_i$ have different distributions. Even though training sequences are not explicitly provided, we believe that by adding a rejection option to this setting~\cite{gruner96}, we can obtain the optimal tradeoff between confidence width for detecting $C$, the erasure probability $\epsilon$,  and the exponent of the undetected error probability $\lambda$. 
\end{enumerate}

\section{Proof of Theorem \ref{Thm:optimal resolution}}\label{proof of Thm 1}

\subsection{Preliminaries for the Proofs}

Before presenting the proof of Theorem \ref{Thm:optimal resolution}, we will find it convenient to collect the following preliminary definitions and preparatory results.

\paragraph{Notation} To simplify   notation, for any $\rho\in(0,r)$, let us define    two convex combinations of $P_1(x)$ and $P_2(x)$ as follows: 
\begin{align}
\tilPjCleft(x)&:=\frac{(C-j)P_1(x)+(n+\rho n-C)P_2(x)}{n+\rho n-j},\quad\mbox{and} \nn\\
 \tilPjCright(x)&:=\frac{(C+\rho n)P_1(x)+(j-C)P_2(x)}{j+\rho n}.\label{eqn:conv1}
\end{align} 
We will use these distributions in which $j$ is constrained to be in  the interval $[n]$. 

In the following, we employ the sequence 
\begin{equation}
	\kappa_n:=\frac{3}{4}+ \frac{\log \log n}{2\log n} , \label{eqn:kappa_n}
\end{equation}
which clearly converges to $3/4$ as $n\to\infty$.

\paragraph{Tools for Bounding Probabilities} In anticipation of applying the central limit and Berry-Esseen theorems to bound the the erasure probability, for any pair of distributions $(Q_1,Q_2)\in\calP(\calX)^2$, let us define the following variance-like quantity 
\begin{align}
&V(Q_1,Q_2,j,r,\rho) \nn\\*
&:= \left\{ \begin{array}{ll}
\frac{n+\rho n-j}{n+2\rho n}\mathrm{Var}_{Q_1}\Big[ \log\frac{(n-j+N)Q_1(X)}{(n+\rho n-j)Q_1(X)+(N-\rho n)Q_2(X)}\Big] &\\ +\frac{r-\rho}{1+2\rho}\mathrm{Var}_{Q_2}\Big[\log\frac{(n-j+N)Q_2(X)}{(n+\rho n-j)Q_1(X)+(N-\rho n)Q_2(X)} \Big], &  \\
\hspace{2in} \text{ if }j\in [1 , C); &\\
\frac{j+\rho n}{n+2\rho n}\mathrm{Var}_{Q_1}\Big[ \log\frac{(j+N)Q_1(X)}{(j+\rho n)Q_1(X)+(N-\rho n)Q_2(X)}\Big] &\\
+\frac{r-\rho}{1+2\rho}\mathrm{Var}_{Q_2}\Big[\log\frac{(j+N)Q_2(X)}{(j+\rho n)Q_1(X)+(N-\rho n)Q_2(X)} \Big],&\\
\hspace{2in} \text{ if } j\in [C ,n].
\end{array}	  \right.   \label{eqn:var}
\end{align}
In \eqref{eqn:var}, $\mathrm{Var}_Q[\fndot]$ means that  the random variable $X$  that appears in the variance operator has distribution $Q$. We  can analogously define the third absolute moment  as $T(Q_1,Q_2,j,r,\rho)$.  We now use the fact that $j \in [n]$ to show that the variances and third absolute moments are sufficiently well-behaved so that we can apply the Berry-Esseen theorem in the following.

\begin{lemma}\label{Lem:V>0, T finite}
For any pair  of distributions  $(P_1,P_2)\in\calP(\calX)^2$ with $\supp(P_1)=\supp(P_2)=|\calX|$ such that $P_1\ne P_2$, any $\rho\in(0,r)$, any $C\in[n]$ and $j\notin [C\pm(\bar{\Delta}n-n^{\kappa_n})]$ with some $\bar{\Delta}>0$,    we have the following points:
\begin{itemize}
	\item $0<\underline{a}_1(\rho)< V(\tilPjCleft,P_2,j,r,\rho), V(\tilPjCright,P_1,j,r,\rho)< \overline{a}_1(\rho)$, and $T(\tilPjCleft,P_2,j,r,\rho)$, $T(\tilPjCright,P_1,j,r,\rho)<\overline{a}_2(\rho)$ where $\underline{a}_1(\rho)$, $\overline{a}_1(\rho)$, and $\overline{a}_2(\rho)$ are positive and finite for every fixed $\rho$ and $n$.
	\item  $\limsup_{n\to\infty}\overline{a}_1(\rho)$, $\limsup_{n\to\infty}\overline{a}_2(\rho)$ and $\liminf_{n\to\infty}\underline{a}_1(\rho)$ are positive and finite for every fixed $\rho$. 
	\item  For any $t\in(0,1/2)$ and any $j\notin [C\pm(\bar{\Delta}_t n^{1-t/2}-n^{\kappa_n})]$ with some $\bar{\Delta}_t>0$,   $\liminf_{n\to\infty}n^t\underline{a}_1(\rho)$ is positive and finite for every fixed $\rho$. 
\end{itemize}
\end{lemma}
The proof of Lemma \ref{Lem:V>0, T finite} is provided in Appendix \ref{proof of V>0, T finite}.

\paragraph{Bounds on Probabilities of Atypical Events} 
 The sequence $\kappa_n$ defined in \eqref{eqn:kappa_n} is used to define various properties of  atypical events in the following.  Given any distribution $P\in\calP(\calX)$, define the following {\em typical set}
\begin{align}
\calB(P):=\left\{x^n\in\calX^n: \max_{a\in\calX}\big| T_{x^n}(a)-P(a) \big| \leq \sqrt{\frac{\log n}{n}}\right\}, \label{eqn:typ_B}
\end{align}
where the lengths  of the sequences contained in instantiations of the set $\calB(\fndot)$ in the proofs below are implicit and not necessarily $n$. For any $j\in[n]$ and $i<j-n^{\kappa_n}$, let us define the atypical event 
\begin{align}\label{eqn:atypical set}
&\calA_{i,j}^\rho:=\big\{\tilX_1^{i+\rho n}\notin\calB(P_1) \text{ or } X_{i+1}^j\notin\calB(P_1) \text{ or } \nn\\
& \tilX_2^{n-j+\rho n}\!\notin\!\calB(P_2) \text{ or } Y_1^{N-\rho n}\!\notin\!\calB(P_1) \text{ or } Y_2^{N-\rho n}\!\notin\!\calB(P_2)\big\}.
\end{align}
By Hoeffding's inequality, we can show (see Appendix \ref{prob of atypical set}) that  for any $\rho\in[n^{\kappa_n-1},r-n^{\kappa_n-1}]$,
\begin{align}
\bbP_j\{\calA_{i,j}^\rho \}=O\bigg(\frac{1}{n^{2\kappa_n}}\bigg). \label{eqn:atypical}
\end{align}

\paragraph{Monotonicity of Functions} Fix any $\zeta\in (0,1)$, any $\rho\in(0,r)$ and recall that $C=\alpha n$. 
Define the functions 
\begin{align}
\!\!\!g_1(\alpha,\zeta)&\!:=\! \mathrm{GJS}\bigg(\!\tilP_{\alpha n-\zeta n}^-,P_2,\frac{n\!+\!\rho n\!-\!(\alpha n\!-\!\zeta n)}{rn-\rho n}\bigg),\;\mbox{and} \label{eqn:defg1}\\*
\!\!\!g_2(\alpha,\zeta)&\!:=\! \mathrm{GJS}\bigg(\tilP_{\alpha n+\zeta n}^+,P_1,\frac{\alpha n+\rho n+\zeta n}{rn-\rho n}\bigg). \label{eqn:defg2}
\end{align}
\begin{lemma}\label{Lem:GJS monotone in C}
	For any $\zeta\in(0,1)$, the function $g_1(\alpha,\zeta)$ is monotonically increasing in $\alpha\in[\zeta,1]$ and $g_2(\alpha,\zeta)$ is monotonically decreasing in $\alpha\in[0,1-\zeta]$.
\end{lemma}
The proof of Lemma \ref{Lem:GJS monotone in C} is provided in Appendix \ref{proof of g1g2}.

\begin{figure}[t]
		\centering
		\subfloat[Illustration of $\Delta_{n,\rmL}^{C,\rho}(\Delta)$ and $\Delta_{n,\rmR}^{C,\rho}(\Delta)$.]{
				\centering
				\includegraphics[width=1\columnwidth]{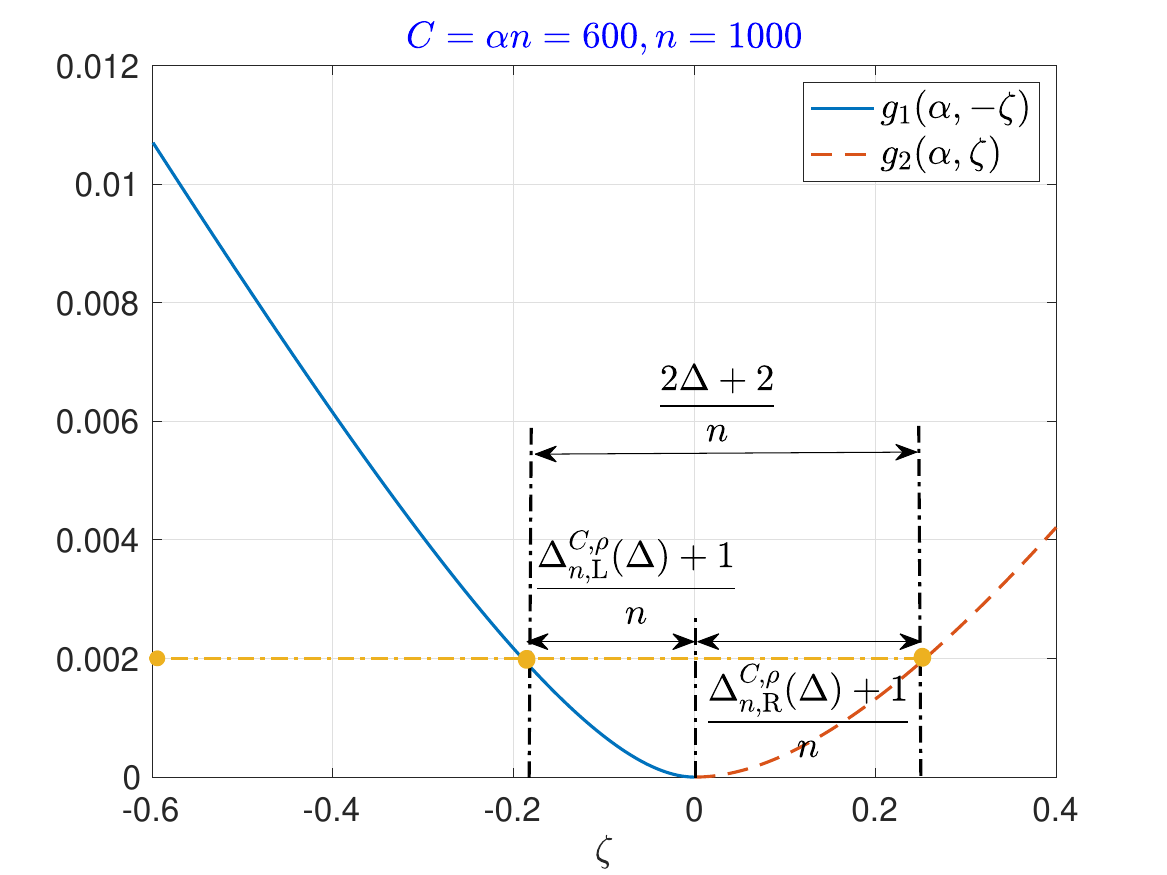}
				\vspace{-2.5cm}
				\label{Fig:illu Delta_n LR}
	}\\
		\subfloat[Illustration of $i_\rho(C)$.]{
				\centering
				\includegraphics[width=1\columnwidth]{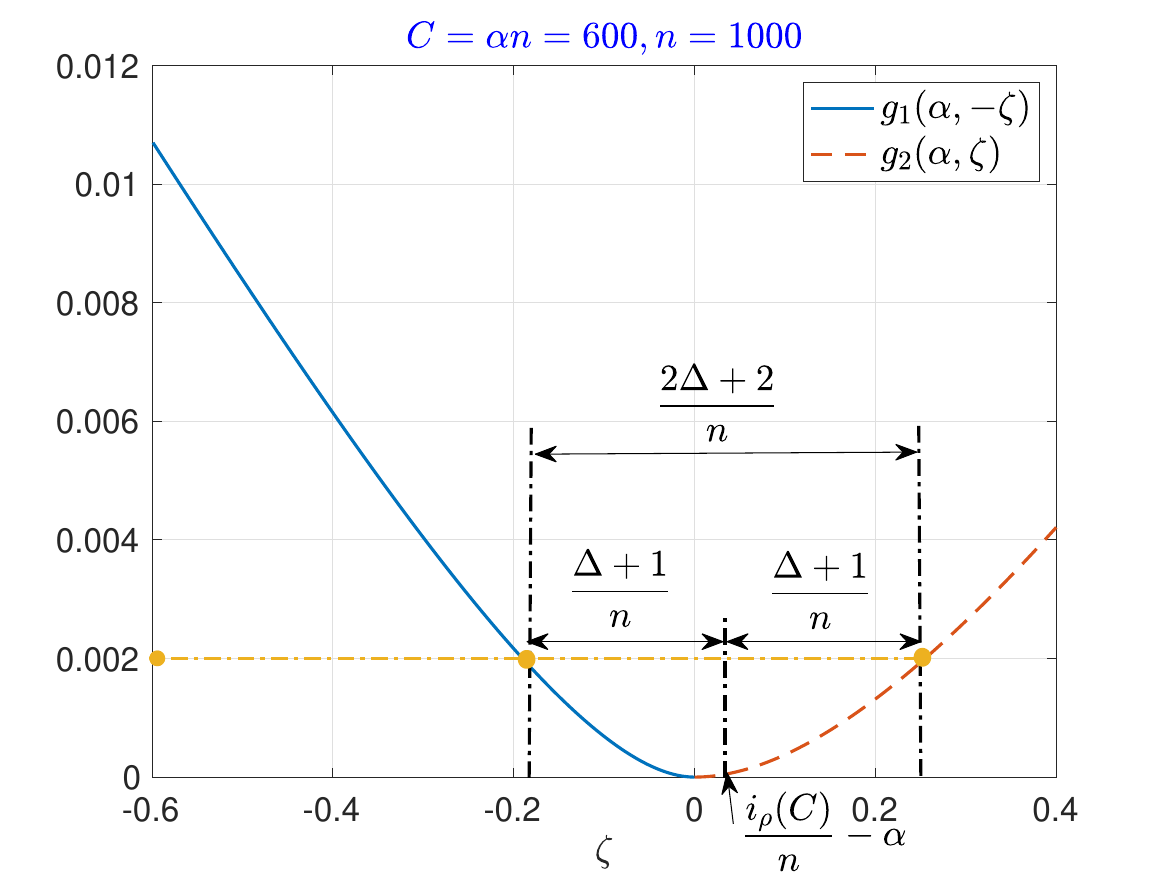}
				\vspace{-2.5cm}
				\label{Fig:CPD_irhoC}
	}
		\caption{$g_1(\alpha,-\zeta)$ and $g_2(\alpha,\zeta)$ versus $\zeta$ when $n=1000$, $C=\alpha n=600$, $r=10$, $\rho=0.1$, $P_1=\mathrm{Bern}(0.6)$ and $P_2=\mathrm{Bern}(0.2)$. Given any $\Delta$, (a) $\Delta_{n,\rmL}^{C,\rho}(\Delta)$ and $\Delta_{n,\rmR}^{C,\rho}(\Delta)$ are the values that sum up to $2\Delta$ and satisfy \eqref{Eq:def Delta^C L}; (b) $i_\rho(C)$ is the middle point between $C-\Delta_{n,\rmL}^{C,\rho}(\Delta)-1$ and $C+\Delta_{n,\rmR}^{C,\rho}(\Delta)+1$.}
		\label{Fig:illu Delta_n LR irhoC}
	\end{figure}

Given any $\Delta\in[0:n/2-1]$, $C\in[n]$ and $\rho\in(0,r)$, as shown in Figure \subref*{Fig:illu Delta_n LR}, we define $\Delta_{n,\rmL}^{C,\rho}(\Delta)\in[0,2\Delta]$ and $\Delta_{n,\rmR}^{C,\rho}(\Delta)\in[0,2\Delta]$ to be the unique numbers such that $\Delta_{n,\rmL}^{C,\rho}(\Delta)+\Delta_{n,\rmR}^{C,\rho}(\Delta)=2\Delta$ and
	\begin{align}
		 g_1\bigg(\frac{C}{n},\frac{\Delta_{n,\rmL}^{C,\rho}(\Delta)+1}{n} \bigg)=g_2\bigg(\frac{C}{n},\frac{\Delta_{n,\rmR}^{C,\rho}(\Delta)+1}{n} \bigg). \label{Eq:def Delta^C L}
	\end{align}

	Specifically, by letting $C-\Delta_{n,\rmL}^{C,\rho}(\Delta)-1=1$, we denote the unique solution to \eqref{Eq:def Delta^C L} as $\Delta_{n,\rmL}^\rho(\Delta)$; by letting $C+\Delta_{n,\rmR}^{C,\rho}(\Delta)+1=n$ and replacing $\Delta_{n,\rmL}^{C,\rho}(\Delta)$ with $2\Delta-\Delta_{n,\rmR}^{C,\rho}(\Delta)$ in \eqref{Eq:def Delta^C L}, we denote the unique solution to \eqref{Eq:def Delta^C L} as $\Delta_{n,\rmR}^\rho(\Delta)$. Note that $\Delta_{n,\rmL}^{C,\rho}(\Delta), \Delta_{n,\rmR}^{C,\rho}(\Delta)$ exist only when $C\in[\Delta_{n,\rmL}^\rho(\Delta)+2:n-\Delta_{n,\rmR}^\rho(\Delta)-1]$. 
	\begin{corollary}\label{coro:Delta_n L R increase}
		For any $C\in[n]$ and any $\rho\in(0,r)$, if $\Delta_{n,\rmL}^{C,\rho}(\Delta)$ and $\Delta_{n,\rmR}^{C,\rho}(\Delta)$ exist, they both are monotonically increasing in $\Delta\in[0:n/2-1]$. Furthermore, for any $\Delta\in[0:n/2-1]$, $\Delta_{n,\rmL}^{C,\rho}(\Delta)=\Theta(\Delta)$ and $\Delta_{n,\rmR}^{C,\rho}(\Delta)=\Theta(\Delta)$. 
	\end{corollary}
	The proof of Corollary \ref{coro:Delta_n L R increase} is provided in Appendix \ref{pf of coro:Delta_n L R increase}.

\subsection{Achievability Proof of \eqref{Eq:LDR Delta^*} } \label{Sec: achiev proof large}

	In this subsection, we  prove that for any $\xi>0$,  $\rmG_{\min}^{-1}(\lambda)+\xi$ is a $(r,\lambda,\epsilon,0)$-achievable NCW. That is, here we consider the achievability part for the large deviations regime in which $t=0$.

	Recall the definitions of $\bT_j^\rho$ and $\LD\Big(\bT_j^\rho,\frac{j+\rho n}{n+2\rho n} ,\frac{r-\rho}{1+2\rho} \Big)$ from \eqref{eqn:defLD} and under \eqref{Eq:tilX_2}. We assume that  the estimator is given by~\eqref{Eq:estimator}, with $\rho=n^{\kappa_n-1}$ ($\rho\in(0,r)$ for $n$ sufficiently large), $\Delta_n=n(\rmG_{\min}^{-1}(\lambda)+\xi)=\Theta(n)$ and  the threshold $\lambda$ is replaced by $\tilde{\lambda}=\lambda+\sigma_n$, where $\sigma_n= \frac{1}{n} (|\calX|\log ((N+n+1)^2(N+1)^2)+\log N)=O(\frac{\log n}{n})$ is a vanishing sequence.  Clearly $\limsup_{n\to\infty}{\Delta_n}/{n}\leq \rmG_{\min}^{-1}(\lambda)+\xi$.
	
	Let the true change-point be, as usual, denoted by $C \in[n]$. Then for all pairs of distributions $(\tilP_1,\tilP_2)\in\calP(\calX)^2$, the undetected error probability according to the probability measure $\tilde{\bbP}_C$  can be uniformly (in $(\tilP_1,\tilP_2)$) upper bounded as follows:
	\begin{align}
		&\tilde{\bbP}_C\{\calE_C\} =\tilde{\bbP}_C \bigg\{\exists i\notin[C\pm\Delta_n]\cap[n], \nn\\ 
		&\quad  \min_{j\notin[i\pm \Delta_n ]\cap[n]}(1\!+\! 2\rho)\LD\Big(\bT_j^\rho,\frac{j\!+\!\rho n}{n\!+\!2\rho n} ,\frac{r\!-\!\rho}{1\!+\!2\rho} \Big)\!>\!\tilde{\lambda} \bigg\}\\
		&=\tilde{\bbP}_C \bigg\{\exists i\notin[C\pm\Delta_n]\cap[n], \forall j\notin[i\pm \Delta_n ]\cap[n], \nn\\
		&\qquad \quad  (1+2\rho)\LD\Big(\bT_j^\rho,\frac{j+\rho n}{n+2\rho n} ,\frac{r-\rho}{1+2\rho} \Big)>\tilde{\lambda}  \bigg\}\\
		&\leq \tilde{\bbP}_C \bigg\{\LD\Big(\bT_C^\rho,\frac{C+\rho n}{n+2\rho n} ,\frac{r-\rho}{1+2\rho} \Big)>\frac{\tilde{\lambda}}{1+2\rho} \bigg\}\\
		&=\sum_{ {(x^n,y_1^N,y_2^N): \LD\big(\bT_C^\rho,\frac{C+\rho n}{n+2\rho n} ,\frac{r-\rho}{1+2\rho} \big)>\frac{\tilde{\lambda}}{1+2\rho} }} \nn\\
		&\qquad \quad \tilP_1^C(x^C)\tilP_2^{n-C}(x_{C+1}^n)\tilP_1^N(y_1^N)\tilP_2^N(y_2^N)\\
		&=\sum_{\bT^\rho_C:\LD\big(\bT_C^\rho,\frac{C+\rho n}{n+2\rho n} ,\frac{r-\rho}{1+2\rho} \big)>\frac{\tilde{\lambda}}{1+2\rho}} \nn \\
		&\qquad  \exp\bigg\{-(n+2\rho n)\LD\Big(\bT_C^\rho,\frac{C+\rho n}{n+2\rho n} ,\frac{r-\rho}{1+2\rho} \Big) \bigg\} \label{eqn:usedefLD}\\
		&< |\calP_{C+\rho n}||\calP_{n-C+\rho n}||\calP_{N-\rho n}|^2\exp\{-n\tilde{\lambda} \}\\
		&\leq \exp\{-n(\tilde{\lambda}-\delta_n ) \} \label{eqn:type_counting}\\
		&= \exp(-n\lambda),
	\end{align}
	where \eqref{eqn:usedefLD} uses the fact that $P^n(T_{x^n})=\exp\{-nD(T_{x^n}\|P)\}$ \cite{Csi97} and the  definition of $\LD(\fndot)$ in~\eqref{eqn:defLD} and~\eqref{eqn:type_counting} follows from the type counting lemma~\cite{Csi97}.
	Thus the constraint that the undetected error probability based on distributions $(\tilP_1, \tilP_2)$  is upper bounded by $\exp(-n\lambda)$ is satisfied. 
	
	Given any fixed pair of distributions $(P_1,P_2)\in\calP(\calX)^2$, we now bound the erasure probability
	\begin{align}
		\bbP_C\{\calE_\rme \}
		&=\bbP_C\bigg\{\max_{i\in[n]}\min_{j\notin[i\pm\Delta_n ]\cap[n]}(1+2\rho) \nn\\
		&\qquad \quad \cdot \LD\Big(\bT_j^\rho,\frac{j+\rho n}{n+2\rho n} ,\frac{r-\rho}{1+2\rho} \Big)\leq \tilde{\lambda}\bigg\}. \label{eqn:erasure_prob}
	\end{align}

To bound the erasure probability, we first need the following definitions.

	According to \eqref{eqn:erasure_prob}, we define the following random variables dependent on types $\bT_j^\rho$ of training and test sequences,
	\begin{align}
		I_\rho^*&:=\argmax_{i\in[n]}\min_{j\notin[i\pm\Delta_n]}\LD\Big(\bT_j^\rho,\frac{j+\rho n}{n+2\rho n} ,\frac{r-\rho}{1+2\rho} \Big), \label{Eq:I^*}\\
		J_\rho^*(i)&\!:=\!\argmin_{j\notin[i\pm\Delta_n]}\LD\Big(\bT_j^\rho,\frac{j+\rho n}{n+2\rho n} ,\frac{r-\rho}{1+2\rho} \Big), ~\forall i\in[n] \label{Eq:J^*}.
	\end{align}

Given any $C\in[n]$, let $\bP_j^-:=(P_1,\tilPjCleft,P_1,P_2)$ for $j\leq C$ and $\bP_j^+:=(\tilPjCright,P_2,P_1,P_2)$ for $j>C$.  Then we define
\begin{align}\label{Eq:i_rho(C)}
	i_\rho(C):=\argmax_{i\in[n]} \min_{j\notin[i\pm\Delta_n]}\LD\Big(\bP_{j},\frac{j+\rho n}{n+2\rho n} ,\frac{r-\rho}{1+2\rho} \Big),
\end{align}
where $\bP_j=\bP_j^-$ for $j\leq C$ and $\bP_j=\bP_j^+$ for $j> C$.

\begin{remark}[Values and properties of $i_\rho(C)$]\label{Rmk:value of irhoC} 
	Since  $\LD(\bP_{j}^-,\frac{j+\rho n}{n+2\rho n} ,\frac{r-\rho}{1+2\rho})$ is strictly decreasing in $j$ and $\LD(\bP_{j}^+,\frac{j+\rho n}{n+2\rho n} ,\frac{r-\rho}{1+2\rho} )$ is strictly increasing in $j$ (a fact that can be proved similarly as that of Lemma~\ref{Lem:Gmin increase}) and from Lemma \ref{Lem:GJS monotone in C}, we can obtain the values of $i_\rho(C)$ for different $C$ as follows:
	\begin{itemize}
		\item According to \eqref{Eq:def Delta^C L}, for any $C\in[\Delta_{n,\rmL}^\rho(\Delta_n)+2:n-\Delta_{n,\rmR}^\rho(\Delta_n)-1]$, we can obtain that $i_\rho(C)$ satisfies
		\begin{align}\label{Eq: i_rho(C) GJS equal}
			g_1\bigg(\frac{i_\rho(C)}{n},\frac{\Delta_n+1}{n} \bigg)=g_2\bigg(\frac{i_\rho(C)}{n},\frac{\Delta_n+1}{n} \bigg);
		\end{align}
		this is also illustrated in Figure \subref*{Fig:CPD_irhoC}.
		\item  For any $C\in[\Delta_{n,\rmL}^\rho(\Delta_n)+1]$, $i_\rho(C)=\Delta_n+1$. For any $C\in[n-\Delta_{n,\rmR}^\rho(\Delta_n):n]$, $i_\rho(C)=n-\Delta_n$.
	\end{itemize}
From Lemma \ref{Lem:GJS monotone in C}, we can deduce that $i_\rho(C)$ is monotonically increasing on $[\Delta_{n,\rmL}^\rho(\Delta_n)+2:n-\Delta_{n,\rmR}^\rho(\Delta_n)-1]$.
\end{remark}


	Recall that $\alpha=C/n$ and $\iota(\alpha)$ defined in \eqref{Eq:def iota}. Since $\lim_{n\to\infty}\rho=0$, we can see that $\lim_{n\to\infty}i_\rho(C)/n=\iota(\alpha)$.

	
We now present a lemma which states that for any true change-point $C$ and $\rho\in[n^{\kappa_n-1},r-n^{\kappa_n-1}]$, $I_\rho^*$ is close to $i_\rho(C)$ defined in \eqref{Eq:i_rho(C)} with high probability for $n$ large enough. This allows us to restrict our attention to a small interval of width $\approx n^{3/4}$ around $i_\rho(C)$ in the subsequent analyses. 

\begin{lemma}\label{lem:L(j)<L(i)}
	Let $C\in[n]$ be the true change-point. Given any $\rho\in[n^{\kappa_n-1},r-n^{\kappa_n-1}]$, we have for any pair of $(i,j)$ such that $i\in[C], j\in[i-n^{\kappa_n}] $ or $i\in[C+1:n], j\in[i+n^{\kappa_n}:n]$, 
	\begin{align}
		&\bbP_C\bigg\{\LD\Big(\bT_j^\rho,\frac{j+\rho n}{n+2\rho n} ,\frac{r-\rho}{1+2\rho} \Big)<\LD\Big(\bT_i^\rho,\frac{i+\rho n}{n+2\rho n} ,\frac{r-\rho}{1+2\rho} \Big) \bigg\} \nn\\
		&=O\bigg(\frac{1}{n^{3/2}}\bigg).
	\end{align}
\end{lemma}
The proof of Lemma \ref{lem:L(j)<L(i)} is provided in Appendix \ref{pf of lem:L(j)<L(i)}. With Lemma \ref{lem:L(j)<L(i)}, we can then prove the following lemma.

\begin{lemma}\label{lem:I_C close to i_C}
	Let $C\in[n]$ be the true change-point. For any $\rho\in[n^{\kappa_n-1},r-n^{\kappa_n-1}]$, we have
	\begin{align}
		\bbP_{C}\big\{I_\rho^*\notin[i_\rho(C)\pm n^{\kappa_n}] \big\}=O\bigg(\frac{1}{\sqrt{n}}\bigg).
	\end{align}
\end{lemma}
The proof of Lemma \ref{lem:I_C close to i_C} is provided in Appendix \ref{pf:pf of lem:I_C close to i_C}.

	\begin{figure}[t]
		\centering
		\includegraphics[height=0.11\textheight,width=1.05\columnwidth]{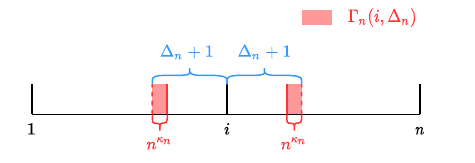}
		\caption{Illustration of $\Gamma_n(i,\Delta_n)$ for any $i\in[n]$.}
		\label{fig:R_c_border}
	\end{figure}
	
 The next preparatory result  implies that the minimizer of the set $\{ \LD(\bT_j^\rho,\frac{j+\rho n}{n+2\rho n} ,\frac{r-\rho}{1+2\rho} ): j \notin [i_\rho(C)\pm \Delta_n  ]\}$ is close to the boundary (end points) of $[i_\rho(C)\pm \Delta_n]$  with high probability.  With this observation, we can again restrict the subsequent analyses to two small intervals close to the boundary of $[i_\rho(C)\pm \Delta_n]$. To be precise, we define for any $i\in[n]$
\begin{align}
	\Gamma_n^{\rmL}(i,\Delta_n) & :=[n]\cap (i-\Delta_n-1,i-\Delta_n-1+n^{\kappa_n}] \label{Eq:Gamma L}\\*
	\Gamma_n^{\rmR}(i,\Delta_n) & :=[n]\cap  [i+\Delta_n+1-n^{\kappa_n},i+\Delta_n+1) \label{Eq:Gamma R}\\*
	\Gamma_n(i,\Delta_n) &:=  \Gamma_n^{\rmL}(i,\Delta_n)\cup\Gamma_n^{\rmR}(i,\Delta_n). \label{Eq:Gamma}
\end{align}
 The set $\Gamma_n(i_\rho(C),\Delta_n)$ represents the union of  two small intervals $\Gamma_n^{\rmL}(i_\rho(C),\Delta_n)$ and $\Gamma_n^{\rmR}(i_\rho(C),\Delta_n)$ each of width  $n^{\kappa_n}$ adjacent to the boundary of $[i_\rho(C)\pm \Delta_n]$ (and restricted to be in $[n]$) as shown in Fig.~\ref{fig:R_c_border}. 
 

	\begin{lemma}\label{Lem:second min=C+-(delta+1)}
		Let $C\in[n]$ be the true change-point. For any $\rho\in[n^{\kappa_n-1},r-n^{\kappa_n-1}]$ and any $\Delta_n\geq n^{\kappa_n}$,
		\begin{align}
			&\bbP_{C}\bigg\{\argmin_{j\notin[i_\rho(C)\pm(\Delta_n-n^{\kappa_n})]}\LD\Big(\bT_j^\rho,\frac{j+\rho n}{n+2\rho n} ,\frac{r-\rho}{1+2\rho} \Big) \nn\\* 
			&\qquad \quad \notin\Gamma_n(i_\rho(C),\Delta_n) \bigg\} 
			=  O\bigg(\frac{1}{\sqrt{n}}\bigg).
		\end{align}
	\end{lemma}
	By applying the union bound over all $O(n)$ values of $j\notin[i\pm \Delta_n]$ and Lemma \ref{lem:L(j)<L(i)}, 
	Lemma \ref{Lem:second min=C+-(delta+1)} is proved.
	 
	With Lemma \ref{Lem:V>0, T finite}, we can then apply the central limit theorem, more precisely a uniform version of it such as the Berry-Esseen theorem, to bound~\eqref{eqn:erasure_prob}. 

	Let us define the following events 
	\begin{align}
		\calF_1&:=\bigg\{\min_{j\notin [i_\rho(C)\pm(\Delta_n-n^{\kappa_n})]}(1+2\rho) \nn\\
		&\qquad \quad \cdot \LD\Big(\bT_j^\rho,\frac{j+\rho n}{n+2\rho n} ,\frac{r-\rho}{1+2\rho} \Big)\leq \tilde{\lambda} \bigg\}, \quad\mbox{and}\\*
		\calF_2&:=\bigg\{\argmin_{j\notin[i_\rho(C)\pm(\Delta_n-n^{\kappa_n})]}\LD\Big(\bT_j^\rho,\frac{j+\rho n}{n+2\rho n} ,\frac{r-\rho}{1+2\rho} \Big) \nn\\
		 &\qquad \quad\in \Gamma_n(i_\rho(C),\Delta_n) \bigg\}.
	\end{align} 
	It can then be verified from the definitions of these events and that of $\Gamma_n(C,\Delta_n)$ that
	\begin{align}
		&\calF_1\cap \calF_2 \nn\\
		&=\bigg\{\min_{j\in\Gamma_n(i_\rho(C),\Delta_n) } (1+2\rho)\LD\Big(\bT_j^\rho,\frac{j+\rho n}{n+2\rho n} ,\frac{r-\rho}{1+2\rho} \Big) \leq \tilde{\lambda}\bigg\}.
	\end{align}
	With Lemma \ref{lem:I_C close to i_C} and Lemma \ref{Lem:second min=C+-(delta+1)}, we can bound the erasure probability as follows:
	\begin{align}
		&\bbP_C\{\calE_\rme \}\nn\\
		 &=\bbP_C\bigg\{\min_{j\notin[i\pm\Delta_n ]\cap[n]}(1+2\rho)\LD\Big(\bT_j^\rho,\frac{j+\rho n}{n+2\rho n} ,\frac{r-\rho}{1+2\rho} \Big)\leq \tilde{\lambda},\nn\\
		 &\qquad \qquad \forall i\in[n] \bigg\}  \nn\\
		&=\bbP_C\bigg\{ \min_{j\notin[I_\rho^* \pm \Delta_n ]\cap[n]}(1\! +\! 2\rho)\LD\Big(\bT_j^\rho,\frac{j+\rho n}{n+2\rho n} ,\frac{r-\rho}{1+2\rho} \Big)\! \leq\!  \tilde{\lambda} \bigg\}\\
		&\leq \bbP_C\bigg\{ \min_{j\notin[I_\rho^* \pm \Delta_n ]\cap[n]}(1+2\rho)\LD\Big(\bT_j^\rho,\frac{j+\rho n}{n+2\rho n} ,\frac{r-\rho}{1+2\rho} \Big)\leq \tilde{\lambda},\nn\\
		&\qquad \quad I_\rho^*\! \in\! [i_\rho(C)\! \pm\!  n^{\kappa_n}] \bigg\}\! +\! \bbP_C\big\{I_\rho^*\! \notin\! [i_\rho(C)\! \pm\!  n^{\kappa_n}] \big\} \label{Eq:use lem I_C close to i_C}\\
		&\leq \bbP_C\{\calF_1\}+O\bigg(\frac{1}{\sqrt{n}}\bigg)\\
		\label{Eq:in or out Rc_border}&\leq  (\bbP_C\{\calF_1\cap \calF_2\} +\bbP_C\{\calF_2^{\rmc}\})+O\bigg(\frac{1}{\sqrt{n}}\bigg)\\
		\label{Eq:apply lemma 3}&= \bbP_C\{\calF_1\cap \calF_2\}+O\bigg(\frac{1}{\sqrt{n}}\bigg)\\
		&\leq \bbP_C\bigg\{ \min_{\Gamma_n^\rmL(i_\rho(C),\Delta_n)}(1\! +\! 2\rho)\LD\Big(\bT_j^\rho,\frac{j+\rho n}{n+2\rho n} ,\frac{r-\rho}{1+2\rho} \Big)\leq \tilde{\lambda} \bigg\}\nn\\
		&~ +\bbP_C\bigg\{ \min_{\Gamma_n^\rmR(i_\rho(C),\Delta_n)}(1\! +\! 2\rho)\LD\Big(\bT_j^\rho,\frac{j+\rho n}{n+2\rho n} ,\frac{r-\rho}{1+2\rho} \Big)\! \leq \! \tilde{\lambda} \bigg\} \nn\\
		&~ +O\bigg(\frac{1}{\sqrt{n}} \bigg), \label{eqn:use_ub}
	\end{align}
	where \eqref{Eq:use lem I_C close to i_C} follows from Lemma \ref{lem:I_C close to i_C}, \eqref{Eq:apply lemma 3} follows from Lemma \ref{Lem:second min=C+-(delta+1)} and \eqref{eqn:use_ub} follows from the union bound.
	Note that in \eqref{eqn:use_ub} the first term is equal to 0 if $i_\rho(C)< \Delta_{n}+2$ (i.e. $C<\Delta_{n,\rmL}^\rho(\Delta_n)+2$ according to Remark \ref{Rmk:value of irhoC}) and the second term is equal to 0 for $i_\rho(C)> n-\Delta_{n}-1$ (i.e. $C>n-\Delta_{n,\rmR}^\rho(\Delta_n)-1$ according to Remark \ref{Rmk:value of irhoC}). For simplicity, we let $\tau_n=\Delta_{n,\rmL}^\rho(\Delta_n)+2$ and $\tau_n'=n-\Delta_{n,\rmR}^\rho(\Delta_n)-1$ in the following analyses.

	We then define 
	\begin{align}
		&C_{\rmL}^*:=\argmin_{j\in\Gamma_n^{\rmL}(i_\rho(C),\Delta_n)}\LD\Big(\bT_j^\rho,\frac{j+\rho n}{n+2\rho n} ,\frac{r-\rho}{1+2\rho} \Big), \quad\mbox{and} \nn\\
		&C_{\rmR}^*:=\argmin_{j\in\Gamma_n^{\rmR}(i_\rho(C),\Delta_n)}\LD\Big(\bT_j^\rho,\frac{j+\rho n}{n+2\rho n} ,\frac{r-\rho}{1+2\rho} \Big).
	\end{align}

	In the following, we provide an upper bound on $\bbP_C \big\{ (1+2\rho)\LD(\bT_{C_{\rmL}^*}^\rho,\frac{C_{\rmL}^*+\rho n}{n+2\rho n} ,\frac{r-\rho}{1+2\rho}) \leq \tilde{\lambda}\big\}$ by first conditioning on typical random vectors of sufficiently long lengths and then approximating the function $\LD(\fndot)$ with sums of independent log-likelihood  terms. Finally, we bound $\bbP_C \big\{ (1+2\rho)\LD(\bT_{C_{\rmL}^*}^\rho,\frac{C_{\rmL}^*+\rho n}{n+2\rho n} ,\frac{r-\rho}{1+2\rho}) \leq \tilde{\lambda}\big\}$  using the uniform version of central limit theorem. 
	
	Recall the definition of the atypical event $\calA_{i,j}^{\rho}$ in \eqref{eqn:atypical set}. On the event $(\calA_{C_\rmL^*,C}^{\rho})^{\rmc}$, all partial test ($\tilX_1^{C_\rmL^*+\rho n}, X_{C_\rmL^*+1}^C, \tilX_2^{n-C+\rho n}$) and full training ($Y_1^{N-\rho n}, Y_2^{N-\rho n}$) sequences are typical according to the definition in \eqref{eqn:typ_B}. Thus,  by applying a Taylor expansion to the function $\LD(\fndot)$ and using the bound in~\eqref{eqn:atypical},  we have \eqref{Eq:typical set C_L begin}--\eqref{Eq:typical set C_L} on the top of next page,
	\begin{figure*}
	\begin{align}
		\nn&\bbP_C\Big\{(1+2\rho)\LD\Big(\bT_{C_{\rmL}^*}^\rho,\frac{C_{\rmL}^*+\rho n}{n+2\rho n} ,\frac{r-\rho}{1+2\rho} \Big) \leq \tilde{\lambda}\Big\}\\*
		& \le \bbP_C\Big\{(1+2\rho)\LD\Big(\bT_{C_{\rmL}^*}^\rho,\frac{C_{\rmL}^*+\rho n}{n+2\rho n} ,\frac{r-\rho}{1+2\rho} \Big) \leq \tilde{\lambda},\;(\calA_{C_\rmL^*,C}^\rho)^{\rmc} \Big\} +\bbP_C\big\{\calA_{C_\rmL^*,C}^\rho \big\} \label{Eq:typical set C_L begin} \\
		\nn&=\bbP_C\bigg\{\frac{1}{n+2\rho n} \bigg( \sum_{i\in[C_\rmL^*+1:n]}\log\frac{(n-C_\rmL^*+N)\tilP_{C_\rmL^*}^-(X_i)}{(n-C_\rmL^*+\rho n)\tilP_{C_\rmL^*}^-(X_i)+(N-\rho n)P_2(X_i)} \nn\\
		&\qquad \quad +\sum_{i\in[N-\rho n:N]}\log\frac{(n-C_\rmL^*+N)\tilP_{C_\rmL^*}^-(Y_{2,i})}{(n-C_\rmL^*+\rho n)\tilP_{C_\rmL^*}^-(Y_{2,i})+(N-\rho n)P_2(Y_{2,i})} \bigg) \nn \\*
		&\qquad \quad  +\frac{1}{n+2\rho n} \sum_{i\in[N-\rho n]} \log\frac{(n-C_\rmL^*+N)P_2(Y_{2,i})}{(n-C_\rmL^*+\rho n)\tilP_{C_\rmL^*}^-(Y_{2,i})+(N-\rho n)P_2(Y_{2,i})} +O\bigg(\frac{\log (C-C_\rmL^*)}{C-C_\rmL^*}\bigg)\leq \frac{\tilde{\lambda}}{1+2\rho}, \; (\calA_{C_\rmL^*,C}^\rho)^{\rmc}  \bigg\}\nn\\
		&\qquad +\bbP_C\{ \calA_{C_\rmL^*,C}^\rho\}\label{eqn:taylor_ach}\\
		\label{Eq:typical set C_L}\nn&\le\bbP_C\bigg\{\frac{1}{n+2\rho n} \bigg(\sum_{i\in[C_\rmL^*+1:n]} \log\frac{(n-C_\rmL^*+N)\tilP_{C_\rmL^*}^-(X_i)}{(n-C_\rmL^*+\rho n)\tilP_{C_\rmL^*}^-(X_i)+(N-\rho n)P_2(X_i)} \nn\\
		&\qquad \quad +\sum_{i\in[N-\rho n:N]} \log\frac{(n-C_\rmL^*+N)\tilP_{C_\rmL^*}^-(Y_{2,i})}{(n-C_\rmL^*+\rho n)\tilP_{C_\rmL^*}^-(Y_{2,i})+(N-\rho n)P_2(Y_{2,i})} \bigg) \nn\\
		&\qquad \quad +\frac{1}{n+2\rho n} \sum_{i\in[N-\rho n]} \log\frac{(n-C_\rmL^*+N)P_2(Y_{2,i})}{(n-C_\rmL^*+\rho n)\tilP_{C_\rmL^*}^-(Y_{2,i})+(N-\rho n)P_2(Y_{2,i})}+O\bigg(\frac{\log n}{n}\bigg)\!\leq\! \frac{\tilde{\lambda}}{1+2\rho} \bigg\} \! +\!O\bigg(\frac{1}{n^{2\kappa_n}}\bigg).
	\end{align}\hrulefill
	\end{figure*}
	where \eqref{Eq:typical set C_L} follows from the bound on the probability of the atypical set in \eqref{eqn:atypical} and  the fact that $ C-C_{\rmL}^*=\Theta(\Delta_n)=\Theta(n)$ from Corollary \ref{coro:Delta_n L R increase}. 
	 Now since the random variables involved in \eqref{Eq:typical set C_L} are independent, by applying the uniform version of the central limit theorem, we have
	 \begin{align}
	 	\nn &\limsup_{n\to\infty}\max_{C\in[\tau_n,n]}  \bbP_C\Big\{\LD\Big(\bT_{C_{\rmL}^*}^\rho,\frac{{C_{\rmL}^*+\rho n}}{n+2\rho n},\frac{r-\rho}{1+2\rho}\Big) \leq \frac{\tilde{\lambda}}{1+2\rho}\Big\} \\
	 	\label{Eq:C_L apply CLT}&\leq\limsup_{n\to\infty} \max_{C\in[\tau_n,n]} \max_{C_\rmL^*\in\Gamma_n^{\rmL}(i_\rho(C),\Delta_n)} \Phi\Bigg(\bigg(\frac{\tilde{\lambda}}{1+2\rho}\nn\\
	 	&\qquad -\frac{r-\rho}{1+2\rho}\mathrm{GJS}\bigg(\tilP_{C_\rmL^*}^-,P_2,\frac{n-C_\rmL^*+\rho n}{(r-\rho)n}\bigg)+O\bigg(\frac{\log n }{n}\bigg) \bigg) \nn\\*
	 	&\qquad \times \sqrt{\frac{(n+2\rho n)^2}{(n-C_\rmL^*+rn)V(\tilP_{C_\rmL^*}^-,P_2,C_\rmL^*,r,\rho)}} \Bigg) \\
	 	&\le\limsup_{n\to\infty} \max_{C\in[\tau_n,n]}  \max_{C_\rmL^*\in\Gamma_n^{\rmL}(i_\rho(C),\Delta_n)} \Phi\Bigg(\bigg(\tilde{\lambda} \nn\\
	 	&\qquad -(r-\rho)\mathrm{GJS}\bigg(\tilP_{C_\rmL^*}^-,P_2,\frac{n-C_\rmL^*+\rho n}{(r-\rho)n}\bigg)+O\bigg(\frac{\log n }{n}\bigg) \bigg) \nn\\
	 	&\qquad \times \sqrt{ \frac{n}{c_1(\rho)} } \Bigg) \label{eqn:c_1}\\
	 	&= \Phi\Bigg(\limsup_{n\to\infty}\max_{C\in[\tau_n,n]}\max_{C_\rmL^*\in\Gamma_n^{\rmL}(i_\rho(C),\Delta_n)} \bigg(\tilde{\lambda}\nn\\
	 	&\qquad -(r-\rho)\mathrm{GJS}\bigg(\tilP_{C_\rmL^*}^-,P_2,\frac{n-C_\rmL^*+\rho n}{(r-\rho)n}\bigg)+O\bigg(\frac{\log n }{n}\bigg) \bigg) \nn\\
	 	&\qquad \times \sqrt{ \frac{n}{c_1(\rho)} } \Bigg)\label{eqn:below c_1},
	 \end{align}
	 where \eqref{eqn:c_1}, in which $0<c_1(\rho)<\infty$ is a constant, uses the bounds on $V(\tilP_{C_\rmL^*}^-,P_2,C_\rmL^*,r,\rho)$ as stated in Lemma \ref{Lem:V>0, T finite}, and the final step follows from the continuity and monotonicity of~$\Phi$.

	In a symmetric fashion, we obtain
	\begin{align}
		\nn&\limsup_{n\to\infty}\max_{C\in[1,\tau_n']}  \bbP_C\Big\{\LD\Big(\bT_{C_{\rmR}^*},\frac{{C_{\rmR}^*+\rho n}}{n+2\rho n},\frac{r-\rho}{1+2\rho}\Big) \leq \frac{\tilde{\lambda}}{1+2\rho} \Big\}\\
		&\leq \Phi\Bigg(\limsup_{n\to\infty}\max_{C\in[1,\tau_n']} \max_{C_\rmR^*\in\Gamma_n^{\rmR}(i_\rho(C),\Delta_n)} \bigg(\tilde{\lambda} \nn\\
		&\qquad -(r-\rho)\mathrm{GJS}\bigg(\tilP_{C_\rmR^*}^+,P_1,\frac{C_\rmR^*+\rho n}{(r-\rho) n}\bigg)+O\bigg(\frac{\log n }{n}\bigg) \bigg) \nn\\
		&\qquad \times \sqrt{\frac{n }{c_2(\rho)}} \Bigg)\label{eqn:c_2}
	\end{align}
	for some $0<c_2(\rho)<\infty$.

	To simplify the limit superiors inside the $\Phi(\fndot)$ functions in \eqref{eqn:below c_1} and \eqref{eqn:c_2}, it would be useful to leverage the  monotonicity properties of $\rmG_{\min}(\cdot)$ (cf. \eqref{Eq: def Gmin}) in Corollary \ref{coro:Gmin increase}. 

	
	Recall that $\alpha=C/n$, $\iota(\alpha)$ is defined in \eqref{Eq:def iota}, and $\Delta_{\rmL}(\cdot)$, $\Delta_{\rmR}(\cdot)$ defined in \eqref{Eq:def Delta_L} and \eqref{Eq:def Delta_R}. Since $\lim_{n\to\infty}\rho=0$, we have $\lim_{n\to\infty}\tau_n/n=\lim_{n\to\infty}(\Delta_{n,\rmL}^\rho(\Delta_n)+2)/n=\Delta_{\rmL}(\rmG_{\min}^{-1}(\lambda)+\xi)$, $\lim_{n\to\infty}\tau_n'/n=\lim_{n\to\infty}(n-\Delta_{n,\rmR}^\rho(\Delta_n)-1)/n=1-\Delta_{\rmR}(\rmG_{\min}^{-1}(\lambda)+\xi)$ and $\lim_{n\to\infty}C_{\rmL}^*/n=\lim_{n\to\infty}(i_\rho(C)-\Delta_n)/n=\iota(\alpha)-(\rmG_{\min}^{-1}(\lambda)+\xi)$. Using the continuity of $\mathrm{GJS}$ in the distributions and the third argument, and $\lim_{n\to\infty}\rho=0$, we see that 
	\begin{align}
		&\liminf_{n\to\infty}\min_{C\in[\tau_n,n]}\min_{C_\rmL^*\in\Gamma_n^{\rmL}(i_\rho(C),\Delta_n)} \nn\\
		&\qquad  (r-\rho)\mathrm{GJS}\bigg(\tilP_{C_\rmL^*}^-,P_2,\frac{n-C_\rmL^*+\rho n}{(r-\rho)n}\bigg) \nn\\*
		&=\min_{\alpha\in[\Delta_\rmL(\rmG_{\min}^{-1}(\lambda)+\xi) ,1]}\nn\\
		&\qquad  r \mathrm{GJS}\bigg(\frac{(\alpha-\iota(\alpha)+\rmG_{\min}^{-1}(\lambda)+\xi) P_1+(1-\alpha)P_2}{1-\iota(\alpha)+\rmG_{\min}^{-1}(\lambda)+\xi},P_2, \nn\\*
		&\qquad \qquad \qquad \frac{1-\iota(\alpha)+\rmG_{\min}^{-1}(\lambda)+\xi}{r}\bigg) \label{eqn:gjs1}\\
		&=\min_{{\alpha\in[\Delta_\rmL(\rmG_{\min}^{-1}(\lambda)+\xi)  ,  1-\Delta_\rmR(\rmG_{\min}^{-1}(\lambda)+\xi)] }} \nn\\
		&\qquad r \mathrm{GJS}\bigg(\frac{(\alpha-\iota(\alpha)+\rmG_{\min}^{-1}(\lambda)+\xi) P_1+(1-\alpha)P_2}{1-\iota(\alpha)+\rmG_{\min}^{-1}(\lambda)+\xi},P_2, \nn\\*
		&\qquad \qquad \qquad \frac{1-\iota(\alpha)+\rmG_{\min}^{-1}(\lambda)+\xi}{r}\bigg) \label{Eq: shrink alpha range} \\
		&=\min_{\alpha\in[\Delta_\rmL(\rmG_{\min}^{-1}(\lambda)) ,1-\Delta_\rmR(\rmG_{\min}^{-1}(\lambda))]}\nn\\
		&\qquad  r \mathrm{GJS}\bigg(\frac{(\alpha-\iota(\alpha)+\rmG_{\min}^{-1}(\lambda)) P_1+(1-\alpha)P_2}{1-\iota(\alpha)+\rmG_{\min}^{-1}(\lambda)},P_2, \nn\\
		&\qquad \qquad \qquad \frac{1-\iota(\alpha)+\rmG_{\min}^{-1}(\lambda)}{r}\bigg)+\xi_1' \label{Eq: G increase in Delta}\\
		&=\rmG_{\min}(\rmG_{\min}^{-1}(\lambda))+\xi_1'\\
		&=\lambda+\xi_1',
	\end{align}
	where 
	 \eqref{Eq: shrink alpha range} follows since $\iota(\alpha)$ is fixed for all $\alpha\in[1-\Delta_{\rmR}(\rmG_{\min}^{-1}(\lambda)+\xi),1]$ and from the proof of Lemma \ref{Lem:Gmin increase}, \eqref{Eq: G increase in Delta} follows from Corollary \ref{coro:Gmin increase} and $\xi_1'> 0$ is a funciton of $\xi $ satisfying $\xi_1'\to 0^+$ as $\xi\to 0^+$. 
	
	Similarly,
	\begin{align}
		&\liminf_{n\to\infty}\min_{C\in[1,\tau_n']} \min_{C_\rmR^*\in\Gamma_n^{\rmR}(i_\rho(C),\Delta_n)} \nn\\*
		&\qquad  (r-\rho)\mathrm{GJS}\bigg(\tilP_{C_\rmR^*}^+,P_1,\frac{C_\rmR^*+\rho n}{(r-\rho) n}\bigg) \nn\\
		&=\min_{\alpha\in[0,1-\Delta_{\rmR}(\rmG_{\min}^{-1}(\lambda)+\xi)]} \nn\\*
		&\qquad r\mathrm{GJS}\bigg(\frac{\alpha P_1+(\iota(\alpha)+\rmG_{\min}^{-1}(\lambda)+\xi-\alpha)P_2}{\iota(\alpha)+\rmG_{\min}^{-1}(\lambda)+\xi},P_1, \nn\\
		&\qquad \qquad \qquad \frac{\iota(\alpha)+\rmG_{\min}^{-1}(\lambda)+\xi}{r}\bigg)\\
		&=\min_{\alpha\in[\Delta_\rmL(\rmG_{\min}^{-1}(\lambda)) ,1-\Delta_\rmR(\rmG_{\min}^{-1}(\lambda))]} 
		\nn\\
		&\qquad r\mathrm{GJS}\bigg(\frac{\alpha P_1+(\iota(\alpha)+\rmG_{\min}^{-1}(\lambda)-\alpha)P_2}{\iota(\alpha)+\rmG_{\min}^{-1}(\lambda)},P_1, \nn\\
		&\qquad \qquad \qquad  \frac{\iota(\alpha)+\rmG_{\min}^{-1}(\lambda)}{r}\bigg)+\xi_2' \label{Eq: achieve GJS2}\\
		&=\lambda+\xi_2',
	\end{align}
	where  $\xi_2'> 0$ is a funciton of $\xi $ satisfying $\xi_2'\to 0^+$ as $\xi\to 0^+$.
	
	Using these observations, definitions and the continuity of $\Phi(\fndot)$, we have
	\begin{align}
		&\limsup_{n\to\infty}\max_{C\in[\tau_n,n]}  \bbP_C\Big\{\LD\Big(\bT_{C_{\rmL}^*}^\rho,\frac{{C_{\rmL}^*+\rho n}}{n+2\rho n},\frac{r-\rho}{1+2\rho}\Big) \leq \frac{\tilde{\lambda}}{1+2\rho}\Big\} \nn\\
		&\leq \Phi\bigg(\limsup_{n\to\infty}\big(\tilde{\lambda}-\lambda-\xi_1')\sqrt{\frac{n}{c_1(\rho)}} \bigg) \label{eqn:bound_G1}
	\end{align}
	and
	\begin{align}
		&\limsup_{n\to\infty}\max_{C\in[1,\tau_n']}  \bbP_C\Big\{\LD\Big(\bT_{C_{\rmR}^*},\frac{{C_{\rmR}^*+\rho n}}{n+2\rho n},\frac{r-\rho}{1+2\rho}\Big) \leq \frac{\tilde{\lambda}}{1+2\rho} \Big\} \nn\\
		&\leq\Phi\bigg(\limsup_{n\to\infty}\big(\tilde{\lambda}-\lambda-\xi_2')\sqrt{\frac{n}{c_2(\rho)}} \bigg). \label{eqn:bound_G2} 
	\end{align}

	Recall that $\tilde{\lambda} = \lambda+o(1)$. From Lemma \ref{Lem:V>0, T finite}, we have $0<\liminf_{n\to\infty}c_1(\rho),\liminf_{n\to\infty}c_2(\rho)<\infty$.
	Then, for any $\lambda\in(0,\rmG_{\min}( 1/2))$, $\lim_{n\to\infty}\tilde{\lambda} =\lambda$,
	which guarantees that both the upper bounds in \eqref{eqn:bound_G1} and \eqref{eqn:bound_G2} vanish. Then the  sequence of estimators as defined in~\eqref{Eq:estimator} yields a sequence of erasure probabilities $\{\epsilon_n\}_{n=1}^\infty$  and a sequence of confidence widths $\{\Delta_n\}_{n=1}^\infty$ such that 
	$\limsup_{n\to\infty}\epsilon_n=0 \leq \epsilon$ (for any $\epsilon\in [0,1)$) and $\limsup_{n\to\infty}\Delta_n/n=\rmG_{\min}^{-1}(\lambda)+\xi$.
	Therefore,  $\rmG_{\min}^{-1}(\lambda)+\xi$ is a $(r,\lambda,\epsilon,0)$-achievable NCW.  Since $\xi>0$ is arbitrary, taking $\xi\to 0^+$, we see that 
	\begin{align}\label{Eq: lambda lower in LDR}
		\bar{\Delta}^*(r,\lambda,\epsilon) \leq \rmG_{\min}^{-1}(\lambda).
	\end{align}

	\subsection{Converse Proof of \eqref{Eq:LDR Delta^*}}\label{Sec: converse LDR}
	
	Given any vector  $\bm{\eta}:=[\eta_{1},\ldots,\eta_{n}]\in[0,1]^{n}$,  let
	\begin{align}
	\eta_{\min}:=\min_{l\in[n]}\eta_{l},\quad\mbox{and}\quad  \eta_{\mathrm{sum}}:=\sum_{l\in[n]}\eta_{l}.
	\end{align}
	
For any $\rho\in(0,r)$ and any $j\in[n]$, recall the tuple of types $\bT_j^\rho=(T_{\tilX_1^{j+\rho n}},T_{\tilX_2^{n-j+\rho n}},T_{Y_1^{N-\rho n}},T_{Y_2^{N-\rho n}} )$ defined under \eqref{Eq:tilX_2}. 
		We denote an estimator that  only makes use of $\bT_j^\rho$ as a type-based estimator $\gamma^{\mathrm{type}}_{\rho,j}:\calX^{n+2N}\to[n]\cup\{\rme\} $. The following lemma shows that for any arbitrary estimator $\gamma$, given any $j\in[n]$, there always exists a type-based estimator $\gamma^{\mathrm{type}}_{\rho,j}$ that has a similar asymptotic performance as $\gamma$ under the case when the change-point is $j$.
	
	\begin{lemma}\label{Lem:type based test optimal}
		Given any arbitrary estimator $\gamma:\calX^{n+2N}\to [n]\cup\{\rme\}$, any $\bm{\eta}\in[0,1]^n$ and any $\rho\in(0,r)$, for any pair of distributions $(P_1,P_2)\in\calP(\calX)^2$ and any $j\in[n]$, there exists a type-based estimator $\gamma^{\mathrm{type}}_{\rho,j}$ such that
		\begin{align}
		&\bbP_j\big\{\gamma(X^n,Y_1^N,Y_2^N)\notin [j\pm\Delta]\cup\{\rme\}  \big\} \nn\\ 
		&\geq \eta_{\min}\bbP_j\big\{\gamma^{\mathrm{type}}_{\rho,j}(\bT_j^\rho)\notin  [j\pm\Delta]\cup\{\rme\}\big\},
		\end{align}
		and
		\begin{align}
		\bbP_{j}\{\gamma(X^n,Y_1^N,Y_2^N)=\rme \}\geq (1-\eta_{\mathrm{sum}})\bbP_{j}\{\gamma^{\mathrm{type}}_{\rho,j}(\bT_j^\rho)=\rme \}.
		\end{align}
	\end{lemma}
	The proof of Lemma \ref{Lem:type based test optimal} is provided in Appendix \ref{proof of Lemma type based test optimal}. 
	
	Let $\delta_n=\frac{1}{n}|\calX|\log((n+rn+1)^2 (N+1)^2) = O(\frac{\log n}{ n})$. In Section \ref{Sec: achiev proof large}, we have proven that given any $C\in[n]$, the estimator in \eqref{Eq:estimator} satisfies $\tilde{\bbP}_C\{\calE_C\}\leq \exp(-n\lambda)$ for all $(\tilP_1,\tilP_2)\in\calP(\calX)^2$. The following lemma shows that for each $C\in[n]$, given any type-based estimator $\gamma^{\mathrm{type}}_{\rho,C}$ that also statisfies $\tilde{\bbP}_C\{\calE_C\}\leq \exp(-n\lambda)$ for all $(\tilP_1,\tilP_2)\in\calP(\calX)^2$, the erasure probability of it can be approximately lower bounded by that of the estimator in \eqref{Eq:estimator}.

	\begin{lemma}\label{Lem:proposed test type-based optimal}
		For any $\lambda\in\bbR_+$, any $\rho\in(0,r)$, and any tuple of type-based estimators $\{\gamma^{\mathrm{type}}_{\rho,j}\}_{j\in[n]}$ such that for all pairs of distributions $(\tilP_1,\tilP_2)\in\calP(\calX)^2$,
		\begin{align}\label{Eq:error prob less than exp(-n lambda)}
		 \max_{C\in[n]}\tilde{\bbP}_C\big\{\gamma^{\mathrm{type}}_{\rho,C}(\bT_C^\rho)\notin [C\pm\Delta]\cup\{\rme\}  \big\}\leq \exp(-n\lambda),
		\end{align}
		we have that for any particular pair of distributions $(P_1,P_2)\in\calP(\calX)^2$ and for any $C\in[n]$,
		\begin{align}
		 &\bbP_C\{\gamma^{\mathrm{type}}_{\rho,C}(\bT_C^\rho)=\rme \}\nn\\
		 &\geq   \bbP_{C}\bigg\{\min\limits_{j\notin [i\pm\Delta]} (1+2\rho)\bigg(\LD\Big(\bT_j^\rho,\frac{j+\rho n}{n+2\rho n},\frac{r-\rho}{1+2\rho}\Big)+\delta_n \bigg) \nn\\
		 &\qquad \qquad  \leq \lambda, \forall i\in[n] \bigg\}.
		\end{align}		
	\end{lemma}
	The proof of Lemma \ref{Lem:proposed test type-based optimal} is provided in Appendix \ref{proof of Lem our test optimal}. 
	
	The following corollary shows that given any  estimator $\gamma$ with  undetected error probability  upper bounded by $\exp(-n\lambda)$, the erasure probability of $\gamma$ can be approximately lower bounded in terms of the tail probability of the function $\LD(\fndot)$, defined in~\eqref{eqn:defLD}.  Let $\Xi_n:=[n^{\kappa_n-1},r-n^{\kappa_n-1}]$.
	\begin{corollary}\label{Coro:proposed estimator optimal}
		For any $\lambda\in\bbR_+$ and any estimator $\gamma$ such that for all pairs for $(\tilP_1,\tilP_2)\in\calP(\calX)^2$,
		\begin{align}\label{Eq:any estimator error decay expo}
		\max_{C\in[n]} \tilde{\bbP}_C\big\{\gamma(X^n,Y_1^N,Y_2^N)\notin [C\pm\Delta]\cup\{\rme\} \big\}\leq \exp(-n\lambda),
		\end{align}
		we have that for any pair of $(P_1,P_2)\in\calP(\calX)^2$ and any $C\in[n]$,
		\begin{align}
		&\bbP_C\{\gamma(X^n,Y_1^N,Y_2^N)=\rme \}  \nn \\
		&\geq\bigg(1-\frac{1}{n}\bigg)\min_{\rho\in\Xi_n} \bbP_{C}\bigg\{\min\limits_{j\notin [i\pm\Delta]}(1+2\rho) \nn\\
		&\qquad  \times \bigg(\LD\Big(\bT_j^\rho,\frac{j+\rho n}{n+2\rho n},\frac{r-\rho}{1+2\rho}\Big) + \delta_n  \bigg)+\frac{2\log n}{n}\leq \lambda, \nn\\
		&\qquad ~ \forall i\in[n] \bigg\}.
		\end{align}
	\end{corollary}
	Combining Lemma \ref{Lem:type based test optimal} and \ref{Lem:proposed test type-based optimal} and setting $\eta_l={1}/{n^2}$ for all $l\in[n]$, we obtain Corollary~\ref{Coro:proposed estimator optimal}. 
	

	Let $\hatC_\rmL:=i_\rho(C)-\Delta-n^{\kappa_n}-1$ and $\hatC_\rmR:=i_\rho(C)+\Delta+n^{\kappa_n}+1$.
	For any $C\in[n]$ and any estimator $\gamma$ satisfying \eqref{Eq:any estimator error decay expo}, the erasure probability can be lower bounded as follows:
	\begin{align}
	&\bbP_C\{\calE_\rme \} \nn\\
	&\geq \bigg(1-\frac{1}{n}\bigg)\min_{\rho\in\Xi_n}\bbP_{C}\bigg\{\min\limits_{j\notin [i\pm\Delta]}\LD\Big(\bT_j^\rho,\frac{j+\rho n}{n+2\rho n},\frac{r-\rho}{1+2\rho}\Big) \nn\\
	&\qquad +\frac{2\log n}{(1+2\rho)n}\leq \frac{\lambda}{1+2\rho}-\delta_n, \forall i\in[n]\bigg\} \label{eqn:erasure_lb} \nn\\
	&=\bigg(1-\frac{1}{n}\bigg)\min_{\rho\in\Xi_n}\bbP_{C}\bigg\{\min\limits_{j\notin [I_\rho^*\pm\Delta]}\LD\Big(\bT_j^\rho,\frac{j+\rho n}{n+2\rho n},\frac{r-\rho}{1+2\rho}\Big) \nn\\
	&\qquad +\frac{2\log n}{(1+2\rho)n}\leq \frac{\lambda}{1+2\rho}-\delta_n \bigg\} \\
	&\geq \bigg(1-\frac{1}{n}\bigg)\min_{\rho\in\Xi_n}\bbP_{C}\bigg\{\min\limits_{j\notin [I_\rho^*\pm\Delta]}\LD\Big(\bT_j^\rho,\frac{j+\rho n}{n+2\rho n},\frac{r-\rho}{1+2\rho}\Big)\nn\\
	&\qquad +\frac{2\log n}{(1+2\rho)n}\leq \frac{\lambda}{1+2\rho}-\delta_n, |I_\rho^*-i_\rho(C)|\leq n^{\kappa_n} \bigg\} \\
	\label{Eq:min j in smaller range}&\geq \bigg(1-\frac{1}{n}\bigg)\min_{\rho\in\Xi_n}\bbP_{C}\bigg\{\min\limits_{j\notin [i_\rho(C)\pm(\Delta+n^{\kappa_n})]} \nn\\
	&\qquad \LD\Big(\bT_j^\rho,\frac{j+\rho n}{n+2\rho n},\frac{r-\rho}{1+2\rho}\Big)+\frac{2\log n}{(1+2\rho)n}\leq \frac{\lambda}{1+2\rho}-\delta_n, \nn\\
	&\qquad |I_\rho^*-i_\rho(C)|\leq n^{\kappa_n} \bigg\}\\
	&\geq \bigg(1-\frac{1}{n}\bigg)\min_{\rho\in\Xi_n}\bigg(\bbP_{C}\bigg\{\min\limits_{j\notin [i_\rho(C)\pm(\Delta+n^{\kappa_n})]} \nn\\
	&\qquad \LD\Big(\bT_j^\rho,\frac{j+\rho n}{n+2\rho n},\frac{r-\rho}{1+2\rho}\Big)\!+\!\frac{2\log n}{(1+2\rho)n}\leq \frac{\lambda}{1+2\rho}\!-\!\delta_n\bigg\}\! \nn\\
	&\qquad  - \!\bbP_{C}\{ |I_\rho^*-i_\rho(C)|\!>\! n^{\kappa_n} \}\bigg) \label{eqn:probrue}\\
	\label{Eq:lower bound max}&\geq\bigg(1-\frac{1}{n}\bigg)\max\Bigg\{\min_{\rho\in\Xi_n}\bbP_{C}\bigg\{\LD\Big(\bT_{\hat{C}_\rmL}^\rho,\frac{\hat{C}_\rmL+\rho n}{n+2\rho n},\frac{r-\rho}{1+2\rho} \Big)\nn\\
	&\qquad +\frac{2\log n}{(1+2\rho)n}\leq \frac{\lambda}{1+2\rho}-\delta_n \bigg\},  \nn\\
	&\qquad \min_{\rho\in\Xi_n}\bbP_{C}\bigg\{\LD\Big(\bT_{\hat{C}_\rmR}^\rho,\frac{\hat{C}_\rmR+\rho n}{n+2\rho n},\frac{r-\rho}{1+2\rho}\Big)+\frac{2\log n}{(1+2\rho)n} \nn\\*
	&\qquad \leq \frac{\lambda}{1+2\rho}-\delta_n \bigg\} \Bigg\} +O\bigg(\frac{1}{\sqrt{n}}\bigg),
	\end{align}
 where \eqref{Eq:min j in smaller range} follows since $[n]\cap[i_\rho(C)\pm(\Delta+n^{\kappa_n})]^{\rmc}\subset [n]\cap[I_\rho^*\pm\Delta]^{\rmc}$, \eqref{eqn:probrue} follows from the fact that $\bbP_C\{ \calD\cap\calA^\rmc\}\ge\bbP_C\{\calD\} - \bbP_C\{\calA\}$ given any two events $\calD$ and $\calA$ and \eqref{Eq:lower bound max} follows from Lemma \ref{lem:I_C close to i_C}.
	
	Note that for any $\rho\in\Xi_n$, if $i_\rho(C)<\Delta+n^{\kappa_n}+2$ (i.e. $C<\Delta_{n,\rmL}^\rho(\Delta+n^{\kappa_n})+2$ according to Remark \ref{Rmk:value of irhoC}), the first probability term in \eqref{Eq:lower bound max} is equal to 0; on the other hand if $i_\rho(C)>n-\Delta-n^{\kappa_n}-1$ (i.e. $C>n-\Delta_{n,\rmR}^\rho(\Delta+n^{\kappa_n})-1$ according to Remark \ref{Rmk:value of irhoC}), the second probability term in~\eqref{Eq:lower bound max} is equal to $0$. For simplicity, let $\tilde{\tau}_n=\max_{\rho\in\Xi_n}\Delta_{n,\rmL}^\rho(\Delta+n^{\kappa_n})+2$ and $\tilde{\tau}_n'=n-\max_{\rho\in\Xi_n}\Delta_{n,\rmR}^\rho(\Delta+n^{\kappa_n})-1$. Since the left and right-hand sides of \eqref{Eq:def Delta^C L} increases as $\rho$ decreases (a fact that can be proved similarly as that of Lemma \ref{Lem:GJS monotone in C} by using \eqref{Eq: i_rho(C) GJS equal}), we can subsequently deduce that $\Delta_{n,\rmL}^\rho(\Delta+n^{\kappa_n})$ and $\Delta_{n,\rmR}^\rho(\Delta+n^{\kappa_n})$ both increase as $\rho$ decreases and thus, $\tilde{\tau}_n=\Delta_{n,\rmL}^{n^{\kappa_n-1}}(\Delta+n^{\kappa_n})+2$ and $\tilde{\tau}_n'=n-\Delta_{n,\rmR}^{n^{\kappa_n-1}}(\Delta+n^{\kappa_n})-1$.
	
	Next, for any $\rho\in\Xi_n$, we have \eqref{eqn:prob_rules begin}--\eqref{eqn:prob_rules} on the top of next page,
	\begin{figure*}
	\begin{align}
	\nn&\bbP_{C}\bigg\{\LD\Big(\bT_{\hat{C}_\rmL}^\rho,\frac{\hat{C}_\rmL+\rho n}{n+2\rho n},\frac{r-\rho}{1+2\rho} \Big)+\frac{2\log n}{(1+2\rho)n}\leq \frac{\lambda}{1+2\rho}-\delta_n \bigg\}\\*
	&\geq \bbP_{C}\bigg\{\LD\Big(\bT_{\hat{C}_\rmL}^\rho,\frac{\hat{C}_\rmL+\rho n}{n+2\rho n},\frac{r-\rho}{1+2\rho} \Big)+\frac{2\log n}{(1+2\rho)n} \leq \frac{\lambda}{1+2\rho}-\delta_n, (\calA_{\hat{C}_\rmL,C}^\rho)^\rmc \bigg\}  \label{eqn:prob_rules begin}\\
	\label{Eq:lower bound apply CLT}\nn&= \bbP_C\bigg\{\frac{1}{n+2\rho n}\sum_{i\in[\hatC_\rmL+1:n]} \log\frac{(n-\hatC_\rmL+N)\tilP_{\hatC_\rmL}^-(X_i)}{(n-\hatC_\rmL+\rho n)\tilP_{\hatC_\rmL}^-(X_i)+(N-\rho n)P_2(X_i)} \nn \\
	&\qquad \quad +\frac{1}{n+2\rho n}\sum_{i\in[\rho n]} \log\frac{(n-\hatC_\rmL+N)\tilP_{\hatC_\rmL}^-(Y_{2,i})}{(n-\hatC_\rmL+\rho n)\tilP_{\hatC_\rmL}^-(Y_{2,i})+(N-\rho n)P_2(Y_{2,i})} \\*
	&\qquad \quad +\frac{1}{n+2\rho n}\sum_{i\in[\rho n+1:N]}\log\frac{(n-\hatC_\rmL+N)P_2(Y_{2,i})}{(n-\hatC_\rmL+\rho n)\tilP_{\hatC_\rmL}^-(Y_{2,i})+(N-\rho n)P_2(Y_{2,i})}  +O\bigg(\frac{\log\tau_n'}{\tau_n'}\bigg)\leq \frac{\lambda}{1+2\rho} ,~ (\calA_{\hat{C}_\rmL,C}^\rho)^\rmc \bigg\}\\
	&\ge \bbP_C\bigg\{\frac{1}{n+2\rho n}\sum_{i\in[\hatC_\rmL+1:n]}  \log\frac{(n-\hatC_\rmL+N)\tilP_{\hatC_\rmL}^-(X_i)}{(n-\hatC_\rmL+\rho n)\tilP_{\hatC_\rmL}^-(X_i)+(N-\rho n)P_2(X_i)} \nn\\
	&\qquad \quad +\frac{1}{n+2\rho n}\sum_{i\in[\rho n]}  \log\frac{(n-\hatC_\rmL+N)\tilP_{\hatC_\rmL}^-(Y_{2,i})}{(n-\hatC_\rmL+\rho n)\tilP_{\hatC_\rmL}^-(Y_{2,i})+(N-\rho n)P_2(Y_{2,i})} \nn\\*
	&\qquad \quad +\frac{1}{n+2\rho n}\sum_{i\in[\rho n+1:N]} \log\frac{(n-\hatC_\rmL+N)P_2(Y_{2,i})}{(n-\hatC_\rmL+\rho n)\tilP_{\hatC_\rmL}^-(Y_{2,i})+(N-\rho n)P_2(Y_{2,i})}  +O\bigg(\frac{\log\tau_n'}{\tau_n'}\bigg)\leq \lambda\bigg\}+O\bigg(\frac{1}{n^{2\kappa_n}}\bigg). \label{eqn:prob_rules}
	\end{align} \hrulefill
\end{figure*}
where~\eqref{Eq:lower bound apply CLT} follows from the same steps as those leading to~\eqref{eqn:taylor_ach} in the achievability part and~\eqref{eqn:prob_rules} again follows from the fact that $\bbP\{\calD\cap\calA^\rmc\}\ge\bbP\{\calD\} - \bbP\{\calA\}$ and from the probability of the atypical event in \eqref{eqn:atypical}. Since the random variables involved in \eqref{eqn:prob_rules} are independent, by applying the uniform version of central limit theorem, we have
	\begin{align}
	\nn&\liminf\limits_{n\to\infty}\max_{C\in[\tilde{\tau}_n,n]}\min_{\rho\in\Xi_n}\bbP_{C}\bigg\{\LD\Big(\bT_{\hat{C}_\rmL}^\rho,\frac{\hat{C}_\rmL+\rho n}{n+2\rho n},\frac{r-\rho}{1+2\rho} \Big) \nn\\
	&\qquad \qquad \qquad\qquad\quad +\frac{2\log n}{(1+2\rho)n}\leq \frac{\lambda}{1+2\rho}-\delta_n \bigg\}\\
	&\geq \liminf\limits_{n\to\infty} \max_{C\in[\tilde{\tau}_n,n]}\min_{\rho\in\Xi_n}\Phi\Bigg(\bigg(\lambda \nn\\
	&\qquad -(r-\rho)\mathrm{GJS}\bigg(\tilP_{\hatC_\rmL}^-,P_2,\frac{n-\hatC_\rmL+\rho n}{(r-\rho)n}\bigg)\nn\\*
	&\qquad +O\bigg(\frac{\log n}{n} \bigg) \bigg) \nn\\
	&\qquad \times \sqrt{\frac{n^2}{(n-\hatC_\rmL+rn)V(\tilP_{\hatC_\rmL}^-,P_2,\hat{C}_\rmL,r,\rho)}} \Bigg)\\
	\label{Eq:converse c1}&\geq \liminf\limits_{n\to\infty}  \max_{C\in[\tilde{\tau}_n,n]}\min_{\rho\in\Xi_n}\Phi\Bigg(\bigg(\lambda \nn\\
	&\qquad -(r-\rho)\mathrm{GJS}\bigg(\tilP_{\hatC_\rmL}^-,P_2,\frac{n-\hatC_\rmL+\rho n}{(r-\rho)n}\bigg) \nn\\
	&\qquad +O\bigg(\frac{\log n}{n} \bigg) \bigg)\sqrt{\frac{n}{\hat{c}_1(\rho)}} \Bigg)\\
	&=\Phi\Bigg(\liminf\limits_{n\to\infty}\max_{C\in[\tilde{\tau}_n,n]}\min_{\rho\in\Xi_n}\bigg(\lambda \nn\\
	&\qquad -(r-\rho)\mathrm{GJS}\bigg(\tilP_{\hatC_\rmL}^-,P_2,\frac{n-\hatC_\rmL+\rho n}{(r-\rho)n}\bigg) \nn\\
	&\qquad +O\bigg(\frac{\log n}{n} \bigg) \bigg)\sqrt{\frac{n}{\hat{c}_1(\rho)}} \Bigg),
	\end{align}
	where \eqref{Eq:converse c1}, in which $0<\hat{c}_1(\rho)<\infty$ is a constant, uses the bounds on $V(\tilP_{\hat{C}_\rmL}^-,P_2,\hat{C}_\rmL,r,\rho)$ as stated in Lemma~\ref{Lem:V>0, T finite}, and the final step follows from the continuity and monotonicity of $\Phi(\fndot)$.
	In a symmetric fashion, we obtain
	\begin{align}
	\nn&\liminf\limits_{n\to\infty}\max_{C\in[1,\tilde{\tau}_n']} \min_{\rho\in\Xi_n}\bbP_{C}\bigg\{\LD\Big(\bT_{\hat{C}_\rmR}^\rho,\frac{\hat{C}_\rmR+\rho n}{n+2\rho n},\frac{r-\rho}{1+2\rho} \Big) \nn\\
	&\qquad \qquad \qquad \qquad\quad  +\frac{2\log n}{(1+2\rho)n}\leq \frac{\lambda}{1+2\rho}-\delta_n \bigg\}\\
	\label{Eq: same before this eq}&\geq \Phi\Bigg(\liminf\limits_{n\to\infty}   \max_{C\in[1,\tilde{\tau}_n']} \min_{\rho\in\Xi_n}\bigg(\lambda \nn\\*
	&\qquad \quad -(r-\rho)\mathrm{GJS}\bigg(\tilP_{\hatC_\rmR}^+,P_1,\frac{\hatC_\rmR+\rho n}{(r-\rho)n}\bigg)+O\bigg(\frac{\log n}{n} \bigg) \bigg) \nn\\*
	&\qquad \quad \times\sqrt{\frac{n}{\hat{c}_2(\rho)}} \Bigg)
	\end{align}
	for some $0<\hat{c}_2(\rho)<\infty$.
	
Recall the definitions of $\Delta_{\rmL}(\cdot),\Delta_{\rmR}(\cdot)$ in \eqref{Eq:def Delta_L} and \eqref{Eq:def Delta_R}. Let $\Delta=n(\rmG_{\min}^{-1}(\lambda)-\xi)$ for some arbitrarily small constant $\xi>0$ and let $\iota_\rho(\alpha)=\lim_{n\to\infty}i_\rho(C)/n$. 
Note that $\lim_{n\to\infty}\tilde{\tau}_n/n=\lim_{n\to\infty}(\Delta_{n,\rmL}^{n^{\kappa_n-1}}(\Delta+n^{\kappa_n})+2)/n=\Delta_{\rmL}(\rmG_{\min}^{-1}(\lambda)-\xi)$ and $\lim_{n\to\infty}\tilde{\tau}_n'/n=\lim_{n\to\infty}(n-\Delta_{n,\rmR}^{n^{\kappa_n-1}}(\Delta+n^{\kappa_n})-1)/n=1-\Delta_{\rmR}(\rmG_{\min}^{-1}(\lambda)-\xi)$. Using the continuity of $\mathrm{GJS}$ in the distributions and the third argument, we have \eqref{Eq:lim GJS CL begin}--\eqref{Eq:lim GJS CL end},
\begin{figure*}[t]
	\begin{align}
		&\limsup\limits_{n\to\infty}\min_{C\in[\tilde{\tau}_n,n]} \max_{\rho\in\Xi_n} (r-\rho)\mathrm{GJS}\bigg(\tilP_{\hatC_\rmL}^-,P_2,\frac{n-\hatC_\rmL+\rho n}{(r-\rho)n}\bigg) \label{Eq:lim GJS CL begin}\\*
		&=\min_{\alpha\in[\Delta_\rmL(\rmG_{\min}^{-1}(\lambda)-\xi) ,1]} \max_{\rho\in(0,r)} (r-\rho) \mathrm{GJS}\bigg(\frac{(\alpha-\iota_\rho(\alpha)+\rmG_{\min}^{-1}(\lambda)-\xi) P_1+(1-\alpha+\rho)P_2}{1-\iota_\rho(\alpha)+\rmG_{\min}^{-1}(\lambda)-\xi+\rho}, \nn\\ 
		&\hspace{2.5in} P_2,\frac{1-\iota_\rho(\alpha)+\rmG_{\min}^{-1}(\lambda)-\xi+\rho}{r-\rho}\bigg) \label{Eq:GJS decrease in rho}\\
		&=\min_{\alpha\in[\Delta_\rmL(\rmG_{\min}^{-1}(\lambda)-\xi) ,1]} r \mathrm{GJS}\bigg(\frac{(\alpha-\iota(\alpha)+\rmG_{\min}^{-1}(\lambda)-\xi) P_1+(1-\alpha)P_2}{1-\iota(\alpha)+\rmG_{\min}^{-1}(\lambda)-\xi},P_2,\frac{1-\iota(\alpha)+\rmG_{\min}^{-1}(\lambda)-\xi}{r}\bigg) \label{Eq:max over rho GJS}\\
		&=\min_{\substack{\alpha\in[\Delta_\rmL(\rmG_{\min}^{-1}(\lambda)-\xi), \\ 1-\Delta_\rmR(\rmG_{\min}^{-1}(\lambda)-\xi)]}}r \mathrm{GJS}\bigg(\frac{(\alpha-\iota(\alpha)+\rmG_{\min}^{-1}(\lambda)-\xi) P_1+(1-\alpha)P_2}{1-\iota(\alpha)+\rmG_{\min}^{-1}(\lambda)-\xi},P_2,\frac{1-\iota(\alpha)+\rmG_{\min}^{-1}(\lambda)-\xi}{r}\bigg) \label{Eq: shrink alpha range converse}\\
		&=\min_{\substack{\alpha\in[\Delta_\rmL(\rmG_{\min}^{-1}(\lambda)), \\ 1-\Delta_\rmR(\rmG_{\min}^{-1}(\lambda))]}} r \mathrm{GJS}\bigg(\frac{(\alpha-\iota(\alpha)+\rmG_{\min}^{-1}(\lambda)) P_1+(1-\alpha)P_2}{1-\iota(\alpha)+\rmG_{\min}^{-1}(\lambda)},P_2,\frac{1-\iota(\alpha)+\rmG_{\min}^{-1}(\lambda)}{r}\bigg) -\xi_1'' \label{Eq: G increase in Delta converse}\\
		&=\lambda-\xi_1''. \label{Eq:lim GJS CL end}
	\end{align}
\hrulefill
\end{figure*}
	where \eqref{Eq:max over rho GJS} follows since the function of $(\alpha,\rho)$ in \eqref{Eq:GJS decrease in rho} decreases as $\rho$ increases for any fixed $\alpha$ (a fact that can be proved similarly as that of Lemma \ref{Lem:GJS monotone in C} by using \eqref{Eq: i_rho(C) GJS equal}), \eqref{Eq: shrink alpha range converse} follows since $\iota(\alpha)$ is fixed for all $\alpha\in[1-\Delta_{\rmR}(\rmG_{\min}^{-1}(\lambda)-\xi),1]$ and from the proof of Lemma \ref{Lem:Gmin increase}, \eqref{Eq: G increase in Delta converse} follows from Corollary \ref{coro:Gmin increase} and  $\xi_1''>0$ is a function of $\xi$ satisfying $\xi_1''\to 0^+$ as $\xi\to0^+$.
	Similarly,
	\begin{align}
	&\limsup_{n\to\infty}  \min_{C\in[1, \tilde{\tau}_n']}\max_{\rho\in\Xi_n} (r-\rho)\mathrm{GJS}\bigg(\tilP_{\hat{C}_\rmR}^+,P_1,\frac{\hat{C}_\rmR+\rho n}{(r-\rho)n}\bigg)\nn\\*
	&=\min_{ \substack{\alpha\in[\Delta_\rmL(\rmG_{\min}^{-1}(\lambda)),\\ 1-\Delta_\rmR(\rmG_{\min}^{-1}(\lambda))]} } r\mathrm{GJS}\bigg(\frac{\alpha P_1+(\iota(\alpha)+\rmG_{\min}^{-1}(\lambda)-\alpha)P_2}{\iota(\alpha)+\rmG_{\min}^{-1}(\lambda)}, \nn\\
	&\hspace{1.2in} P_1,\frac{\iota(\alpha)+\rmG_{\min}^{-1}(\lambda)}{r}\bigg)-\xi_2''\\
	&=\lambda-\xi_2'',
	\end{align}
	where $\xi_2''>0$ satisfies $\xi_2''\to 0^+$ as $\xi\to0^+$.
	
	Using these bounds and the continuity of $\Phi(\fndot)$, we have 
	\begin{align}\label{Eq:lim P_L lower bound}
		\nn&\liminf\limits_{n\to\infty}\max_{C\in[\tilde{\tau}_n,n]} \min_{\rho\in\Xi_n}\bbP_{C}\bigg\{\LD\Big(\bT_{\hat{C}_\rmL}^\rho,\frac{\hat{C}_\rmL+\rho n}{n+2\rho n},\frac{r-\rho}{1+2\rho} \Big)\nn\\
		&\hspace{1.2in} +\frac{2\log n}{(1+2\rho)n}\leq \frac{\lambda}{1+2\rho}-\delta_n \bigg\}\\
		&\geq \Phi\Bigg(\liminf_{n\to\infty}~ \xi_1'' \sqrt{\frac{n}{\max_{\rho\in\Xi_n}\hat{c}_1(\rho)}} \Bigg)
	\end{align}
 	and
	\begin{align}\label{Eq:lim P_R lower bound}
		\nn&\liminf\limits_{n\to\infty} \max_{C\in[1,\tilde{\tau}_n']} \min_{\rho\in\Xi_n}\bbP_{C}\bigg\{\LD\Big(\bT_{\hat{C}_\rmR}^\rho,\frac{\hat{C}_\rmR+\rho n}{n+2\rho n},\frac{r-\rho}{1+2\rho} \Big)\nn\\*
		&\hspace{1.2in}+\frac{2\log n}{(1+2\rho)n}\leq \frac{\lambda}{1+2\rho}-\delta_n \bigg\}\\
		&\geq \Phi\Bigg(\liminf_{n\to\infty}~\xi_2'' \sqrt{\frac{n}{\max_{\rho\in\Xi_n}\hat{c}_2(\rho)}} \Bigg).
	\end{align}
From Lemma \ref{Lem:V>0, T finite}, we have that $0<\limsup_{n\to\infty}\max_{\rho\in\Xi_n}\hat{c}_1(\rho),\limsup_{n\to\infty}\max_{\rho\in\Xi_n}\hat{c}_2(\rho)<\infty$. Since $\xi_1'',\xi_2''>0$, we can see that both of the lower bounds in  \eqref{Eq:lim P_L lower bound} and \eqref{Eq:lim P_R lower bound} tend to one,  which means that the erasure probability of all estimators, which is lower bounded as in~\eqref{eqn:erasure_lb}, cannot be made less  than $\epsilon$ for any $\epsilon\in [0,1)$. This constitutes a strong converse. Thus, the optimal NCW cannot be smaller than or equal to $\rmG_{\min}^{-1}(\lambda)-\xi$, i.e., $\bar{\Delta}^*(r,\lambda,\epsilon) >\rmG_{\min}^{-1}(\lambda)-\xi$. Since $\xi>0$ can be made arbitrarily small, it means that
	\begin{align}
		\bar{\Delta}^*(r,\lambda,\epsilon)\geq \rmG_{\min}^{-1}(\lambda),
	\end{align}
	completing the proof of the converse of \eqref{Eq:LDR Delta^*}.

	\subsection{Achievability Proof of \eqref{Eq:MDR Delta^*}} \label{Sec: proof of Eq:MDR Delta^*}
%
	Throughout the proof, we let $\rho=n^{\kappa_n-1}$ and let $\bar{\Delta}_t$ be equal to the right-hand side of \eqref{Eq:MDR Delta^*}. 
We seek to prove that for any $\xi>0$, $\bar{\Delta}_t+\xi$ is an $(r,\lambda,\epsilon,t)$-achievable NCW. That is, here we consider the achievability part for the moderate deviations regime in which $t\in (0,1/2)$.  Let $\Delta_{n}=n^{1-t/2}(\bar{\Delta}_t+\xi)=\Theta(n^{1-t/2})$ and clearly $\limsup_{n\to\infty}\Delta_{n}/n^{1-t/2}\leq \bar{\Delta}_t+\xi$. Since $t\in(0,1/2)$ and $\kappa_n\to 3/4 $ (see its definition in \eqref{eqn:kappa_n}), from Corollary \ref{coro:Delta_n L R increase}, we have that  $C-C_\rmL^*=\Theta(\Delta_{n,\rmL}^{C,\rho}(\Delta_n))=\Theta(n^{1-t/2})$, $C_\rmR^*-C=\Theta(\Delta_{n,\rmR}^{C,\rho}(\Delta_n))=\Theta(n^{1-t/2})$, $\tau_n=\Theta(\Delta_n)=\Theta(n^{1-t/2})$ and $\tau'=n-\Theta(\Delta_n)=\Theta(n)$. 
	
	The achievability proof of \eqref{Eq:MDR Delta^*} is the same as that for~\eqref{Eq:LDR Delta^*} up to~\eqref{Eq:C_L apply CLT} with $\tilde{\lambda}$ replaced by $\tilde{\lambda}n^{-t}$ and $\lambda$ replaced by $\lambda n^{-t}$. We highlight the salient differences in the following steps. In essence, for the moderate deviations case, since the normalized gaps $\frac{1}{n}(C-C_\rmL^*)$ and  $\frac{1}{n}(C_\rmR^*-C)$ vanish, we are able to exploit this to approximate  the GJS functions that appear in \eqref{eqn:gjs1}--\eqref{Eq: achieve GJS2} by   chi-square divergences by invoking Taylor's theorem. 
	
	By applying the central limit theorem and Lemma \ref{Lem:V>0, T finite}, we have
	\begin{align}
	\nn&\limsup_{n\to\infty}\max_{C\in[\tau_n,n]}\bbP_C\bigg\{\LD\Big(\bT_{C_{\rmL}^*}^\rho,\frac{{C_{\rmL}^*}+\rho n}{n+2\rho n},\frac{r-\rho}{1+2\rho}\Big) \leq \frac{\tilde{\lambda}n^{-t}}{1+2\rho} \bigg\}\\*
	&\leq \Phi\Bigg(\limsup_{n\to\infty}\max_{C\in[\tau_n,n]} \max_{C_\rmL^*\in\Gamma_n^{\rmL}(i_\rho(C),\Delta_n)} \nn\\
	&\qquad \quad  \bigg(\frac{1}{n^t}\bigg(\tilde{\lambda}-(r-\rho) n^t \mathrm{GJS}\bigg(\tilP_{C_\rmL^*}^-,P_2,\frac{n-C_\rmL^*+\rho n}{(r-\rho)n}\bigg) \bigg) \nn\\
	&\qquad \quad +O\bigg(\frac{\log n}{n^{1-t/2}}\bigg) \bigg) \sqrt{ \frac{n}{c_1(\rho)}} \Bigg),
	\end{align} 
	where $c_1(\rho)$ is defined under \eqref{eqn:below c_1}.
	In a completely similar and symmetric fashion, we have
	\begin{align}
	\nn&\limsup_{n\to\infty}\max_{C\in[1,\tau_n']}\bbP_C\bigg\{\LD\Big(\bT_{C_{\rmR}^*}^\rho,\frac{{C_{\rmR}^*}+\rho n}{n+2\rho n},\frac{r-\rho}{1+2\rho}\Big) \leq \frac{\tilde{\lambda}n^{-t}}{1+2\rho}\bigg\}\\
	&\leq \Phi\Bigg(\limsup_{n\to\infty}\max_{C\in[1,\tau_n']}\max_{C_\rmR^*\in\Gamma_n^{\rmR}(i_\rho(C),\Delta_n)} \nn\\
	&\qquad \quad \bigg(\frac{1}{n^t}\bigg(\tilde{\lambda}-(r-\rho)n^t\mathrm{GJS}\bigg(\tilP_{C_\rmR^*}^+,P_1,\frac{C_\rmR^*+\rho n}{(r-\rho)n}\bigg)\bigg) \nn\\
	&\qquad \quad +O\bigg(\frac{\log n}{n^{1-t/2}} \bigg) \bigg)\sqrt{\frac{n }{c_2(\rho)}} \Bigg),
	\end{align}
	where $c_2(\rho)$ is defined under \eqref{eqn:c_2}.
	
	Since $\frac{1}{n}(C -C_\rmL^*)=\Theta(n^{-t/2})$ for any $C_\rmL^*\in\Gamma_n^{\rmL}(i_\rho(C),\Delta_n)$ and $\frac{1}{n}(C_\rmR^*-C)=\Theta(n^{-t/2})$ for any $C_\rmR^*\in\Gamma_n^{\rmR}(i_\rho(C),\Delta_n)$ are vanishing, we have the following Taylor expansion of $(r-\rho)\mathrm{GJS}\big(\fndot ,P_2,( n-C_\rmL^*+\rho n)/( (r-\rho)n)\big)$ around $P_2$:
	\begin{align}\label{Eq:taylor expand GJS_L}
		&(r-\rho)\mathrm{GJS}\bigg(\tilP_{C_\rmL^*}^-,P_2,\frac{n-C_\rmL^*+\rho n}{(r-\rho)n}\bigg) \nn\\
		&=\frac{(r-\rho)(C-C_\rmL^*)^2}{2(n-C_\rmL^*+\rho n)(n-C_\rmL^*+rn)}\upchi_2(P_1\|P_2) \nn\\
		&\quad +O\bigg(\frac{ (C-C_\rmL^*)^3  }{ (n-C_\rmL^*+\rho n)(n-C_\rmL^*+rn)^2 }\bigg).
	\end{align}
	Similarly, Taylor expanding of $(r-\rho)\mathrm{GJS}\big(\fndot,P_1, (C_\rmR^*+\rho n)/((r-\rho)n)\big)$ around $ P_1$:
	\begin{align}\label{Eq:taylor expand GJS_R}
		&(r-\rho)\mathrm{GJS}\bigg(\tilP_{C_\rmR^*}^+,P_1,\frac{C_\rmR^*+\rho n}{(r-\rho)n}\bigg) \nn\\
		&=\frac{(r-\rho)(C_\rmR^*-C)^2}{2(C_\rmR^*+\rho n)(C_\rmR^*+rn)}\upchi_2(P_2\|P_1) \nn\\
		&\quad +O\bigg(\frac{ (C_\rmR^*-C)^3  }{ (C_\rmR^*+\rho n)(C_\rmR^*+r n)^2} \bigg).
	\end{align}
 Note that the remainder terms in the $O(\fndot)$ notation in~\eqref{Eq:taylor expand GJS_L} and~\eqref{Eq:taylor expand GJS_R} are of order $O(n^{-3t/2})$. 
	
     Let $\bar{\Delta}^{\alpha}_{t,\rmL}(\bar{\Delta}_t+\xi)=\lim_{n\to\infty}\frac{\Delta_{n,\rmL}^{\alpha n,\rho}(\Delta_n)}{n^{1-t/2}}$ and $\bar{\Delta}^{\alpha}_{t,\rmR}(\bar{\Delta}_t+\xi)=\lim_{n\to\infty}\frac{\Delta_{n,\rmR}^{\alpha n,\rho}(\Delta_n)}{n^{1-t/2}}$.
     Using the continuity of $\mathrm{GJS}$ in the distributions and the third argument, $\lim_{n\to\infty}\rho=0$ and $\lim_{n\to\infty}\rho/n^{-t/2}=0$, we see that 
   	\begin{align}
   		\nn&\liminf_{n\to\infty}\min_{C\in[\tau_n,n]}\min_{C_\rmL^*\in\Gamma_n^{\rmL}(i_\rho(C),\Delta_n)} \nn\\*
   		&\qquad \qquad (r-\rho) n^t \mathrm{GJS}\bigg(\tilP_{C_\rmL^*}^-,P_1,\frac{n-C_\rmL^*+\rho n}{(r-\rho)n}\bigg) \\
   		&=\min_{\alpha\in[0,1]}\frac{r(\bar{\Delta}^{\alpha}_{t,\rmL}(\bar{\Delta}_t+\xi))^2\upchi_2(P_1\|P_2)}{2(1-\alpha)(1-\alpha+r)} \label{Eq:before GJS left moderate}\\
   		&=:\min_{\alpha\in[0,1]}h_1(\alpha), \label{Eq:GJS left moderate}
   	\end{align}
   and 
   \begin{align}
   	 &\liminf_{n\to\infty}\min_{C\in[1,\tau_n']}\min_{C_\rmR^*\in\Gamma_n^{\rmR}(i_\rho(C),\Delta_n)} \nn\\
   	 &\qquad \qquad (r-\rho) n^t \mathrm{GJS}\bigg(\tilP_{C_\rmR^*}^+,P_1,\frac{C_\rmR^*+\rho n}{(r-\rho)n}\bigg) \nn\\
   	 &=\min_{\alpha\in[0,1]}\frac{r(\bar{\Delta}^{\alpha}_{t,\rmR}(\bar{\Delta}_t+\xi))^2\upchi_2(P_2\|P_1)}{2\alpha(\alpha+r)} \label{Eq:GJS right before}\\
   	 &=:\min_{\alpha\in[0,1]} h_2(\alpha). \label{Eq:GJS right moderate}
   \end{align}
	Note that in \eqref{Eq:GJS left moderate} and \eqref{Eq:GJS right moderate}, $h_1(\alpha)=h_2(\alpha)$ for any $\alpha\in[0,1]$ from the definitions of $\Delta_{n,\rmL}^{C,\rho}(\Delta_n), \Delta_{n,\rmR}^{C,\rho}(\Delta_n)$ in \eqref{Eq:def Delta^C L} as well as the values and property of $i_\rho(C)$ in Remark \ref{Rmk:value of irhoC}. Since $\bar{\Delta}^{\alpha}_{t,\rmL}(\bar{\Delta}_t+\xi)=2(\bar{\Delta}+\xi)-\bar{\Delta}^{\alpha}_{t,\rmR}(\bar{\Delta}_t+\xi)$, by letting $h_1(\alpha)=h_2(\alpha)$, we can obtain 
	\begin{align}
		&\bar{\Delta}^{\alpha}_{t,\rmL}(\bar{\Delta}_t+\xi)\nn\\
		&\!\!\!\!\!=\frac{2(\bar{\Delta}_t+\xi)\sqrt{(1-\alpha)(1-\alpha)+r)\upchi_2(P_2\|P_1)}}{\sqrt{\alpha(\alpha+r)\upchi_2(P_1\|P_2)}+\sqrt{(1-\alpha)(1-\alpha+r)\upchi_2(P_2\|P_1)} }. \label{Eq: Delta_t,L}
	\end{align}
By plugging \eqref{Eq: Delta_t,L} back to \eqref{Eq:GJS left moderate}, we have \eqref{Eq:moderate GJS CL begin}--\eqref{Eq:moderate GJS CL end},
\begin{figure*}
	\begin{align}
		\nn&\liminf_{n\to\infty}\min_{C\in[\tau_n,n]}\min_{C_\rmL^*\in\Gamma_n^{\rmL}(i_\rho(C),\Delta_n)} (r-\rho) n^t \mathrm{GJS}\bigg(\tilP_{C_\rmL^*}^-,P_1,\frac{n-C_\rmL^*+\rho n}{(r-\rho)n}\bigg) \label{Eq:moderate GJS CL begin} \\
		&=\min_{\alpha\in[0,1]}\frac{2r(\bar{\Delta}_t+\xi)^2\upchi_2(P_1\|P_2)\upchi_2(P_2\|P_1)}{\big(\sqrt{\alpha(\alpha+r)\upchi_2(P_1\|P_2)}+\sqrt{(1-\alpha)(1-\alpha+r)\upchi_2(P_2\|P_1)}~ \big)^2}\\
		&=\min_{\alpha\in[0,1]}\frac{2r\bar{\Delta}_t^2\upchi_2(P_1\|P_2)\upchi_2(P_2\|P_1)}{\big(\sqrt{\alpha(\alpha+r)\upchi_2(P_1\|P_2)}+\sqrt{(1-\alpha)(1-\alpha+r)\upchi_2(P_2\|P_1)}~ \big)^2}+\tilde{\xi}_1\\
		&=\lambda+\tilde{\xi}_1. \label{Eq:moderate GJS CL end}
	\end{align} \hrulefill
\end{figure*}
	where $\tilde{\xi}_1>0$ is a function of $\xi$ satisfying $\tilde{\xi}_1\to 0^+$ as $\xi\to 0^+$;
	and
	\begin{align}
 &\liminf_{n\to\infty}\min_{C\in[1,\tau_n']}\min_{C_\rmR^*\in\Gamma_n^{\rmR}(i_\rho(C),\Delta_n)}  \nn\\
 &\qquad ~~ (r-\rho) n^t \mathrm{GJS}\bigg(\tilP_{C_\rmR^*}^+,P_1,\frac{C_\rmR^*+\rho n}{(r-\rho)n}\bigg)
		=\lambda+\tilde{\xi}_2, \label{Eq:liminf GJS right}
	\end{align}
	where $\tilde{\xi}_2>0$ is a function of $\xi$ satisfying $\tilde{\xi}_2\to 0^+$ as $\xi\to 0^+$.	
	Using these bounds and the continuity of $\Phi(\fndot)$, we have
	\begin{align}
 &\!\!\!\!\limsup_{n\to\infty}\max_{C\in[\tau_n,n]}\bbP_C\Big\{\LD\Big(\bT_{C_{\rmL}^*}^\rho,\frac{{C_{\rmL}^*}+\rho n}{n+2\rho n},\frac{r-\rho}{n+2\rho n}\Big) \leq \frac{\tilde{\lambda}n^{-t}}{1+2\rho}\Big\}\nn\\*
 & \!\!\!\!\leq \Phi\bigg(\limsup_{n\to\infty}\bigg(\tilde{\lambda}-\lambda-\tilde{\xi}_1 \bigg)\sqrt{\frac{n}{n^{2t} c_1(\rho)}} \bigg),		\label{Eq:MDR liminf P up bound1} \qquad\mbox{and} \\
		&\!\!\!\! \limsup_{n\to\infty}\max_{C\in[1,n-\tau_n]}\bbP_C\Big\{\LD\Big(\bT_{C_{\rmR}^*}^\rho,\frac{{C_{\rmR}^*+\rho n}}{n+2\rho n},\frac{r-\rho}{1+2\rho}\Big) \leq \frac{\tilde{\lambda}n^{-t}}{1+2\rho} \Big\} \nn\\*
		&\!\!\!\! \leq \Phi\bigg(\limsup_{n\to\infty}\bigg(\tilde{\lambda}-\lambda-\tilde{\xi}_2\bigg)\sqrt{\frac{n}{n^{2t}c_2(\rho)}}\bigg).\label{Eq:Eq:MDR liminf P up bound2} 
	\end{align}
	
	Recall that $\tilde{\lambda} = \lambda+o(n^{-t})$. From Lemma \ref{Lem:V>0, T finite}, we have $0<\liminf_{n\to\infty}c_1(\rho)$, $\liminf_{n\to\infty}c_2(\rho)<\infty$. Since $\tilde{\xi}_1,\tilde{\xi}_2>0$ and $t\in(0,1/2)$, we have that for any $\lambda\in(0,\infty)$, the $\limsup$'s in both~\eqref{Eq:MDR liminf P up bound1} and~\eqref{Eq:Eq:MDR liminf P up bound2} are equal to $-\infty$, 
	which guarantees that both the upper bounds in \eqref{Eq:MDR liminf P up bound1} and \eqref{Eq:Eq:MDR liminf P up bound2} vanish.
	Then the  sequence of estimators as defined in~\eqref{Eq:estimator} yields a sequence of erasure probabilities $\{\epsilon_n\}_{n=1}^\infty$  and a sequence of confidence widths $\{\Delta_n\}_{n=1}^\infty$ such that 
	$\limsup_{n\to\infty}\epsilon_n=0 \leq \epsilon$ (for any $\epsilon\in [0,1)$) and $\limsup_{n\to\infty}\Delta_n/n^{1-t/2}=\bar{\Delta}_t+\xi$.  
	Therefore, $\bar{\Delta}_t+\xi$ is a $(r,\lambda,\epsilon,t)$-achievable NCW. Since $\xi>0$ is arbitrary, taking $\xi\to 0^+$, we see that
	\begin{align}\label{Eq: delta upper in MDR}
		&\bar{\Delta}_t^*(r,\lambda,\epsilon)\leq \bar{\Delta}_t \nn\\
		&= \max_{\alpha\in[0,1]}\frac{\sqrt{\lambda}}{\sqrt{2r\upchi_2(P_1\|P_2)\upchi_2(P_2\|P_1)}}
			\Big(\sqrt{\alpha(\alpha+r)\upchi_2(P_1\|P_2)}\nn\\
		&\qquad \qquad +\sqrt{(1-\alpha)(1-\alpha+r)\upchi_2(P_2\|P_1)}~ \Big).
 	\end{align}

	\subsection{Converse Proof of \eqref{Eq:MDR Delta^*}}
	
	The converse proof of \eqref{Eq:MDR Delta^*} follows along the same lines as that of \eqref{Eq:LDR Delta^*} in Section \ref{Sec: converse LDR} up to and including~\eqref{Eq: same before this eq} with $\lambda$ replaced by $\lambda n^{-t}$. Given any $\xi>0$ and recalling the definition of $\bar{\Delta}_t$ in Section \ref{Sec: proof of Eq:MDR Delta^*}, we  let 
	\begin{equation}
	\Delta=n^{1-t/2}\big(\bar{\Delta}_t-\xi\big),
	\end{equation} throughout this converse proof. Then from Corollary \ref{coro:Delta_n L R increase}, $\tilde{\tau}_n=\Delta_{n,\rmL}^{n^{\kappa_n-1}}(\Delta+n^{\kappa_n})+2=\Theta(\Delta+n^{\kappa_n})=\Theta(n^{1-t/2})$ and $\tilde{\tau}_n'=n-\Delta_{n,\rmR}^{n^{\kappa_n-1}}(\Delta+n^{\kappa_n})-1=n-\Theta(n^{1-t/2})=\Theta(n)$.
	Since $\frac{1}{n} (C-\hat{C}_\rmL)=\frac{1}{n}  (\Delta_{n,\rmL}^{C,\rho}(\Delta+n^{\kappa_n})+1)=\Theta(n^{-t/2})$ and $\frac{1}{n} (\hat{C}_\rmR-C)=\frac{1}{n}  (\Delta_{n,\rmR}^{C,\rho}(\Delta+n^{\kappa_n})+1)=\Theta(n^{-t/2})$, we can reuse the Taylor expansions of GJS as in~\eqref{Eq:taylor expand GJS_L} and~\eqref{Eq:taylor expand GJS_R}. 
	
	Using the continuity of $\mathrm{GJS}$ in the distributions and the third argument, and following the monotonicity used in \eqref{Eq:max over rho GJS} and similar techniques from \eqref{Eq:GJS right before}-\eqref{Eq:liminf GJS right}, we see that 
	\begin{align}
	&\limsup_{n\to\infty}\min_{C\in[\tilde{\tau}_n,n]}\max_{\rho\in\Xi_n} (r\! -\! \rho) n^t \mathrm{GJS}\bigg(\tilP_{\hat{C}_\rmL}^-,P_1,\frac{n\! -\! \hat{C}_\rmL\! +\! \rho n}{(r-\rho)n}\bigg) \nn\\
	 &=\lambda-\tilde{\xi}_1', \qquad\mbox{and}\\
	&\limsup_{n\to\infty} \min_{C\in[1,\tilde{\tau}_n']}\max_{\rho\in\Xi_n}  (r-\rho) n^t \mathrm{GJS}\bigg(\tilP_{\hatC_\rmR}^+,P_1,\frac{C_\rmR^*+\rho n}{(r-\rho)n}\bigg) \nn\\
	&=\lambda-\tilde{\xi}_2',
	\end{align}
	where $\tilde{\xi}_1',\tilde{\xi}_2'>0$ are functions of $\xi$ satisfying $\tilde{\xi}_1'\to0^+, \tilde{\xi}_2'\to 0^+$ as $\xi\to0^+$.
	
	Using these bounds and the continuity of $\Phi(\fndot)$, we have 
	\begin{align}\label{Eq:lim P_L lower bound MDR}
	\nn&\liminf\limits_{n\to\infty} \max_{C\in[\tilde{\tau}_n,n]} \min_{\rho\in\Xi_n}\bbP_{C}\bigg\{\LD\Big(\bT_{\hat{C}_\rmL}^\rho,\frac{\hat{C}_\rmL+\rho n}{n+2\rho n},\frac{r-\rho}{1+2\rho} \Big) \nn\\
	&\hspace{1.2in} +\frac{2\log n}{(1+2\rho)n}\leq \frac{\lambda n^{-t}}{1+2\rho}-\delta_n \bigg\}\\
	&\geq \Phi\Bigg(\liminf_{n\to\infty}~\tilde{\xi}_1' \sqrt{\frac{n}{n^{2t}  \max_{\rho\in\Xi_n}\hat{c}_1(\rho)}} \Bigg)
	\end{align}
	and
	\begin{align}\label{Eq:lim P_R lower bound MDR}
	\nn&\liminf\limits_{n\to\infty}\max_{C\in[1, \tilde{\tau}_n']}\min_{\rho\in\Xi_n} \bbP_{C}\bigg\{\LD\Big(\bT_{\hat{C}_\rmR}^\rho,\frac{\hat{C}_\rmR+\rho n}{n+2\rho n},\frac{r-\rho}{1+2\rho} \Big) \nn\\
	&\hspace{1.2in} +\frac{2\log n}{(1+2\rho)n}\leq \frac{\lambda n^{-t}}{1+2\rho}-\delta_n \bigg\}\\*
	&\geq \Phi\Bigg(\liminf_{n\to\infty}~\tilde{\xi}_2' \sqrt{\frac{n}{n^{2t} \max_{\rho\in\Xi_n} \hat{c}_2(\rho)}} \Bigg).
	\end{align}
	From Lemma \ref{Lem:V>0, T finite}, we have   $ 0<\limsup_{n\to\infty}n^t \max_{\rho\in\Xi_n}\hat{c}_1(\rho)$, $ \limsup_{n\to\infty}n^t \max_{\rho\in\Xi_n}\hat{c}_2(\rho)<\infty$. 
	Since $\tilde{\xi}_1',\tilde{\xi}_2'>0$ and $t\in(0,1/2)$, we have that
	\begin{align}
		\liminf_{n\to\infty}~\tilde{\xi}_1' \sqrt{\frac{n}{n^{2t} \hat{c}_1(\rho)}}&=\liminf_{n\to\infty}\frac{\tilde{\xi}_1'n^{(1-t)/2}}{\sqrt{n^t \hat{c}_1(\rho)}}=\infty, \quad\mbox{and}\\
		\liminf_{n\to\infty}~\tilde{\xi}_2' \sqrt{\frac{n}{n^{2t}  \hat{c}_2(\rho)}}&=\liminf_{n\to\infty}\frac{\tilde{\xi}_2'n^{(1-t)/2}}{\sqrt{n^t \hat{c}_2(\rho)}}=\infty,
	\end{align} 
	which means that at both of the two bounds in \eqref{Eq:lim P_L lower bound MDR} and \eqref{Eq:lim P_R lower bound MDR} tend to one; that is, the erasure probability $\max_{C\in [n]}\bbP_C\{\calE_\rme\}$ cannot be made less than $\epsilon$ for any $\epsilon\in[0,1)$. Thus, the $t$-optimal NCW cannot be smaller than or equal to $\bar{\Delta}_t-\xi$, i.e., $\bar{\Delta}_t^*(r,\lambda,\epsilon)>\bar{\Delta}_t-\xi$. Since $\xi>0$ can be made arbitrarily small, it means that
	\begin{align}\label{Eq: delta lower in MDR}
		&\bar{\Delta}_t^*(r,\lambda,\epsilon)\geq\bar{\Delta}_t\nn\\
		&=\max_{\alpha\in[0,1]}\frac{\sqrt{\lambda}}{\sqrt{2r\upchi_2(P_1\|P_2)\upchi_2(P_2\|P_1)}}
		\Big(\sqrt{\alpha(\alpha+r)\upchi_2(P_1\|P_2)}\nn\\
		&\qquad \qquad +\sqrt{(1-\alpha)(1-\alpha+r)\upchi_2(P_2\|P_1)}~ \Big),
	\end{align}
	completing the proof of the converse of \eqref{Eq:MDR Delta^*}.


\appendix

\subsection{Proof of Lemma \ref{Lem:Gmin increase}}\label{proof of monotonicity}
	Let $\beta=1/r$. Since $\mathrm{GJS}(tP_1+(1-t)P_2,P_2,\beta)$ and $\mathrm{GJS}((1-t)P_1+tP_2,P_1,\beta)$ are symmetric, it suffices to prove that for any $\beta>0$ and any pair of $(P_1,P_2)$, $\mathrm{GJS}(tP_1+(1-t)P_2,P_2,\beta)$ is a strictly increasing function of $t\in[0,1]$.
	
	The first and second derivatives  of $\mathrm{GJS}(tP_1+(1-t)P_2,P_2,\beta)$ with respect to $t$ are  
	\begin{align}
	\nn&\frac{\partial \mathrm{GJS}(tP_1+(1-t)P_2,P_2,\beta)}{\partial t}\\*
	\nn&=\sum_x(P_1(x)-P_2(x))\log\frac{(1+\beta)(tP_1(x)+(1-t)P_2(x))}{\beta tP_1(x)+(\beta-\beta t+1)P_2(x)}\nn\\*
	&~ -\sum_x(tP_1(x)+(1-t)P_2(x))\frac{\beta(P_1(x)-P_2(x))}{\beta tP_1(x)+(\beta-\beta t+1)P_2(x)}\nn\\*
	&~ -\sum_xP_2(x)\frac{\beta(P_1(x)-P_2(x))}{\beta tP_1(x)+(\beta-\beta t+1)P_2(x)}
 \end{align}
 and
 \begin{align}
	\nn&\frac{\partial^2 \mathrm{GJS}(tP_1+(1-t)P_2,P_2,\beta) }{\partial t^2}\\
	\nn&=\sum_x\frac{(P_1(x)-P_2(x))^2}{tP_1(x)+(1-t)P_2(x)} \nn\\
	&\quad -2\sum_x\frac{\beta(P_1(x)-P_2(x))^2}{\beta tP_1(x)+(\beta-\beta t+1)P_2(x)} \nn\\*
	&\quad+\sum_x\frac{\beta^2(P_1(x)-P_2(x))^2(tP_1(x)+(1-t)P_2(x))}{(\beta tP_1(x)+(\beta-\beta t+1)P_2(x))^2} \nn\\*
	&\quad  +\sum_x \frac{\beta^2(P_1(x)-P_2(x))^2}{(\beta tP_1(x)+(\beta-\beta t+1)P_2(x))^2}\\
	\label{Eq: GJS(t) second deriv}&=\sum_x\bigg(\frac{P_1(x)-P_2(x)}{\sqrt{tP_1(x)+(1-t)P_2(x)}} \nn\\
	&\quad -\frac{\sqrt{tP_1(x)+(1-t)P_2(x)}\beta(P_1(x)-P_2(x))}{\beta tP_1(x)+(\beta-\beta t+1)P_2(x)} \bigg)^2 \nn\\*
	&\quad+\sum_x \frac{\beta^2(P_1(x)-P_2(x))^2}{(\beta tP_1(x)+(\beta-\beta t+1)P_2(x))^2},
	\end{align}
	where \eqref{Eq: GJS(t) second deriv} is strictly positive when $P_1\neq P_2$.
	Thus, we have
	\begin{align}
	&\frac{\partial \mathrm{GJS}(tP_1+(1-t)P_2,P_2,\beta)}{\partial t} \nn\\
	&> \frac{\partial \mathrm{GJS}(tP_1+(1-t)P_2,P_2,\beta)}{\partial t}\bigg|_{t=0}=0,
	\end{align}
	which implies that $\mathrm{GJS}(tP_1+(1-t)P_2,P_2,\beta)$ is a strictly increasing convex function of $t\in[0,1]$. In a completely symmetric manner,  $\mathrm{GJS}((1-t)P_1+tP_2,P_1,\beta)$ is also a strictly increasing convex function of $t\in[0,1]$.
	
%

\subsection{Proof of Corollary \ref{coro:Gmin increase}} \label{proof of coro:Gmin increase}
We have the following deduction based on the existing lemmas and corollaries,
\begin{itemize}
	\item  From Corollary \ref{coro:Delta L R increase}, the minimization interval $[\Delta_\rmL(\bar{\Delta}) ,1-\Delta_\rmR(\bar{\Delta})]$ shrinks as $\bar{\Delta}$ increases.
	
	\item From Corollary \ref{coro:Delta_n L R increase}, we can deduce that with fixed $\alpha$, as $\bar{\Delta}$ increases, both $\alpha-\iota(\alpha)+\bar{\Delta}$ and $1-\iota(\alpha)+\bar{\Delta}$ increase and thus, the function $	\mathrm{GJS}(\frac{(\alpha-\iota(\alpha)+\bar{\Delta})P_1+(1-\alpha)P_2}{1-\iota(\alpha)+\bar{\Delta}},P_2,\frac{1-\iota(\alpha)+\bar{\Delta}}{r})$ increases based on Lemma \ref{Lem:GJS increase in a} and Lemma \ref{Lem:Gmin increase}.
\end{itemize}
Consequently, by combining these two points, Corollary \ref{coro:Gmin increase} is proved.

\subsection{Proof of Lemma \ref{Lem:V>0, T finite}}\label{proof of V>0, T finite}
\subsubsection{Upper Bound}
For any $j\in[1,C-(\bar{\Delta}n-n^{\kappa_n}))$ and any $\rho\in(0,r)$, let $\beta=j/n$ and we recall that  $V(\tilPjCleft,P_2,j,r,\rho)$ is given by

\begin{align}
	&V(\tilPjCleft,P_2,j,r,\rho)\nn\\
	&=\frac{n+\rho n-j}{n+2\rho n} \nn\\
	&\quad \times\mathrm{Var}_{\tilPjCleft}\Big[ \log\frac{(n-j+N)\tilPjCleft(X)}{(n+\rho n-j)\tilPjCleft(X)+(N-\rho n)P_2(X)}\Big] \nn\\
	&\quad +\frac{N-\rho n}{n+2\rho n} \nn\\
	&\quad \times\mathrm{Var}_{P_2}\Big[\log\frac{(n-j+N)P_2(X)}{(n+\rho n-j)\tilPjCleft(X)+(N-\rho n)P_2(X)} \Big] \\
	&=\frac{1+\rho-\beta}{1+2\rho}\mathrm{Var}_{\frac{(\alpha-\beta)P_1+(1-\alpha+\rho)P_2}{1-\beta+\rho}}\bigg[ \nn\\
	&\hspace{1in} \log\frac{(1-\beta+r)\frac{(\alpha-\beta)P_1+(1-\alpha+\rho)P_2}{1-\beta+\rho}(X)}{(\alpha-\beta)P_1(X)+(1-\alpha+r)P_2(X)}\bigg] \nn\\
	&+\frac{r-\rho}{1+2\rho}\mathrm{Var}_{P_2}\bigg[  \log\frac{(1-\beta+r)P_2(X)}{(\alpha-\beta)P_1(X)+(1-\alpha+r)P_2(X)} \bigg]. \label{eqn:var_decomp}
\end{align}

Since $\supp(P_1)=\supp(P_2)=|\calX|$, we 
note that for any $j\in[1,C-(\bar{\Delta} n-n^{\kappa_n}))$ (i.e., $\beta\in[1/n,\alpha-(\bar{\Delta}-n^{\kappa_n-1}))$ ) with some $\bar{\Delta}\in\bbR_+$ and any $\rho \in(0,r)$, we have that $\tilPjCleft(x)>0$ and $P_2(x)>0$ uniformly in $j$ and thus 
\begin{align}
&\!\!\!\!\!\!\!-A_1(\rho)\! \le \log\frac{(1-\beta+r)\frac{(\alpha-\beta)P_1+(1-\alpha+\rho)P_2}{1-\beta+\rho}(X)}{(\alpha-\beta)P_1(X)+(1-\alpha+r)P_2(X)}\! \le  \! A_2(\rho), \nn\\
&\hspace{0.8\columnwidth}\mbox{a.s.}
\end{align}	
for some   $A_1(\rho),A_2(\rho)\in\bbR_+$. Then we have
\begin{align}
\nn&\frac{1+\rho-\beta}{1+2\rho}\mathrm{Var}_{\frac{(\alpha-\beta)P_1+(1-\alpha+\rho)P_2}{1-\beta+\rho}}\bigg[ \nn\\
&\hspace{0.6in} \log\frac{(1-\beta+r)\frac{(\alpha-\beta)P_1+(1-\alpha+\rho)P_2}{1-\beta+\rho}(X)}{(\alpha-\beta)P_1(X)+(1-\alpha+r)P_2(X)}\bigg]\\
&=\frac{1+\rho-\beta}{1+2\rho}\mathbb{E}_{\frac{(\alpha-\beta)P_1+(1-\alpha+\rho)P_2}{1-\beta+\rho}}\bigg[\nn\\
&\hspace{0.3in} \bigg( \log\frac{(1-\beta+r)\frac{(\alpha-\beta)P_1+(1-\alpha+\rho)P_2}{1-\beta+\rho}(X)}{(\alpha-\beta)P_1(X)+(1-\alpha+r)P_2(X)}\bigg)^2 \bigg]\\
&\hspace{0.3in}  -\frac{1+\rho-\beta}{1+2\rho}\mathbb{E}_{\frac{(\alpha-\beta)P_1+(1-\alpha+\rho)P_2}{1-\beta+\rho}}\bigg[ \nn\\
&\hspace{0.3in} \log\frac{(1-\beta+r)\frac{(\alpha-\beta)P_1+(1-\alpha+\rho)P_2}{1-\beta+\rho}(X)}{(\alpha-\beta)P_1(X)+(1-\alpha+r)P_2(X)} \bigg]^2\\
&<\frac{1+\rho -\beta}{1+2\rho }\big(A_2^2(\rho)+A_1^2(\rho)\big)=:A_3(\rho),
\end{align}
for some $A_3(\rho)\in\bbR_+$.

Similarly, the second term in \eqref{eqn:var_decomp} is smaller than some constant $A_4(\rho)\in\bbR_+$.  Hence, for any $j\in[1,C-(\bar{\Delta}n-n^{\kappa_n}))$  and any $\rho\in(0,r)$, $V(\tilPjCleft,P_2,j,r,\rho)<A_3(\rho)+A_4(\rho)$. 	Similarly, for any $j\in(C+(\bar{\Delta}n-n^{\kappa_n}),n]$  and any $\rho\in(0,r)$, $V(\tilPjCright,P_1,j,r,\rho)<A_5(\rho)$ for some   $A_5(\rho)\in\bbR_+$. Let $\overline{a}_1(\rho):=\max\{A_3(\rho)+A_4(\rho), A_5(\rho)\}$.

Since $0<\overline{a}_1(\rho)<\infty$ holds for any $\rho\in(0,r)$, when $n\to\infty$, we can similarly upper bound $\lim_{n\to\infty}V(\tilPjCleft,P_2,j,r,\rho)$ by  $\limsup_{n\to\infty}\overline{a}_1(\rho)\in\bbR_+$ for any $\rho\in(0,r)$.


In a similar manner, we can also show that for any $\rho\in(0,r)$, $T(\tilPjCleft,P_2,j,r,\rho),T(\tilPjCright,P_1,j,r,\rho)<\overline{a}_2(\rho)$ for some   $\overline{a}_2(\rho)\in\bbR_+$ and $\limsup_{n\to\infty}\overline{a}_2(\rho)<\infty$.

\subsubsection{Lower Bound}
On the other hand, when $P_1\ne P_2$, for any $\rho\in(0,r)$
\begin{itemize}
	\item $V(\tilPjCleft,P_2,j,r,\rho)=V(\tilPjCright,P_1,j,r,\rho)=0$ if and only if $j=C$,
	
	\item $\displaystyle \lim_{n\to\infty}V(\tilPjCleft,P_2,j,r,\rho)=\lim_{n\to\infty}V(\tilPjCright,P_1,j,r,\rho)=0$ if and only if $\lim_{n\to\infty}|j-C|/n=0$,
	

	\item By taking Taylor expansion of $V(\tilPjCleft,P_2,j,r,\rho)$ around $\tilPjCleft=P_2$, for any $j\in [1,C-(\bar{\Delta}_t n^{1-t/2}-n^{\kappa_n}))$ with some $\bar{\Delta}_t\in\bbR_+$, any $t\in(0,1/2)$ and any $\rho\in(0,r)$, we have
	\begin{align}
		n^tV(\tilPjCleft,P_2,j,r,\rho)&=\Theta(n^t(\alpha-\beta)^2)\nn\\
		&=\Omega((\bar{\Delta}_t-n^{\kappa_n-1+t/2})^2).
	\end{align}
	Thus, $\lim_{n\to\infty}n^tV(\tilPjCleft,P_2,j,r,\rho)=\lim_{n\to\infty}n^tV(\tilPjCright,P_1,j,r,\rho)=0$ iff $\bar{\Delta}_t=0$.
\end{itemize} 
	
 For $j\notin [C\pm(\bar{\Delta} n-n^{\kappa_n})]$ with some constant $\bar{\Delta}>0$, we have $\min\{V(\tilPjCleft,P_2,j,r,\rho),V(\tilPjCright,P_1,j,r,\rho)\}\ge\underline{a}_1(\rho)>0$ for some  $\underline{a}_1(\rho)\in\bbR_+$ and we also have $\liminf_{n\to\infty}\underline{a}_1(\rho)>0$ for any $\rho\in(0,r)$. For $j\notin[C\pm(\bar{\Delta}_t n^{1-t/2}-n^{\kappa_n})]$,  we have $\liminf_{n\to\infty}n^t\underline{a}_1(\rho)>0$ for any $\rho\in(0,r)$.

\subsection{Probability of Atypical Set}\label{prob of atypical set}
For any $i,j\in[n]$ and $i<j$, when $j$ is the true change-point, the probability of  $\tilX_1^{i+\rho n}\notin\calB(P_1)$ can be bounded using Hoeffding's inequality  and the union bound as 
\begin{equation}
\bbP_j\{\tilX_1^{i+\rho n}\notin\calB(P_1) \}\le \frac{2|\calX|}{(i+\rho n)^2}.
\end{equation}
Reusing this calculation on the other typical sequences, the probability of the atypical set $\calA_{i,j}$ can be bounded as 
\begin{align}
&\bbP_j\{\calA_{i,j}^\rho n\} \nn\\
&\leq \bbP_j\{\tilX_1^{i+\rho n}\notin\calB(P_1) \}+\bbP_j\{X_{i+1}^j\notin\calB(P_1)\}\nn\\*
&\quad+\bbP_j\{\tilX_2^{n-j+\rho n}\notin\calB(P_2) \}+\bbP_j\{ Y_1^{N-\rho n}\notin\calB(P_1)  \} \nn\\
&\quad +\bbP_j\{Y_2^{N-\rho n}\notin\calB(P_2)\}\\
\label{Eq:typical set hoeffding ineq}&\leq \frac{2|\calX|}{(i+\rho n)^2}+	\frac{2|\calX|}{(j-i)^2}+\frac{2|\calX|}{(n-j+\rho n)^2} \nn\\
&\quad +\frac{2|\calX|}{(N-\rho n)^2}+\frac{2|\calX|}{(N-\rho n)^2}.
\end{align}

\subsection{Proof of Lemma \ref{Lem:GJS monotone in C}}\label{proof of g1g2}

First of all, the functions $g_1(\alpha,\zeta)$, $g_2(\alpha,\zeta)$ can be rewritten as 
\begin{align}
	&g_1(\alpha,\zeta)\nn\\
	&=\mathrm{GJS}\bigg(\frac{\zeta P_1+(1+\rho-\alpha)P_2}{1+\rho-\alpha+\zeta},P_2,\frac{1+\rho-\alpha+\zeta}{r-\rho}\bigg),\\
	&g_2(\alpha,\zeta)=\mathrm{GJS}\bigg(\frac{(\alpha+\rho) P_1+\zeta P_2}{\alpha+\rho+\zeta},P_1,\frac{\alpha+\rho+\zeta}{r-\rho}\bigg).
\end{align}

Fix any $\zeta\in(0,r)$. Define the function  
\begin{equation}
t(\alpha)=\frac{\zeta}{1+\rho-\alpha+\zeta}\in \bigg[\frac{\zeta}{1+\rho},\frac{\zeta}{\rho +\zeta}\bigg]\subset(0,1).
\end{equation}
Note that $t(\alpha)$ is a strictly increasing function of $\alpha\in[\zeta,1]$.  In the following, we write $t$ instead of $t(\alpha)$ for brevity.

Then we can further rewrite $g_1(\alpha,\zeta)$ as $\mathrm{GJS}\big(tP_1+(1-t)P_2,P_2, {\zeta}/ (t(r-\rho))\big)$. 
Let $r'=\zeta/(r-\rho)$. Next, it suffices to prove that $\mathrm{GJS}\big(tP_1+(1-t)P_2,P_2, {r'}/{t}\big)$ is an increasing function of $t\in(0,1)$.
\begin{align}
&\mathrm{GJS}\bigg(tP_1+(1-t)P_2,P_2,\frac{r'}{t}\bigg) \nn\\*
&=D\bigg(P_2\Big\|\frac{\frac{r'}{t}(tP_1+(1-t)P_2)+P_2}{1+\frac{r'}{t}}\bigg) \nn\\
&\quad +\frac{r'}{t}D\bigg(tP_1+(1-t)P_2 \Big\| \frac{\frac{r'}{t}(tP_1+(1-t)P_2)+P_2}{1+\frac{r'}{t}}\bigg). \label{eqn:gjs2}
\end{align}
Let $\beta= r't/(t+r')$, which increases as $t$ increases. Then 
\begin{align}
&D\bigg(P_2\Big\|\frac{\frac{r'}{t}(tP_1+(1-t)P_2)+P_2}{1+\frac{r'}{t}}\bigg)\nn\\
&=D(P_2\|\beta P_1+(1-\beta)P_2). \label{eqn:div}
\end{align}
The derivatives of $D(P_2\|\beta P_1+(1-\beta)P_2)$ with respect to $\beta$ are given by
\begin{align}
&\frac{\partial D(P_2\|\beta P_1+(1-\beta)P_2)}{\partial \beta} \nn\\
&\quad =\sum_x P_2(x)\frac{-(P_1(x)-P_2(x))}{\beta P_1(x)+(1-\beta)P_2(x)},
\end{align}
and
\begin{align}
&\frac{\partial^2 D(P_2\|\beta P_1+(1-\beta)P_2)}{\partial \beta^2} \nn\\
&=\sum_x P_2(x)\frac{(P_1(x)-P_2(x))^2}{(\beta P_1(x)+(1-\beta)P_2(x))^2}> 0.
\end{align}
Thus, we have
\begin{align}
&\frac{\partial D(P_2\|\beta P_1+(1-\beta)P_2)}{\partial \beta} \nn\\
&> \frac{\partial D(P_2\|\beta P_1+(1-\beta)P_2)}{\partial \beta}\bigg|_{\beta=0}=0, 
\end{align}
which implies that $D(P_2\|\beta P_1+(1-\beta)P_2)$ is an increasing function of $\beta\in(0,1)$. Thus, the KL divergence on the left of the equality in~\eqref{eqn:div}  is an increasing function of $t\in(0,1)$.

The first derivative of the second term in~\eqref{eqn:gjs2} with respect to $t$ is given by
\begin{align}
\nn&\frac{\partial \frac{r'}{t}D\bigg(tP_1+(1-t)P_2 \Big\| \frac{\frac{r'}{t}(tP_1+(1-t)P_2)+P_2}{1+\frac{r'}{t}}\bigg)}{\partial t}\\*
\nn&=\frac{r'}{t}\sum_x(tP_1(x)+(1-t)P_2(x)) \nn\\
&\quad \times\bigg(\frac{tP_1(x)+(1-t)P_2(x)+(t+r')(P_1(x)-P_2(x))}{(t+r')(tP_1(x)+(1-t)P_2(x))} \nn\\*
&\quad -\frac{r'P_1(x)+(1-r')P_2(x)}{r'tP_1(x)+(r'-r't+t)P_2(x)}\bigg)\nn\\
&\quad -\frac{r'}{t^2}\sum_xP_2(x)\log\frac{(t+r')(tP_1(x)+(1-t)P_2(x))}{r'tP_1(x)+(r'-r't+t)P_2(x)}\\
\label{Eq:log x<=x-1}\nn&\geq \frac{r'}{t}\sum_x(tP_1(x)+(1-t)P_2(x)) \nn\\
&\quad \times \bigg(\frac{tP_1(x)+(1-t)P_2(x)+(t+r')(P_1(x)-P_2(x))}{(t+r')(tP_1(x)+(1-t)P_2(x))} \nn\\*
&\quad-\frac{r'P_1(x)+(1-r')P_2(x)}{r'tP_1(x)+(r'-r't+t)P_2(x)}\bigg) \nn\\
&\quad  -\frac{r'}{t^2}\sum_xP_2(x)\bigg(\frac{(t+r')(tP_1(x)+(1-t)P_2(x))}{r'tP_1(x)+(r'-r't+t)P_2(x)}-1\bigg)\\
&=\frac{r'}{t(t+r')}+\frac{r'}{t^2}-r'\sum_x(tP_1(x)+(1-t)P_2(x)) \nn\\
&\quad \times \frac{\frac{1}{t}(r'P_1(x)+(1-r')P_2(x))+\frac{t+r'}{t^2}P_2(x)}{r'tP_1(x)+(r'-r't+t)P_2(x)}\\
&=\frac{r'}{t(t+r')}-\frac{r'}{t}\sum_x\frac{(tP_1(x)+(1-t)P_2(x))P_2(x)}{r'tP_1(x)+(r'-r't+t)P_2(x)}, \label{eqn:last_exp}
\end{align}
where \eqref{Eq:log x<=x-1} follows since $\log y\leq y-1$ for any $y\in\bbR$.
If we want to prove  the expression in \eqref{eqn:last_exp} is nonnegative, it suffices to prove
\begin{align}
1\geq \sum_x\frac{(t+r')(tP_1(x)+(1-t)P_2(x))P_2(x)}{r'tP_1(x)+(r'-r't+t)P_2(x)}. \label{eqn:suffices}
\end{align}
Note that when $t=0$, equality in~\eqref{eqn:suffices} holds. 

Next we calculate the derivatives of  the function on the right-hand-side of \eqref{eqn:suffices}  as follows:
\begin{align}
&\frac{\partial \sum_x\frac{(t+r')(tP_1(x)+(1-t)P_2(x))P_2(x)}{r'tP_1(x)+(r'-r't+t)P_2(x)}}{\partial t} \nn\\*
&=\sum_x\frac{1}{\left(r'tP_1(x)+(r'-r't+t)P_2(x)\right)^2}\Big(P_2(x)\nn\\
&\quad \times \left(P_2(x)-P_1(x)\right)t\big(\left(\left(P_2(x)-P_1(x)\right)r'-P_2(x)\right)t \nn\\
&\quad -2P_2(x)r\big)\Big),
\end{align}
and 
\begin{align}
&\frac{\partial^2 \sum_x\frac{(t+r')(tP_1(x)+(1-t)P_2(x))P_2(x)}{r'tP_1(x)+(r'-r't+t)P_2(x)}}{\partial t^2}\nn\\
&=\sum_x\dfrac{-2P_2^3(x)\left(P_2(x)-P_1(x)\right)r'^2}{\left(r'tP_1(x)+(r'-r't+t)P_2(x)\right)^3}.
\end{align}
When $P_2(x)>P_1(x)$, 
\begin{align}
&\sum_{x:P_2(x)>P_1(x)}\dfrac{-2P_2^3(x)\left(P_2(x)-P_1(x)\right)r'^2}{\left(r'tP_1(x)+(r'-r't+t)P_2(x)\right)^3} \nn\\
&<\sum_{x:P_2(x)>P_1(x)}\frac{-2(P_2(x)-P_1(x))r'^3}{(t+r')^3},
\end{align}
and when $P_2(x)\leq P_1(x)$,
\begin{align}
&\sum_{x:P_2(x)\leq P_1(x)}\dfrac{2P_2^3(x)\left(P_1(x)-P_2(x)\right)r'^2}{\left(r'tP_1(x)+(r'-r't+t)P_2(x)\right)^3} \nn\\
&\leq \sum_{x:P_2(x)\leq P_1(x)}\frac{2(P_1(x)-P_2(x))r'^3}{(t+r')^3},
\end{align}
and we have
\begin{align}
\sum_{x}\dfrac{-2P_2^3(x)\left(P_2(x)-P_1(x)\right)r'^2}{\left(r'tP_1(x)+(r'-r't+t)P_2(x)\right)^3}\leq 0.
\end{align}
Thus, 
\begin{align}
&\frac{\partial \sum_x\frac{(t+r')(tP_1(x)+(1-t)P_2(x))P_2(x)}{r'tP_1(x)+(r'-r't+t)P_2(x)}}{\partial t} \nn\\
&\leq \frac{\partial \sum_x\frac{(t+r')(tP_1(x)+(1-t)P_2(x))P_2(x)}{r'tP_1(x)+(r'-r't+t)P_2(x)}}{\partial t}\bigg|_{t=0}=0,
\end{align}
and 
\begin{align}
&\sum_x\frac{(t+r')(tP_1(x)+(1-t)P_2(x))P_2(x)}{r'tP_1(x)+(r'-r't+t)P_2(x)} \nn\\
&\leq \sum_x\frac{(t+r')(tP_1(x)+(1-t)P_2(x))P_2(x)}{r'tP_1(x)+(r'-r't+t)P_2(x)}\bigg|_{t=0}=1,
\end{align}
which implies that the second term in \eqref{eqn:gjs2} is an increasing function of $t\in(0,1)$.

Finally, we can show that $\mathrm{GJS}\big(tP_1+(1-t)P_2,P_2,r'/t\big)$ is an increasing function of $t\in(0,1)$, and thus $g_1(\alpha, \zeta)$, defined in \eqref{eqn:defg1},  is an increasing function of $\alpha\in[\zeta,1]$.

Following   similar steps, we can also prove that $g_2(\alpha,\zeta)$, defined in \eqref{eqn:defg2}, is a decreasing function of $\alpha\in [0,1-\zeta]$.


\subsection{Proof of Corollary \ref{coro:Delta_n L R increase}}\label{pf of coro:Delta_n L R increase}
The first statement can be easily deduced from Lemma \ref{Lem:GJS increase in a} and Lemma \ref{Lem:Gmin increase}.

For the second statement, by applying the Taylor expansion of $\mathrm{GJS}(\cdot,P_2,\frac{n-C+\Delta_{n,\rmL}^{C,\rho}(\Delta)+1+\rho n}{(r-\rho)n})$ around $P_2$ and that of $\mathrm{GJS}(\cdot,P_1,\frac{C+2\Delta_n-\Delta_{n,\rmL}^{C,\rho}(\Delta_n)+1+\rho n}{(r-\rho) n} )$ around $P_1$, we have
\begin{align}
	&\mathrm{GJS}\bigg(\tilP_{C-\Delta_{n,\rmL}^{C,\rho}(\Delta_n)-1}^-,P_2,\frac{n-C+\Delta_{n,\rmL}^{C,\rho}(\Delta_n)+1+\rho n}{(r-\rho)n}\bigg) \nn\\
	&=O\bigg(\bigg(\frac{\Delta_{n,\rmL}^{C,\rho}(\Delta_n)}{n}\bigg)^2\bigg), \label{Eq:taylor GJS left}\\
	&\mathrm{GJS}\bigg(\tilP_{C+2\Delta_n-\Delta_{n,\rmL}^{C,\rho}(\Delta_n)+1}^+,P_1, \nn\\
	&\qquad \quad \frac{C+2\Delta_n-\Delta_{n,\rmL}^{C,\rho}(\Delta_n)+1+\rho n}{(r-\rho) n}\bigg) \nn\\
	&=O\bigg(\bigg(\frac{2\Delta_n-\Delta_{n,\rmL}^{C,\rho}(\Delta_n)}{n}\bigg)^2\bigg). \label{Eq:taylor GJS 2Delta-left}
\end{align} 
Since the left-hand sides of \eqref{Eq:taylor GJS left} and \eqref{Eq:taylor GJS 2Delta-left} are equal according to \eqref{Eq:def Delta^C L}, to make the right-hand sides to be equal, we must have $\Delta_{n,\rmL}^{C,\rho}(\Delta)=\Theta(\Delta)$. 
Similarly, we also have $\Delta_{n,\rmR}^{C,\rho}(\Delta)=\Theta(\Delta)$.

\subsection{Proof of Lemma \ref{lem:L(j)<L(i)}}\label{pf of lem:L(j)<L(i)}
Recall  the definitions of $\kappa_n$ in \eqref{eqn:kappa_n}, $\calB(P)$ in~\eqref{eqn:typ_B} and $\rho\in[n^{\kappa_n-1},r-n^{\kappa_n-1}]$. According to~\eqref{eqn:atypical},  
for any $i,j\in[n]$ such that $i<j-n^{\kappa_n}$, 
\begin{align}
	&\bbP_j\big\{\tilX_1^{i+\rho n}\notin\calB(P_1) \text{ or } X_{i+1}^j\notin\calB(P_1) \text{ or } \nn\\*
	&\qquad \tilX_2^{n-j+\rho n}\notin\calB(P_2) \text{ or } Y_1^{N-\rho n}\notin\calB(P_1) \text{ or }\nn\\*
	&\qquad  Y_2^{N-\rho n}\notin\calB(P_2) \big\}  = O(n^{-2\kappa_n})\label{eqn:atyp}
\end{align}

Recall the definitions of  $\tilPjCleft$ and $\tilPjCright$ in \eqref{eqn:conv1}. 
Fix any $j\in[1, C-n^{\kappa_n}]$ and let $\bP_j^-:=(P_1,\tilPjCleft,P_1,P_2)$ and $\bP_j^+:=(\tilPjCright,P_2,P_1,P_2)$.   Notice that 
\begin{align}
	&\LD\Big(\bP_j^-,\frac{j+\rho n}{n+2\rho n},\frac{r-\rho}{1+2\rho}\Big) = \Omega\big(n^{2\kappa_n-2}\big) \label{eqn:taylor_ld}
\end{align}	
because $\LD(\bP_j^-,\frac{j+\rho n}{n+2\rho n},\frac{r-\rho}{1+2\rho})$ is strictly decreasing in $j$ (a fact that can be proved similarly as that of Lemma~\ref{Lem:Gmin increase})  and from the Taylor expansion of the function $\LD ( \fndot, \frac{j+\rho n}{n+2\rho n},\frac{r-\rho}{1+2\rho})$ around $P_2$; see \eqref{Eq:taylor expand GJS_L} in which  $C_\rmL^*$ is replaced by $C-n^{\kappa_n}$.  
Similarly for any $j\in[ C+n^{\kappa_n},n]$ we have
\begin{align}
	\LD\Big(\bP_j^+,\frac{j+\rho n}{n+2\rho n},\frac{r-\rho}{1+2\rho}\Big)=\Omega\big(n^{2\kappa_n-2}\big). 
\end{align} 

For any $j\in[1, C-n^{\kappa_n}]$, the Taylor expansion of $\LD(\bT_j^\rho,\frac{j+\rho n}{n+2\rho n} ,\frac{r-\rho}{1+2\rho})$ around $\bT_j^\rho=\bP_j^-$ for any (typical) $(\tilx_1^{j+\rho n}, x_{j+1}^C, \tilx_2^{n-C+\rho n}, y_1^{N-\rho n}, y_2^{N-\rho n})\in\calB(P_1)\times\calB(P_1)\times\calB(P_2)\times\calB(P_1)\times\calB(P_2)$ is given by \eqref{Eq:L(Tj) taylor begin}--\eqref{Eq:LD taylor j<C}.
\begin{figure*}
\begin{align}
	&\LD\Big(\bT_j^\rho,\frac{j+\rho n}{n+2\rho n} ,\frac{r-\rho}{1+2\rho}\Big) \nn\\
	&=\frac{n-j+\rho n}{n+2\rho n}\sum_{x}T_{\tilx_2^{n-j+\rho n}}(x)\log\frac{(n-j+N)\tilPjCleft(x)}{(C-j)P_1(x)+(n-C+N)P_2(x)}   \nn\\*
	&\qquad+\frac{N-\rho n}{n+2\rho n}\sum_{x}T_{y_2^{N-\rho n}}(x)\log\frac{(n-j+N)P_2(x)}{(C-j)P_1(x)+(n-C+N)P_2(x)} \nn\\*
	&\qquad+O\bigg(\frac{\log (j+\rho n)}{j+\rho n}\bigg)+O\bigg(\frac{\log (C-j)}{C-j}\bigg)+O\bigg(\frac{\log (n-C+\rho n)}{n-C+\rho n}\bigg)+O\bigg(\frac{\log (N-\rho n)}{N-\rho n}\bigg) \label{Eq:L(Tj) taylor begin}\\
	&=\frac{1}{n+2\rho n}\bigg(\sum_{i\in[j+1:n]}\log\frac{(n-j+N)\tilPjCleft(x_i)}{(n-j+\rho n)\tilPjCleft(x_i)+(N-\rho n)P_2(x_i)} \nn\\
	&\qquad +\sum_{i\in[N-\rho n+1:N]}\log\frac{(n-j+N)\tilPjCleft(y_{2,i})}{(n-j+\rho n)\tilPjCleft(y_{2,i})+(N-\rho n)P_2(y_{2,i})}\bigg)  \nn\\*
	&\qquad+\frac{1}{n+2\rho n}\sum_{i\in[N-\rho n]}\log\frac{(n-j+N)P_2(y_{2,i})}{(n-j+\rho n)\tilPjCleft(y_{2,i})+ (N-\rho n) P_2(y_{2,i})} + O\Big( \frac{\log n}{ n^{\kappa_n}}\Big).\label{Eq:LD taylor j<C}
\end{align} \hrulefill
\end{figure*}
Similarly, for any $j\in[ C+n^{\kappa_n},n]$, we have the Taylor expansion of $\LD(\fndot)$ around $\bT_j^\rho=(\tilPjCright,P_2,P_1,P_2)$ for any $(\tilx_1^{C+\rho n} , x_{C+1}^j, \tilx_2^{n-j+\rho n}, y_1^{N-\rho n}, y_2^{N-\rho n})\in\calB(P_1)\times  \calB(P_2)\times \calB(P_2)\times \calB(P_1)\times  \calB(P_2)$, given by \eqref{Eq:LD taylor j>C}.
\begin{figure*}
\begin{align}
	&\LD\Big(\bT_j^\rho,\frac{j+\rho n}{n+2\rho n} ,\frac{r-\rho}{1+2\rho}\Big)\nn\\
	\label{Eq:LD taylor j>C}&=\frac{1}{n+2\rho n}\bigg(\sum_{i\in[N-\rho n+1:N]}\log\frac{(j+N)\tilPjCright(y_{1,i})}{(j+\rho n)\tilPjCright(y_{1,i})+(N-\rho n)P_1(y_{1,i})}+\sum_{i\in[j]}\log\frac{(j+N)\tilPjCright(x_i)}{(j+\rho n)\tilPjCright(x_i)+(N-\rho n)P_1(x_i)}\bigg) \nn\\*
	&\qquad +\frac{1}{n+2\rho n}\sum_{i\in[N-\rho n]}\log\frac{(j+N)P_1(y_{1,i})}{(j+\rho n)\tilPjCright(y_{1,i})+(N-\rho n)P_1(y_{1,i})}+O\bigg(\frac{\log n}{n^{\kappa_n}}\bigg). 
\end{align}\hrulefill
\end{figure*}

Recall $\bP_j^-=(P_1,\tilPjCleft,P_1,P_2)$.  Let $c'$ be some positive constant. Using \eqref{Eq:LD taylor j<C},
for any $i\leq C$ and $j\in[i-n^{\kappa_n}]$, we have \eqref{Eq:lemma 4, atypical proba}--\eqref{Eq:LD(C-delta-1) min},
\begin{figure*}
\begin{align}
	&\bbP_C\bigg\{\LD\Big(\bT_j^\rho,\frac{j+\rho n}{n+2\rho n} ,\frac{r-\rho}{1+2\rho} \Big)<\LD\Big(\bT_{i}^\rho,\frac{i+\rho n}{n+2\rho n} ,\frac{r-\rho}{1+2\rho} \Big) \bigg\}\nn\\
	&\le  \bbP_C\bigg\{\frac{1}{n+2\rho n}\bigg(\sum_{k\in[i+1:n]}\log\frac{(n-i+N)\tilP_{i}^-(X_k)}{(n-i+\rho n)\tilP_{i}^-(X_k)+(N-\rho n)P_2(X_k)} \nn\\
	&\qquad +\sum_{k\in[N-\rho n+1:N]}\log\frac{(n-i+N)\tilP_{i}^-(Y_{2,k})}{(n-i+\rho n)\tilP_{i}^-(Y_{2,k})+(N-\rho n)P_2(Y_{2,k})}\bigg)  \nn\\*
	&\qquad +\frac{1}{n+2\rho n}\sum_{k\in[N-\rho n]}\log\frac{(n-i+N)P_2(Y_{2,k})}{(n-i+\rho n)\tilP_{i}^-(Y_{2,k})+(N-\rho n)P_2(Y_{2,k})}\nn\\
	&\qquad-\frac{1}{n+2\rho n}\bigg(\sum_{k\in[j+1:n]}\log\frac{(n-j+N)\tilPjCleft(X_k)}{(n-j+\rho n)\tilPjCleft(X_k)+(N-\rho n)P_2(X_k)} \nn\\
	&\qquad +\sum_{k\in[N-\rho n+1:N]}\log\frac{(n-j+N)\tilPjCleft(Y_{2,k})}{(n-j+\rho n)\tilPjCleft(Y_{2,k})+(N-\rho n)P_2(Y_{2,k})}\bigg)  \nn\\*
	&\qquad-\frac{1}{n+2\rho n}\sum_{k\in[N-\rho n]}\log\frac{(n-j+N)P_2(Y_{2,k})}{(n-j+\rho n)\tilPjCleft(Y_{2,k})+(N-\rho n)P_2(Y_{2,k})}>O\Big(\frac{\log n}{n^{\kappa_n}} \Big) \bigg\} +O\bigg(\frac{1}{n^{2\kappa_n}} \bigg) \label{Eq:lemma 4, atypical proba} \\
	\label{Eq: LD(j)>LD(C border)} &\leq\exp\bigg\{\frac{-2(n+2\rho n)^2\big(\LD\big(\bP_j^-,\frac{j+\rho n}{n+2\rho n} ,\frac{r-\rho}{1+2\rho}\big)-\LD\big(\bP_{i}^-,\frac{i+\rho n}{n+2\rho n} ,\frac{r-\rho}{1+2\rho}\big)+O(\frac{\log n}{n^{\kappa_n}}  ) \big)^2}{(2n+2N-j-i)(M_1(\rho)-M_2(\rho))^2} \bigg\}+ O\bigg(\frac{1}{n^{3/2}} \bigg)\\
	\label{Eq:LDj-LDC taylor}&\le  \exp\big\{-c' \, n^{2\kappa_n-1} \big\} +O\bigg(\frac{1}{n^{3/2}} \bigg) \\
	\label{Eq:LD(C-delta-1) min}&= O\bigg(\frac{1}{n^{3/2}} \bigg).
\end{align} \hrulefill
\end{figure*} 
where \eqref{Eq:lemma 4, atypical proba} follows from  \eqref{eqn:atyp}, \eqref{Eq: LD(j)>LD(C border)} follows from Hoeffding's inequality and $M_1(\rho) \ne M_2(\rho)$ are two constants depending only on $(\rho, r, P_1, P_2)$ such that for any $j\in[1,C-n^{\kappa_n}]$,
\begin{align}
&M_1(\rho)\leq \log\frac{(n-j+N)\tilPjCleft(X)}{(n-j+\rho n)\tilPjCleft(X)+(N-\rho n)P_2(X)}, \nn\\
& \log\frac{(n-j+N)P_2(X)}{(n-j+\rho n)\tilPjCleft(X)+(N-\rho n)P_2(X)} \leq M_2(\rho),~\mbox{a.s.,}
\end{align} 
\eqref{Eq:LDj-LDC taylor} follows since $\LD(\bP_j^-,\frac{j+\rho n}{n+2\rho n} ,\frac{r-\rho}{1+2\rho})$ is strictly decreasing in $j$ and  its Taylor expansion of the first argument around $\bP_{C_{\Delta}^-}^-$ is
\begin{align}
	&\LD \Big(\bP_{i-n^{\kappa_n}}^-,\frac{i-n^{\kappa_n}+\rho n}{n+2\rho n} ,\frac{r-\rho}{1+2\rho} \Big) \nn\\
	&=\frac{r-\rho}{1+2\rho}\mathrm{GJS}\bigg(\tilP_{i-n^{\kappa_n}}^-,P_2,\frac{n-i+n^{\kappa_n}+\rho n}{(r-\rho)n}\bigg)\\*
	&=\frac{r-\rho}{1+2\rho}\mathrm{GJS}\bigg(\tilP_{i}^-,P_2,\frac{n-i+\rho n}{(r-\rho)n}\bigg)+\Theta(n^{\kappa_n-1})\\*
	&=\LD\Big(\bP_{i}^-,\frac{i+\rho n}{n+2\rho n},\frac{r-\rho}{1+2\rho}\Big)+	\Theta(n^{\kappa_n-1}),
\end{align}
and \eqref{Eq:LD(C-delta-1) min} follows since
\begin{align}
	\!\!\!\exp\big\{-c'\,  n^{2\kappa_n-1} \big\} \!=\exp\big\{-c' \, n^{\frac{1}{2}+\frac{\log\log n}{\log n}} \big\} \!<\frac{1}{n^{3/2}},
\end{align}
for $n$ sufficiently large. 

Similarly, using \eqref{Eq:LD taylor j>C}, for any $i> C$ and $j\in[i+n^{\kappa_n}:n]$, we also have
\begin{align}
	&\bbP_C\bigg\{\LD\Big(\bT_j^\rho,\frac{j+\rho n}{n+2\rho n} ,\frac{r-\rho}{1+2\rho} \Big)<\LD\Big(\bT_{i}^\rho,\frac{i+\rho n}{n+2\rho n} ,\frac{r-\rho}{1+2\rho} \Big) \bigg\} \nn\\
	&=O\bigg(\frac{1}{n^{3/2}}\bigg).
\end{align}
This completes the proof of Lemma \ref{lem:L(j)<L(i)}.

\subsection{Proof of Lemma \ref{lem:I_C close to i_C}}\label{pf:pf of lem:I_C close to i_C}

Recall $\Gamma_n(i,\Delta_n)$ defined in \eqref{Eq:Gamma} for any $i\in[n]$. Given any true change-point $C\in[n]$, 
we have that for any $i\in[C\pm\Delta_n]$, using the union bound over all $O(n)$ values of $j \notin [i\pm(\Delta_n+n^{\kappa_n})]$ and Lemma \ref{lem:L(j)<L(i)},
\begin{align}
	&\bbP_C\bigg\{\argmin_{j\notin[i\pm\Delta_n]}\LD\Big(\bT_{j}^\rho,\frac{j+\rho n}{n+2\rho n} ,\frac{r-\rho}{1+2\rho} \Big)\notin\Gamma_n(i,\Delta_n+n^{\kappa_n})\bigg\} \nn\\
	&=O\bigg(\frac{1}{\sqrt{n}}\bigg) \label{Eq:i in C pm Delta, j* at border}.
\end{align}
Similarly, for any $i<C-\Delta_n$, using the union bound over all $O(n)$ values of $j \in[i-\Delta_n-1]$, we have
\begin{align}
	&\bbP_C\bigg\{\min_{j\in[i-\Delta_n-1]}\LD\Big(\bT_{j}^\rho,\frac{j+\rho n}{n+2\rho n} ,\frac{r-\rho}{1+2\rho} \Big) \nn\\
	&\qquad < \LD\Big(\bT_{i+\Delta_n+1}^\rho,\frac{i+\Delta_n+1+\rho n}{n+2\rho n} ,\frac{r-\rho}{1+2\rho} \Big) \bigg\}\nn\\
	&=O\bigg(\frac{1}{\sqrt{n}}\bigg) \label{Eq: L(i-Delta-1) > L(i+Delta+1)}
\end{align}
and also
\begin{align}
	&\bbP_C\bigg\{\argmin_{j\in[i+\Delta_n+1:n]}\LD\Big(\bT_{j}^\rho,\frac{j+\rho n}{n+2\rho n} ,\frac{r-\rho}{1+2\rho} \Big)\notin[C\pm n^{\kappa_n}] \bigg\} \nn\\
	&=O\bigg(\frac{1}{\sqrt{n}}\bigg) \label{Eq:j^*>i+Delta+1, close to C}.
\end{align}
Thus, combining \eqref{Eq: L(i-Delta-1) > L(i+Delta+1)} and \eqref{Eq:j^*>i+Delta+1, close to C}, we have  for any $i<C-\Delta_n$
\begin{align}
	&\bbP_C\bigg\{\argmin_{j\notin[i\pm\Delta]}\LD\Big(\bT_{j}^\rho,\frac{j+\rho n}{n+2\rho n} ,\frac{r-\rho}{1+2\rho} \Big)\notin[C\pm n^{\kappa_n}]\bigg\}\\
	&\leq \bbP_C\bigg\{\argmin_{j\in[i+\Delta_n+1:n]}\LD\Big(\bT_{j}^\rho,\frac{j+\rho n}{n+2\rho n} ,\frac{r-\rho}{1+2\rho} \Big)\notin[C\pm n^{\kappa_n}] \bigg\} \nn\\
	&\quad +O\bigg(\frac{1}{\sqrt{n}}\bigg) \label{Eq: use L(i-Delta-1) > L(i+Delta+1)}\\
	&=O\bigg(\frac{1}{\sqrt{n}}\bigg) \label{Eq:i<C-Delta, j^* close to C},
\end{align}
where \eqref{Eq: use L(i-Delta-1) > L(i+Delta+1)} follows from \eqref{Eq: L(i-Delta-1) > L(i+Delta+1)}. In a completely symmetric manner, \eqref{Eq:i<C-Delta, j^* close to C} also holds for $i>C+\Delta_n$.

We can then show that $I_\rho^*\in [C\pm\Delta_n]$ with high probability for $n$ large enough:
\begin{align}
	&\bbP_C\big\{I_\rho^*\notin [C\pm\Delta_n] \big\} \nn\\
	&=\bbP_C\bigg\{\max_{i\in[C\pm\Delta_n]}\min_{j\notin[i\pm\Delta_n]}\LD\Big(\bT_{j}^\rho,\frac{j+\rho n}{n+2\rho n} ,\frac{r-\rho}{1+2\rho} \Big) \nn\\
	&\qquad \quad <\max_{i\notin[C\pm\Delta_n]}\min_{j\notin[i\pm\Delta_n]}\LD\Big(\bT_{j}^\rho,\frac{j+\rho n}{n+2\rho n} ,\frac{r-\rho}{1+2\rho} \Big) \bigg\}\\
	&\leq \bbP_C\bigg\{\min_{j\notin[C\pm\Delta_n]}\LD\Big(\bT_{j}^\rho,\frac{j+\rho n}{n+2\rho n} ,\frac{r-\rho}{1+2\rho} \Big) \nn\\
	&\qquad \quad <\max_{i\notin[C\pm\Delta_n]}\min_{j\notin[i\pm\Delta_n]}\LD\Big(\bT_{j}^\rho,\frac{j+\rho n}{n+2\rho n} ,\frac{r-\rho}{1+2\rho} \Big) \bigg\}\\
	&\leq \bbP_C\bigg\{\min_{j\notin[C\pm\Delta_n]}\LD\Big(\bT_{j}^\rho,\frac{j+\rho n}{n+2\rho n} ,\frac{r-\rho}{1+2\rho} \Big) \nn\\
	&\qquad \quad <\min_{j\in[C\pm n^{\kappa_n}]}\LD\Big(\bT_{j}^\rho,\frac{j+\rho n}{n+2\rho n} ,\frac{r-\rho}{1+2\rho} \Big) \bigg\}+O\bigg(\frac{1}{\sqrt{n}}\bigg) \label{Eq: use i<C-Delta, j^* close to C}\\
	&=O\bigg(\frac{1}{\sqrt{n}}\bigg) \label{Eq: I^* not in C pm Delta},
\end{align}
where \eqref{Eq: use i<C-Delta, j^* close to C} follows from \eqref{Eq:i<C-Delta, j^* close to C} and \eqref{Eq: I^* not in C pm Delta} follows from Lemma \ref{lem:L(j)<L(i)}. Note that for any $i\in[C\pm\Delta_n]$, $i-\Delta_n\leq C$ and $i+\Delta_n\geq C$.

Next, we can further show that $I_\rho^*$ is close to $i_\rho(C)$ with high probability for $n$ large enough.
\begin{align}
	&\bbP_{C}\big\{I_\rho^*\notin[i_\rho(C)\pm n^{\kappa_n}] \big\} \nn\\
	&=\bbP_C\bigg\{\max_{i\in[n]}\min_{j\notin[i\pm\Delta_n]}\LD\Big(\bT_{j}^\rho,\frac{j+\rho n}{n+2\rho n} ,\frac{r-\rho}{1+2\rho} \Big) \nn\\
	&\qquad \quad >\max_{i'\in[i_\rho(C)\pm n^{\kappa_n}]}\min_{j\notin[i'\pm\Delta_n]}\LD\Big(\bT_{j}^\rho,\frac{j+\rho n}{n+2\rho n} ,\frac{r-\rho}{1+2\rho} \Big) \bigg\}\\
	&\leq \bbP_C\bigg\{\max_{i\in[C\pm\Delta_n]\backslash [i_\rho(C)\pm n^{\kappa_n}] }\min_{j\notin[i\pm\Delta_n]}\LD\Big(\bT_{j}^\rho,\frac{j\! +\! \rho n}{n\! +\! 2\rho n} ,\frac{r\! -\! \rho}{1\! +\! 2\rho} \Big) \nn\\
	&\qquad \quad  >\max_{i'\in[i_\rho(C)\pm n^{\kappa_n}]}\min_{j\notin[i'\pm\Delta_n]}\LD\Big(\bT_{j}^\rho,\frac{j+\rho n}{n+2\rho n} ,\frac{r-\rho}{1+2\rho} \Big) \bigg\} \nn\\
	&\quad +O\bigg(\frac{1}{\sqrt{n}}\bigg) \label{Eq:use I^* not in C pm Delta}\\
	&\leq \sum_{i\in[C\pm\Delta_n]\backslash [i_\rho(C)\pm n^{\kappa_n}]} \!\!\!\bbP_C\bigg\{\min_{j\notin[i\pm\Delta_n]}\LD\Big(\bT_{j}^\rho,\frac{j\! +\! \rho n}{n\! +\! 2\rho n} ,\frac{r\! -\! \rho}{1\! +\! 2\rho} \Big) \nn\\
	&\qquad \quad >\max_{i'\in[i_\rho(C)\pm n^{\kappa_n}]}\min_{j\notin[i'\pm\Delta_n]}\LD\Big(\bT_{j}^\rho,\frac{j\! +\! \rho n}{n\! +\! 2\rho n} ,\frac{r\! -\! \rho}{1\! +\! 2\rho} \Big) \bigg\} \nn\\
	&\quad +O\bigg(\frac{1}{\sqrt{n}}\bigg) \label{Eq: i not close to i_rho(C) union bd}\\
	&\leq \sum_{i\in[C\pm\Delta_n]\backslash [i_\rho(C)\pm n^{\kappa_n}]}\!\!\!\!\bbP_C\bigg\{\min_{j\notin[i\pm\Delta_n]}\LD\Big(\bT_{j}^\rho,\frac{j\! +\! \rho n}{n\! +\! 2\rho n} ,\frac{r\! -\! \rho}{1\! +\! 2\rho} \Big) \nn\\
	&\qquad \quad >\min_{j\notin[i_\rho(C)\pm\Delta_n]}\LD\Big(\bT_{j}^\rho,\frac{j+\rho n}{n+2\rho n} ,\frac{r-\rho}{1+2\rho} \Big) \bigg\} \nn\\
	&\quad +O\bigg(\frac{1}{\sqrt{n}}\bigg) \label{Eq: i=i_rho(C)}\\
	&\leq \sum_{\substack{i\in[i_\rho(C)+ n^{\kappa_n}+1:\\ C+\Delta_n]}}\!\!\!\!\bbP_C\bigg\{\LD\Big(\bT_{i-\Delta_n-1}^\rho,\frac{i-\Delta_n\! -\! 1\! +\! \rho n}{n\! +\! 2\rho n} ,\frac{r\! -\! \rho}{1\! +\! 2\rho} \Big) \nn\\
	&\qquad \quad >\min_{j\in\Gamma_n(i_\rho(C),\Delta_n+n^{\kappa_n})}\LD\Big(\bT_{j}^\rho,\frac{j+\rho n}{n+2\rho n} ,\frac{r-\rho}{1+2\rho} \Big) \bigg\} \nn\\
	&\quad  +\sum_{\substack{i\in[C-\Delta_n: \\ i_\rho(C)- n^{\kappa_n}-1] }} \!\!\!\! \bbP_C\bigg\{\LD\Big(\bT_{i+\Delta_n+1}^\rho,\frac{i\! +\! \Delta_n\! +\! 1\! +\! \rho n}{n\! +\! 2\rho n} ,\frac{r\! -\! \rho}{1\! +\! 2\rho} \Big) \nn\\
	& \qquad \quad  >\min_{j\in\Gamma_n(i_\rho(C),\Delta_n+n^{\kappa_n})}\LD\Big(\bT_{j}^\rho,\frac{j\! +\! \rho n}{n\! +\! 2\rho n} ,\frac{r\! -\! \rho}{1\! +\! 2\rho} \Big) \bigg\} \nn\\
	&\quad  +O\bigg(\frac{1}{\sqrt{n}}\bigg) \label{Eq:middle step I^* close to i_rho},
\end{align}
where \eqref{Eq:use I^* not in C pm Delta} follows from \eqref{Eq: I^* not in C pm Delta}, \eqref{Eq: i not close to i_rho(C) union bd} follows from the union bound over all values of $i\in[C\pm\Delta_n]\backslash [i_\rho(C)\pm n^{\kappa_n}]$, \eqref{Eq: i=i_rho(C)} follows by letting $i'=i_\rho(C)$ and \eqref{Eq:middle step I^* close to i_rho} follows by choosing one of the values of $j\notin[i\pm\Delta_n]$ and from \eqref{Eq:i in C pm Delta, j* at border}.

Recall $\bP_j=\bP_j^-=(P_1,\tilPjCleft,P_1,P_2)$ for $j\leq C$, $\bP_j=\bP_j^+=(\tilPjCright,P_2,P_1,P_2)$ for $j>C$ and the definition of $J_\rho^*(i)$ for any $i\in[n]$ in \eqref{Eq:J^*}. Let $c$ be some positive constant. According to the Taylor expansions of $\LD(\bT_{j}^\rho,\frac{j+\rho n}{n+2\rho n} ,\frac{r-\rho}{1+2\rho})$ around $\bT_{j}^\rho=\bP_j$ (cf. \eqref{Eq:LD taylor j<C} and \eqref{Eq:LD taylor j>C}) and using the Hoeffding's inequality similar to \eqref{Eq: LD(j)>LD(C border)}, for any $i\in[i_\rho(C)+n^{\kappa_n}+1:C+\Delta_n]$, we have \eqref{Eq:Eq:L(i-Delta-1)>L(i_rho border) begin}--\eqref{Eq:L(i-Delta-1)>L(i_rho border)} on the top of next page,
\begin{figure*}
\begin{align}
	&\bbP_C\bigg\{\LD\Big(\bT_{i-\Delta_n-1}^\rho,\frac{i-\Delta_n-1+\rho n}{n+2\rho n} ,\frac{r-\rho}{1+2\rho} \Big) > \min_{j\in\Gamma_n(i_\rho(C),\Delta_n+n^{\kappa_n})}\LD\Big(\bT_{j}^\rho,\frac{j+\rho n}{n+2\rho n} ,\frac{r-\rho}{1+2\rho} \Big) \bigg\} \label{Eq:Eq:L(i-Delta-1)>L(i_rho border) begin}\\
	&\leq \exp\bigg\{\frac{-c n\big(\mathbb{E}_{J^*(i_\rho(C))}\big[\LD\big(\bP_{J^*(i_\rho(C))},\frac{J^*(i_\rho(C))+\rho n}{n+2\rho n} ,\frac{r-\rho}{1+2\rho}\big) \big] -\LD\big(\bP_{i-\Delta_n-1}^-,\frac{i-\Delta_n-1+\rho n}{n+2\rho n} ,\frac{r-\rho}{1+2\rho}\big)+O(\frac{\log n}{n^{\kappa_n}}  ) \big)^2}{(M_1(\rho)-M_2(\rho))^2} \bigg\} \nn\\
	&\qquad + O\bigg(\frac{1}{n^{3/2}} \bigg)\\
	&=\! \exp\bigg\{\frac{-c n\big(\LD\big(\bP^-_{i_\rho(C)-\Delta_n-1},\frac{i_\rho(C)-\Delta_n-1+\rho n}{n+2\rho n} ,\frac{r-\rho}{1+2\rho}\big) \! -\! \LD\big(\bP_{i-\Delta_n-1}^-,\frac{i-\Delta_n-1+\rho n}{n+2\rho n} ,\frac{r-\rho}{1+2\rho}\big)\! +\! O(n^{\kappa_n-1})  \big)^2}{(M_1(\rho)-M_2(\rho))^2} \bigg\}\! +\!  O\bigg(\frac{1}{n^{3/2}} \bigg)\!  \label{Eq:Taylor E[L(P_J^*)]}\\
	&=\exp\bigg\{\frac{-c n\big(\Omega(n^{\kappa_n-1})  \big)^2}{(M_1(\rho)-M_2(\rho))^2} \bigg\}+ O\bigg(\frac{1}{n^{3/2}} \bigg) \label{Eq:Taylor L(i-Delta-1)}\\
	&=O\bigg(\frac{1}{n^{3/2}} \bigg) \label{Eq:L(i-Delta-1)>L(i_rho border)},
\end{align}\hrulefill
\end{figure*}
where \eqref{Eq:Taylor E[L(P_J^*)]} follows from \eqref{Eq: i_rho(C) GJS equal} and from the Taylor expansions of $\LD(\bP_{J^*(i_\rho(C))},\frac{J^*(i_\rho(C))+\rho n}{n+2\rho n} ,\frac{r-\rho}{1+2\rho})$ around $J^*(i_\rho(C))=i_\rho(C)-\Delta_n-1$ for $J^*(i_\rho(C))\in\Gamma_{n}^\rmL(i_\rho(C),\Delta_n+n^{\kappa_n})$ and around $J^*(i_\rho(C))=i_\rho(C)+\Delta_n+1$ for $J^*(i_\rho(C))\in\Gamma_{n}^\rmR(i_\rho(C),\Delta_n+n^{\kappa_n})$:
\begin{align}
	&\mathbb{E}_{J^*(i_\rho(C))}\left[\LD\Big(\bP_{J^*(i_\rho(C))},\frac{J^*(i_\rho(C))+\rho n}{n+2\rho n} ,\frac{r-\rho}{1+2\rho}\Big) \right] \nn\\
	&=\LD\Big(\bP^-_{i_\rho(C)-\Delta_n-1},\frac{i_\rho(C)-\Delta_n-1+\rho n}{n+2\rho n} ,\frac{r-\rho}{1+2\rho}\Big)\nn\\
	&\quad +O(n^{\kappa_n-1}),
\end{align}
and \eqref{Eq:Taylor L(i-Delta-1)} follows since $\LD(\bP_j^-,\frac{j+\rho n}{n+2\rho n},\frac{r-\rho}{1+2\rho})$ is strictly decreasing in $j$ (a fact that can be proved similarly as that of Lemma~\ref{Lem:Gmin increase}) and from the Taylor expansions of $\LD(\bP_{i-\Delta_n-1}^-,\frac{i-\Delta_n-1+\rho n}{n+2\rho n} ,\frac{r-\rho}{1+2\rho})$ around $i=i_\rho(C)$. 

In a completely symmetric manner, we have for any $i\in[C-\Delta_n:i_\rho(C)- n^{\kappa_n}-1]$,
\begin{align}
	&\bbP_C\bigg\{\LD\Big(\bT_{i+\Delta_n+1}^\rho,\frac{i+\Delta_n+1+\rho n}{n+2\rho n} ,\frac{r-\rho}{1+2\rho} \Big) \nn\\
	&\qquad >\min_{j\in\Gamma_n(i_\rho(C),\Delta_n+n^{\kappa_n})}\LD\Big(\bT_{j}^\rho,\frac{j+\rho n}{n+2\rho n} ,\frac{r-\rho}{1+2\rho} \Big) \bigg\} \nn\\
	&=O\bigg(\frac{1}{n^{3/2}} \bigg) \label{Eq:L(i+Delta+1)>L(i_rho border)}.
\end{align}
Finally, combining \eqref{Eq:L(i-Delta-1)>L(i_rho border)} and \eqref{Eq:L(i+Delta+1)>L(i_rho border)}, we can upper bound \eqref{Eq:middle step I^* close to i_rho} as follows:
\begin{align}
	\bbP_{C}\big\{I_\rho^*\notin[i_\rho(C)\pm n^{\kappa_n}] \big\}=O\bigg(\frac{1}{\sqrt{n}}\bigg).
\end{align}

\subsection{Proof of Lemma \ref{Lem:type based test optimal}}\label{proof of Lemma type based test optimal}

	

	For any $\rho\in(0,r)$ and any $i\in[n]$, recall  $\tilX_1^{i+\rho n}, \tilX_{2}^{n-i+\rho n}$ defined in \eqref{Eq:tilX_1} and \eqref{Eq:tilX_2}. Let $\calP_{i+\rho n,n-i+\rho n,2N-2\rho n}(\calX):=\calP_{i+\rho n}(\calX)\times\calP_{n-i+\rho n}(\calX)\times\calP_{N-\rho n}(\calX)^2$ and $\tilde{\bQ}_i^\rho=(Q_1,Q_2,\tilQ_1,\tilQ_2)\in\calP_{i+\rho n,n-i+\rho n,2N-2\rho n}(\calX)$. For any $\tilde{\bQ}_i^\rho$, we use $\tilT_{\tilde{\bQ}_i^\rho}^{n+2N}$ to denote the tuple of sequences for $(X^n,Y_1^N,Y_2^N)$ such that $\tilX_1^{i+\rho n}\in\calT^{i+\rho n}_{Q_1}$, $\tilX_{2}^{n-i+\rho n}\in\calT^{n-i+\rho n}_{Q_2}$, $Y_1^{N-\rho n}\in\calT_{\tilQ_1}^{N-\rho n}$ and $Y_2^{N-\rho n}\in\calT_{\tilQ_2}^{N-\rho n}$.
	
	
	Given any estimator $\gamma$, we define the following sets of test and training sequences:
	\begin{align}
	&\!\calA_k(\gamma)\!=\!\{(x^n,y_1^N,y_2^N)\!:\!\gamma(x^n,y_1^N,y_2^N)\!=\! k \}, ~\forall k\!\in\![n],\\*
	&\!\calA_{\rme}(\gamma)=\{(x^n,y_1^N,y_2^N):\gamma(x^n,y_1^N,y_2^N)=\rme \} \nn\\
	&\!\hspace{0.36in} =\bigg(\bigcup_{k\in[n]}\calA_k\bigg)^{\rmc}.
	\end{align}
	Fix any $\bm{\eta}\in[0,1]^{n}$. 
	Given any $j\in[n]$ and any tuple of types $\tilde{\bQ}_j^\rho$, we can construct the following type-based estimator $\gamma_{\rho,j}^{\mathrm{type}}$:
	\begin{itemize}
		\item if $|\calA_k(\gamma)\cap\calT_{\tilde{\bQ}_j^\rho}^{n+2N}|\geq\eta_{k}|\calT_{\tilde{\bQ}_j^\rho}^{n+2N}|$ and $|\calA_l(\gamma)\cap\calT_{\tilde{\bQ}_j^\rho}^{n+2N}|<\eta_{l}|\calT_{\tilde{\bQ}_j^\rho}^{n+2N}|$ for all  $l<k$,  we define $\gamma_{\rho,j}^{\mathrm{type}}(\tilde{\bQ}_j^\rho):=k$;
		\item if $|\calA_k(\gamma)\cap\calT_{\tilde{\bQ}_j^\rho}^{n+2N}|<\eta_{k}|\calT_{\tilde{\bQ}_j^\rho}^{n+2N}|$ for all $k\in[n]$, we define $\gamma_{\rho,j}^{\mathrm{type}}(\tilde{\bQ}_j^\rho):=\rme$.
	\end{itemize}
	Then for any $j\in[n]$, we have
	\begin{align}
	&\bbP_j\big\{\gamma(X^n,Y_1^N,Y_2^N)\notin  [j\pm\Delta]\cup\{\rme\}  \big\} \nn\\
	&=\sum_{k\notin \{[j\pm\Delta]\cup\{\rme\} \}}\bbP_j\{\calA_k\}\\
	&\geq \sum_{k\notin\{[j\pm\Delta]\cup\{\rme\} \}} \nn\\
	&  \sum_{\substack{\tilde{\bQ}_j^\rho:\\|\calA_k(\gamma)\cap\calT_{\tilde{\bQ}_j^\rho}^{n+2N}|\geq\eta_{k}|\calT_{\tilde{\bQ}_j^\rho}^{n+2N}|,\\|\calA_l(\gamma)\cap\calT_{\tilde{\bQ}_j^\rho}^{n+2N}|<\eta_{l}|\calT_{\tilde{\bQ}_j^\rho}^{n+2N}|, \forall l<k }}\bbP_j\{\calA_k(\gamma)\cap\calT_{\tilde{\bQ}_j^\rho}^{n+2N} \}\\
	&\geq \sum_{k\notin\{[j\pm\Delta]\cup\{\rme\} \}}\eta_{k} \nn\\
	&\quad  \sum_{\substack{\tilde{\bQ}_j^\rho:\\|\calA_k(\gamma)\cap\calT_{\tilde{\bQ}_j^\rho}^{n+2N}|\geq\eta_{k}|\calT_{\tilde{\bQ}_j^\rho}^{n+2N}|,\\|\calA_l(\gamma)\cap\calT_{\tilde{\bQ}_j^\rho}^{n+2N}|<\eta_{l}|\calT_{\tilde{\bQ}_j^\rho}^{n+2N}|, \forall l<k }} \bbP_j\{\calT_{\tilde{\bQ}_j^\rho}^{n+2N} \}\\
	&= \sum_{k\notin\{[j\pm\Delta]\cup\{\rme\} \}}\eta_{k} \bbP_j\{\gamma^{\mathrm{type}}(\tilde{\bQ}_j^\rho)=k \}\\
	&\geq \eta_{\min}\bbP_j\big\{\gamma^{\mathrm{type}}(\tilde{\bQ}_j^\rho)\notin [j\pm\Delta]\cup\{\rme\} \big\}, \label{Eq:any dec err lb}
	\end{align}
	and
	\begin{align}
	&\bbP_j\{\gamma(X^n,Y_1^N,Y_2^N)=e  \}=\bbP_j\{(\cup_{k\in[n]}\calA_k)^{\rmc} \}\\
	&\geq \sum_{\substack{\tilde{\bQ}_j^\rho:|\calA_k(\gamma)\cap\calT_{\tilde{\bQ}_j^\rho}^{n+2N}|\\<\eta_{k}|\calT_{\tilde{\bQ}_j^\rho}^{n+2N}|,\forall k\in[n]} }\bbP_j\{(\cup_{k\in[n]}\calA_k)^{\rmc}\cap\calT_{\tilde{\bQ}_j^\rho}^{n+2N} \}\\
	&=\sum_{\substack{\tilde{\bQ}_j^\rho:|\calA_k(\gamma)\cap\calT_{\tilde{\bQ}_j^\rho}^{n+2N}|\\<\eta_{k}|\calT_{\tilde{\bQ}_j^\rho}^{n+2N}|,\forall k\in[n]}}\Big[\bbP_j\{\calT_{\tilde{\bQ}_j^\rho}^{n+2N}\} \nn\\
	&\qquad \qquad \qquad \qquad -\bbP_j\{(\cup_{k\in[n]}\calA_k)\cap\calT_{\tilde{\bQ}_j^\rho}^{n+2N} \} \Big]\\
	&=\sum_{\substack{\tilde{\bQ}_j^\rho:|\calA_k(\gamma)\cap\calT_{\tilde{\bQ}_j^\rho}^{n+2N}| \\ <\eta_{k}|\calT_{\tilde{\bQ}_j^\rho}^{n+2N}|,\forall k\in[n] }}\Big[\bbP_j\{\calT_{\tilde{\bQ}_j^\rho}^{n+2N}\}-\sum_{k=1}^{n}\bbP_j\{\calA_k\cap\calT_{\tilde{\bQ}_j^\rho}^{n+2N} \} \Big]\\
	&\geq (1-\eta_{\mathrm{sum}})\bbP_j\{\gamma_{\rho,j}^{\mathrm{type}}(\tilde{\bQ}_j^\rho)=\rme \}. \label{Eq:any dec erasure lb}
	\end{align}

\subsection{Proof of Lemma \ref{Lem:proposed test type-based optimal}}\label{proof of Lem our test optimal}

	
	For any $\rho\in(0,r)$ and any $i\in[n]$,  recall the tuple of types $\bT_i^\rho$ defined under \eqref{Eq:tilX_2}. Given any $C\in[n]$ and any type-based estimator $\gamma_{\rho,C}^{\mathrm{type}}$, let
	\begin{align}
		\Omega_i^\rho:=&\{(x^n,y_1^N,y_2^N):\gamma_{\rho,C}^{\mathrm{type}}(\bT_C^\rho)=i\},\\
		\Omega_\rme^\rho:=&\big(\medmath{ \bigcup_{i\in[n]}}\Omega_i^\rho\big)^\rmc=\{(x^n,y_1^N,y_2^N):\gamma_{\rho,C}^{\mathrm{type}}(\bT_C^\rho)=\rme\},
	\end{align} 
	and 
	\begin{align}
		\tilde{\Lambda}_i^\rho:=&\bigg\{(x^n,y_1^N,y_2^N): \nn\\
		& (1+2\rho)\bigg(\LD\Big(\bT_i^\rho,\frac{i +\rho n}{n+2\rho n},\frac{r-\rho}{1+2\rho} \Big)+\delta_n \bigg)> \lambda \bigg\}.
	\end{align}

	Suppose $\gamma_{\rho,C}^{\mathrm{type}}$ satisfies for all pair of $(\tilP_1,\tilP_2)\in\calP(\calX)^2$ and all $\tilde{\bQ}_C^\rho\in\calP_{C+\rho n}(\calX)\times\calP_{n-C+\rho n}(\calX)\times\calP_{N-\rho n}(\calX)^2$ such that $\calT^{n+2N}_{\tilde{\bQ}_C^\rho}\subset \bigcup_{k\notin[C\pm\Delta]}\Omega_k^\rho$, 
	\begin{align}
		&\exp(-n\lambda ) \nn\\
		&\geq\tilde{\bbP}_C\{\gamma_{\rho,C}^{\mathrm{type}}(\bT_C^\rho)\notin  [C\pm\Delta]\cup\{\rme\}  \} \nn\\
		&\geq \tilde{\bbP}_{C}\big\{\calT^{n+2N}_{\tilde{\bQ}_C^\rho}\big\}\\
		&\geq \exp\Big\{-(C+\rho n)D(Q_1^{C+\rho n}\|\tilP_1)\nn\\
		&\qquad -(n-C+\rho n)D(Q_2^{n-C+\rho n}\|\tilP_2) \nn\\*
		&\qquad-(N-\rho n)D(\tilQ_1\|\tilP_1)-(N-\rho n)D(\tilQ_2\|\tilP_2) \nn\\
		&\qquad  -|\calX|\log \big((C+\rho n+1)(n-C+\rho n+1)\nn\\
		&\qquad \cdot(N-\rho n+1)^2 \big) \Big\}.
	\end{align}
	If we choose $\tilP_1=\frac{(C+\rho n)Q_1^{C+\rho n}+(N-\rho n)\tilQ_1}{C+N}$ and $\tilP_2=\frac{(n-C+\rho n)Q_2^{n-C+\rho n}+(N-\rho n)\tilQ_2}{n-C+N}$, then we have
	\begin{align}
		&\exp(-n\lambda) \nn\\
		&\geq\tilde{\bbP}_C\{\gamma_{\rho,C}^{\mathrm{type}}(\bT_C^\rho)\notin  [C\pm\Delta]\cup\{\rme\}  \} \nn\\
		&\geq  \exp\bigg\{-(n+2\rho n)\bigg(\LD\Big(\tilde{\bQ}_C^\rho,\frac{C+\rho n}{n+2\rho n},\frac{r-\rho}{1+2\rho}\Big) \nn\\
		&\qquad +\frac{|\calX|}{n+2\rho n}\log\big((C+\rho n+1)(n-C+\rho n+1)\nn\\
		&\qquad \cdot (N-\rho n+1)^2 \big)\bigg) \bigg\}\\
		&> \exp\bigg\{-(n+2\rho n) \bigg(\LD\Big(\tilde{\bQ}_C^\rho,\frac{C+\rho n}{n+2\rho n},\frac{r-\rho}{1+2\rho}\Big)+\delta_n \bigg) \bigg\}.
	\end{align}
	Therefore, for any $(x^n,y_1^N,y_2^N)\in \bigcup_{k\notin[C\pm\Delta]}\Omega_k$, we have
	\begin{align}
		(1+2\rho) \bigg(\LD\Big(\bT_C^\rho,\frac{C+\rho n}{n+2\rho n},\frac{r-\rho}{1+2\rho}\Big)+\delta_n \bigg)>\lambda,
	\end{align}
	which means $\bigcup_{k\notin[C\pm\Delta]}\Omega_k^\rho \subset \tilde{\Lambda}_C^\rho$.
	
	Let us further define for any $i\in[n]$
	\begin{align}
		\hat{\Lambda}_i^\rho:=&\medmath{\bigcap_{j\notin[i\pm\Delta]}} \tilde{\Lambda}_j^\rho\nn\\
		=&\bigg\{(x^n,y_1^N,y_2^N):\min_{j\notin [i\pm \Delta]}(1+2\rho) \nn\\
		&~ \times\bigg(\LD\Big(\bT_j^\rho,\frac{j+\rho n}{n+2\rho n},\frac{r-\rho}{1+2\rho}\Big)+\delta_n \bigg)> \lambda \bigg\}, 
	\end{align}
	and 
	\begin{align}
		\hat{\Lambda}_\rme^\rho&:=(\medmath{\bigcup_{i\in[n]}}\hat{\Lambda}_i^\rho )^\rmc \nn\\
		&= \bigg\{(x^n,y_1^N,y_2^N):  \min_{j\notin [i\pm \Delta]}(1+2\rho) \nn\\*
		& \times\!  \bigg(\LD\Big(\bT_j^\rho,\frac{j+\rho n}{n+2\rho n},\frac{r-\rho}{1+2\rho}\Big)\! +\! \delta_n\bigg)\! \leq\! \lambda, \forall i\! \in\! [n] \! \bigg\}.
	\end{align}
	Then we have for any $i\in[n]$
	\begin{align}
		\hat{\Lambda}_i^\rho=\medmath{\bigcap_{j\notin[i\pm\Delta]}}\tilde{\Lambda}_j^\rho\supset \medmath{\bigcap_{j\notin[i\pm\Delta]}}\medmath{\bigcup_{k\notin[j\pm\Delta]}}\Omega_k^\rho \supset \Omega_i^\rho.
	\end{align}
	Thus, we have
	\begin{align}
		\hat{\Lambda}_\rme^\rho=(\medmath{\bigcup_{i\in[n]}}\hat{\Lambda}_i^\rho )^\rmc \subset (\medmath{\bigcup_{i\in[n]}}\Omega_i^\rho )^\rmc=\Omega_\rme^\rho.
	\end{align}
	Equivalently, given any pair of $(P_1,P_2)\in\calP(\calX)^2$, we have
	\begin{align}
		\bbP_C\{\Omega_\rme^\rho \}\geq \bbP_C\{\hat{\Lambda}_\rme^\rho\}
	\end{align}
	and so, subsequently,
	\begin{align}
		&\bbP_C\{\gamma_{\rho,C}^{\mathrm{type}}(\bT_C^\rho)=\rme \}\nn\\
		&\geq \bbP_C \bigg\{ \min_{j\notin [i\pm \Delta]}(1+2\rho)\bigg(\LD \Big(\bT_j^\rho,\frac{j+\rho n}{n+2\rho n},\frac{r-\rho}{1+2\rho} \Big)+\delta_n \bigg) \nn\\*
		& \qquad \qquad \leq \lambda, \forall i\in[n] \bigg\}.
	\end{align}

\section*{Acknowledgements} 	
The authors would like to sincerely thank the reviewers and the Associate Editor Prof.\ Ioannis Kontoyiannis for many helpful comments to improve the presentation of the paper. In particular, one of the reviewers pointed out a subtle error in the initial version of the paper, which led to the improved achievability scheme in  \eqref{eqn:Si} and \eqref{Eq:estimator}. 
Comments from Dr.\ Lin Zhou and Dr.\ Anshoo Tandon are also gratefully acknowledged.

\bibliographystyle{IEEEtran}
\bibliography{ref.bib}

\begin{IEEEbiographynophoto}{Haiyun He} (S'18) is currently a Ph.D.\ student in the Department of Electrical and Computer Engineering (ECE) at the National University of Singapore (NUS). From Sep  2017 to Jul  2018, she was first a Research Assistant in ECE at NUS. She received the B.E.\ degree in Beihang University (BUAA) in 2016 and the M.Sc.\ (Electrical Engineering)\ degree in ECE from NUS in 2017. Her research interests include information theory, statistical learning and their applications.
\end{IEEEbiographynophoto}

\begin{IEEEbiographynophoto}{Qiaosheng Zhang}
 received the B.Eng.\ and Ph.D.\ degrees from the Department of Information Engineering, Chinese University of Hong Kong (CUHK) in 2015 and 2019, respectively. He is currently a Research Fellow with the Department of Electrical and Computer Engineering, National University of Singapore (NUS). His research interests include information theory and its applications to machine learning.
\end{IEEEbiographynophoto}

\begin{IEEEbiographynophoto}{Vincent Y.\ F.\ Tan} (S'07-M'11-SM'15) was born in Singapore in 1981. He received the B.A.\ and M.Eng.\ degrees in electrical and information science from Cambridge University in 2005, and the Ph.D.\ degree in electrical engineering and computer science (EECS) from the Massachusetts Institute of Technology (MIT) in 2011. He is currently a Dean’s Chair Associate Professor with the Department of Electrical and Computer Engineering and the Department of Mathematics, National University of Singapore (NUS). His research interests include information theory, machine learning, and statistical signal processing. 

Dr.\ Tan is a member of the IEEE Information Theory Society Board of Governors. He was an IEEE Information Theory Society Distinguished Lecturer in September 2018. He received the MIT EECS Jin-Au Kong Outstanding Doctoral Thesis Prize in 2011, the NUS Young Investigator Award in 2014, the Singapore National Research Foundation (NRF) Fellowship (Class of 2018), and the NUS Young Researcher Award in 2019. He is currently serving as an Associate Editor for the IEEE \textsc{Transactions on Signal Processing} and for the IEEE \textsc{Transactions on Information Theory}.
\end{IEEEbiographynophoto}

\end{document}